\documentclass[pdflatex,sn-basic]{sn-jnl}

\usepackage{graphicx}%
\usepackage{multirow}%
\usepackage{amsmath,amssymb,amsfonts}%
\usepackage{amsthm}%
\usepackage{mathrsfs}%
\usepackage[title]{appendix}%
\usepackage{xcolor}%
\usepackage{textcomp}%
\usepackage{manyfoot}%
\usepackage{booktabs}%
\usepackage{algorithm}%
\usepackage{algorithmicx}%
\usepackage{algpseudocode}%
\usepackage{listings}%

\usepackage{bm} 
\usepackage{setspace}
\usepackage{array}
\usepackage{adjustbox}
\usepackage{makecell}
\usepackage{tabularx}
\usepackage{rotating}
\usepackage{fancyhdr}

\fancypagestyle{firstpage}{
  \fancyhf{} 
  \rhead{\small Accepted for publication in \textit{Archives of Computational Methods in Engineering} in Feb 2024}
  \cfoot{\thepage}
}

\newtheorem{formulation}{Formulation}
\newtheorem{algor}{Algorithm}

\usepackage{geometry}
\begin{document}

\thispagestyle{firstpage}

\title[Diffusion-deformation theories for hydrogels]{A comparative analysis of transient finite-strain coupled diffusion-deformation theories for hydrogels}


\author*[1,2]{\fnm{Jorge-Humberto} \sur{Urrea-Quintero}}\email{jorge.urrea-quintero@tu-braunschweig.de}

\author*[3]{\fnm{Michele} \sur{Marino}}\email{m.marino@ing.uniroma2.it}

\author*[4]{\fnm{Thomas} \sur{Wick}}\email{wick@ifam.uni-hannover.de}

\author*[1]{\fnm{Udo} \sur{Nackenhorst}}\email{nackenhorst@ibnm.uni-hannover.de}

\affil[1]{\orgname{Leibniz University Hannover}, \orgdiv{IBNM - Institute of Mechanics and Computational Mechanics}, \orgaddress{\street{Appelstra\ss e 9a}, \city{Hannover}, \postcode{30167}, \state{Lower Saxony}, \country{Germany}}}

\affil[2]{\orgname{Technische Universtität Braunschweig}, \orgdiv{iRMB - Institute for Computational Modeling in Civil Engineering}, \orgaddress{\street{Pockelsstr. 3}, \city{Braunschweig}, \postcode{38106}, \state{Lower Saxony}, \country{Germany}}}

\affil[3]{\orgdiv{Department of Civil Engineering and Computer Science Engineering}, \orgname{University of Rome Tor Vergata}, \orgaddress{\street{Via del Politecnico 1}, \city{Rome}, \postcode{00133}, \country{Italy}}}

\affil[4]{\orgname{Leibniz University Hannover}, \orgdiv{Institut für Angewandte Mathematik}, \orgaddress{\street{Welfengarten 1}, \city{Hannover}, \postcode{30167}, \state{Lower Saxony}, \country{Germany}}}


\abstract{This work presents a comparative review and classification between some well-known thermodynamically consistent models of hydrogel behavior in a large deformation setting, specifically focusing on solvent absorption/desorption and its impact on mechanical deformation and network swelling. The proposed discussion addresses formulation aspects, general mathematical classification of the governing equations, and numerical implementation issues based on the finite element method. The theories are presented in a unified framework demonstrating that, despite not being evident in some cases, all of them follow equivalent thermodynamic arguments.
A detailed numerical analysis is carried out where Taylor-Hood elements are employed in the spatial discretization to satisfy the inf-sup condition and to prevent spurious numerical oscillations. 
The resulting discrete problems are solved using the FEniCS platform through consistent variational formulations, employing both monolithic and staggered approaches. 
We conduct benchmark tests on various hydrogel structures, demonstrating that major differences arise from the chosen volumetric response of the hydrogel. 
The significance of this choice is frequently underestimated in the state-of-the-art literature but has been shown to have substantial implications on the resulting hydrogel behavior.}




\maketitle

\newpage
\section{Introduction}\label{sec1}

Hydrogels, a class of versatile soft materials with the unique ability to absorb and retain fluid within their three-dimensional network structures, have gained widespread attention in recent years across various industrial applications. Their diverse uses include serving as drug carriers in biomedical applications \citep{Sun2013Alginate,Kabir2018Hydrogels}, absorbents of pollutants in agriculture \citep{Kabir2018Hydrogels}, and smart sensors or actuators in engineering \citep{Chaterji2007SmartHydrogels,Lin2010SensorHydrogel}. These properties can be fine-tuned by manipulating chemical composition, crosslinking density, and fluid content \citep{Caccavo2018hydrogels}. In this work, we aim to provide an in-depth examination of diffusion-deformation hydrogel theories and the implications of selecting underlying constitutive theory in the large deformation setting.

Several comprehensive reviews have emerged in the study of hydrogels, discussing topics ranging from hydrogel deformation theories to the microstructural impact on mechanical behaviors. 
\cite{Liu2015RevHydrogels} provided an exhaustive review of hydrogel deformation theories, blending both theoretical analyses and practical applications, emphasizing mechanics' role. \cite{Huang2020GelsModels_Review} delve into advances in constitutive models for hydrogels and shape memory polymers (SMPs), categorizing six primary hydrogel types and highlighting potential hyperelastic model adaptations. Meanwhile, \cite{Lei2021RevHydrogels} offered insights into hydrogel network models, discussing the microstructural impact on their mechanical behaviors like swelling, elasticity, and fracture. This research underscores the potential synergy between network modeling and continuum mechanics in capturing hydrogel dynamics.

Modeling the diffusion-deformation process in hydrogels necessitates a comprehensive understanding of the material's behavior during swelling and mechanical deformation. This understanding hinges on a mathematical formulation that simultaneously accounts for the diffusion of fluid molecules within the hydrogel's polymer network and the corresponding mechanical deformations induced by swelling. Two primary approaches can be employed to describe these phenomena: mixture theories and macro-scale theories. Mixture theories, representing the porous medium as spatially superimposed interacting layers \citep{ehlers2002porous}, often involve a complex array of model parameters and constitutive choices, making them challenging to calibrate in practical applications. Consequently, this paper focuses on macro-scale poroelastic theories, which view the medium as a homogenous material characterized by a coupled deformation-diffusion response, as developed in the seminal works by Biot \citep[e.g.,][]{Biot1957}. Within this framework, the key equations, such as the mass balance and mechanical equilibrium equations, govern both the conservation of mass and the balance of forces within the hydrogel. An appropriate chemo-mechanical constitutive description for the macroscopic continuum completes the modeling framework. Constitutive laws couple the mechanical response of the polymer network (accounting, for example, for hyperelastic or viscoelastic effects) with changes in fluid distribution within the hydrogel mesh, describing mixing effects and the interaction between diffusion and deformation. 

Over the years, various models have been developed to describe coupled diffusion-deformation effects in elastomeric gels. These models couple a common general poroelastic framework with the effects of different physical fields, resulting in a wide range of stimuli-based responses encountered in various applications. 
Regarding the general poroelastic framework, for instance, \cite{Hong2008GelsTheory} and \cite{Zhang2009Gels} provided a continuum mechanics framework, as well as analytical and finite element solutions, of the coupled problem in the large deformation setting. \cite{Chester2010DiffDeform} provided a comprehensive thermodynamics-consistent formulation of the diffusion-deformation theory under isothermal conditions. In contrast, \cite{Lucantonio2013transientGels} benchmarked the diffusion-deformation theory against some experiments involving localized exposure of the gel boundary to a solvent, where large bending deformations appear during solvent absorption. 
With respect to more complex stimuli-based responses, \cite{Chester2011thermo} extended the theory and accounted for temperature effects as well. \cite{Chester2015Abaqus} summarized the main developments of the thermo-mechanically coupled theory for fluid permeation in elastomeric materials and provided an open-access Abaqus implementation to simulate hydrogel's response in 3D.

A shared mathematical characteristic of these models is the saddle point structure of the underlying variational formulation.
As a result of the finite element implementations, the Ladyzhenskaya-Babuska-Brezzi (LBB), i.e., discrete inf-sup, condition might be violated, requiring equilibrated inf-sup stable finite element spaces. Otherwise, oscillatory distributions of chemical potential as the primary variable controlling species diffusion arise. These challenges can be addressed in several ways. To name a few of these approaches, in \cite{Bouklas2015Nonlinear}, an in-depth analysis of Taylor-Hood elements (see, e.g., \cite{GiRa1986}) for gel modeling is provided, while \cite{Krischok2016MixedFEMGels} proposed an enhanced assumed strain (EAS) method for the coupled problem. \cite{Boger2017minimizationHydrogels} proposed a minimization formulation for the coupled diffusion-deformation problem of polymeric hydrogels at large strains and compared this variational framework to classical saddle point-based structures. Large volume changes with instability patterns in the presence of geometrical constraints were successfully modeled. \cite{Wang2018NumericsGels} developed a numerical platform to simulate the dynamic behaviors of responsive gels. 
The authors of this study addressed some of the challenges that were not previously resolved, such as how to handle time-dependent and coupled mass diffusion and deformation fields in a very short time, as well as numerical instability issues.

Overall, there is no consensus on formulating a constitutive model for hydrogels and its numerical implementation. This study emphasizes the essential building blocks found in common across various hydrogel theories, mainly focusing on the standard poroelastic description. 
It is unclear whether the underlying constitutive choices at the free energy level are equivalent.
The significance of fundamental choices in coupling diffusion and deformation is frequently overlooked. For example, such choices may relate to whether volume changes in the hydrogel are solely attributed to fluid content or involve elastic mechanisms as well. In the latter scenario, the detailed description of the volumetric response plays a crucial role. We aim to analyze the impact of such assumptions on the constitutive model.

Not only do these fundamental constitutive choices lead to differences in the observed physical response, but they also influence the feasibility and appropriateness of different numerical solutions. Some theories lend themselves to a monolithic implementation of the coupled problem, while others favor a staggered approach. Despite their importance, these considerations have been largely neglected in existing literature.

\subsection{General goals}

This paper reviews some of the most notable models that describe the diffusion-deformation process of elastomeric gels. The models were selected based on their consistency with thermodynamic principles, their ability to represent different behaviors, and the easiness to reproduce the results using the open-source computing platform FEniCS \citep{Alnaes2015Fenics}, serving as a benchmark for future studies. We provide a common mathematical framework and classifications from which both staggered and monolithic numerical algorithms are derived for studying the diffusion-deformation behavior of hydrogels despite their various derivation methods. We carefully examine the distinctions and similarities between different constitutive equations and analyze their impact on the deformation of hydrogels and the evolution of associated fields through simulations. This lays the foundation for future theoretical extensions, including additional mechanisms such as chemical reactions, degradation, or damage.

\subsection{Open science context}

Traditionally, researchers have relied on custom numerical solvers or commercial software like Abaqus or COMSOL. While \cite{Chester2015Abaqus} made their Abaqus code publicly available, a recent push has been towards open-source platforms for addressing coupled multiphysics models. Emerging software libraries
such as deal.II\footnote{\url{https://www.dealii.org/}}, OpenFOAM\footnote{\url{https://www.openfoam.com/}}, MOOSE\footnote{\url{https://mooseframework.inl.gov/}}, and FEniCS\footnote{\url{https://fenicsproject.org/}} exemplify this trend, signaling a shift in the way researchers approach the complex challenges of modeling hydrogels.

Unlike commercial software packages, open-source projects, like FEniCS, generally allow the user to have a direct hand in implementing the problem statement, the discretization, the numerical solution, and specific manipulations since all instances can be accessed. To this end, more information about the mathematical-numerical classification and design of algorithms is required. While this provides an advantage, granting users more liberty to develop and test novel numerical algorithms and discretizations, it also constitutes a challenge. The need for more hands-on implementation and debugging is time-consuming.

To include readerships interested in open-source software and as part of the broader scope of our paper, we establish mathematical classifications to derive numerical algorithms for the coupled problem, enabling the precise study of well-posedness and facilitating rigorous numerical analysis. This includes introducing inf-sup stable Taylor-Hood finite elements for spatial discretization and using a strongly $A$-stable, first-order, implicit Euler scheme for temporal discretization. These technical developments are summarized into compact mathematical formulations that serve as the starting point for our FEniCS implementation, allowing for meticulous examination of the numerical algorithms.

\subsection{Main contributions and paper outline}

The main contributions of this paper are summarized as follows:

\begin{enumerate}

    \item Unification in the formulation and notation of some representative diffusion-deformation theories applied to the large deformation of hydrogels.
    
    \item Mathematical classification and subsequent derivation and computational comparison of monolithic and staggered numerical solutions accuracy in a variational setting using the finite element method (FEM).

    \item One- to three-dimensional numerical simulations of well-known prototype problems to study the theories' capabilities and robustness of the FEM implementation.

    \item Comparison of the different constitutive models in a single benchmark problem highlighting their main characteristics.
    
\end{enumerate}

The paper is organized as follows. Section \ref{sec:gels_theory} provides a common background knowledge of gels and the basic formulations of the chosen theories using a unified notation. Section \ref{sec:numerics} is devoted to the mathematical classification and numerical approximation using the FEM in variational settings. 
Section \ref{sec:general-settings} explains the general setting of the numerical simulations campaign.
Section \ref{sec:Sim_results} presents the simulation results for some well-known prototype problems of the diffusion-deformation of hydrogels. Section \ref{sec:benchmark} presents a unified benchmark problem for the diffusion-deformation of hydrogels. Concluding remarks are given in Section \ref{sec:conclusion}.

\section{Nonlinear theory for the diffusion-deformation of elastomeric gels}
\label{sec:gels_theory}

In this section, we summarize some of the well-known theories describing the diffusion-deformation mechanisms in elastomeric gels that undergo large deformations under isothermal conditions. 

As notation rules, we denote gradient in the reference and current configuration by $\text{Grad}(\bullet)$ and $\text{grad}(\bullet)$, respectively, whereas the divergence in the reference and current configuration is denoted by $\text{Div}(\bullet)$ and $\text{div}(\bullet)$, respectively. The time derivative of any field is denoted by $\partial_t (\bullet)$. The operator $\text{tr}(\mathbf{A})$ refers to the trace of the second-order tensor $\mathbf{A}$.
We denote the spatial dimension with $d$, and in this paper, we exclusively work with $d=3$.
Finally, let $I:=(0,T)$ be the time interval with end time value $T>0$ and $\bar{I}$ its closure. 

\subsection{Kinematics of the deformation}

Consider a continuum homogeneous elastomeric body $\mathcal{B}$ living in the Euclidean space $\mathbb{R}^3$ and its boundary $\partial \mathcal{B} = \partial \mathcal{B}_{\mathbf{u}} \cup \partial \mathcal{B}_{\bar{\mathbf{t}}} = \partial \mathcal{B}_{\varphi} \cup \partial \mathcal{B}_{\bar{J}_R}$.
Here, $\partial \mathcal{B}_{\mathbf{u}}$ denotes the displacement boundary (Dirichlet), $\partial \mathcal{B}_{\bar{\mathbf{t}}}$ the traction boundary (Neumann), $\partial \mathcal{B}_{\varphi}$ the fluid concentration-related boundary (Dirichlet), and $\partial \mathcal{B}_{\bar{J}_R}$ the fluid flux boundary (Neumann). The outward normal vector to the domain boundaries in the reference configuration is denoted by $\mathbf{n}_R\in\mathbb{R}^d$. 
The current configuration of this homogeneous body at any instant of time can be described by a one-to-one transformation mapping $\bm{\varphi}_t: \mathcal{B}_R \rightarrow \mathbb{R}^d$, where $\mathcal{B}_R$ refers to the reference configuration. In principle, selecting any state as the reference state $\mathcal{B}_R$ (e.g., the stress-free dry gel) is possible. For convenience, authors have also chosen to define the initial state $\mathcal{B}_R$ as an isotropically swollen configuration from the dry state, leading to better reproduction of experimental conditions  \citep{Bouklas2012Linear-NL,Hajikhani2021chemomechanics}.

The position vector in the reference configuration $\mathbf{X} \in \mathcal{B}_R$ is related to the one in the current configuration $\mathbf{x} \in \mathcal{B}$ by $\mathbf{x} = \bm{\varphi}_t(\mathbf{X}, t)$. The displacement field reads $\mathbf{u}(\mathbf{X},t) = \mathbf{x} - \mathbf{X}$. The transformation map $\bm{\varphi}_t(\mathbf{X}, t)$ can be described in terms of the deformation gradient, 
\begin{equation}
    \mathbf{F} = \text{Grad} (\mathbf{x}) = \mathbf{I} + \text{Grad} (\mathbf{u}),
\end{equation}
with $J = J(\mathbf{X}, t) = \text{det} \mathbf{F} > 0$, the determinant, representing the volume change of a volume element $dv = J dV$ from the reference ($dV$) to the current ($dv$) configuration, and $\mathbf{I}$ being the identity tensor.

As is standard, 
\begin{align}
    \mathbf{C} & = \mathbf{F}^T \mathbf{F}, \\
    \mathbf{b} & = \mathbf{F} \mathbf{F}^T,
\end{align}
denote the right and left Cauchy-Green tensors, respectively. Additionally, the first invariant of $\mathbf{C}$ is given by
\begin{equation} \label{eq:I1}
    I_{1}({\bf F}) = \text{tr} (\mathbf{C}) = \text{tr} (\mathbf{F}^{T}\mathbf{F}) \, .
\end{equation}

\subsubsection{Chemical potential and swelling deformation}

The solvent component of the hydrogel is described by introducing its chemical potential $\mu$, that is, the energy absorbed or released due to a change in its content. Fluid absorption/desorption is kinematically described through an inelastic part of the deformation ${\bf F}_{f}$. The volume change $J_f=\text{det}{\bf F}_f$ associated with fluid absorption/desorption is linked with the referential fluid concentration variable $c_R$, i.e., the number of absorbed fluid molecules per unit volume of the reference configuration, by enforcing:
\begin{equation}\label{eq:kinematic_constraint}
    J_f = 1 + \Omega c_R \, ,
\end{equation}
where $\Omega$ denotes the volume of a mole of fluid molecules. This relationship can also be equivalently described by introducing the polymer volume fraction variable $\phi$, defined as:
\begin{equation}\label{eq:kinematic_constraint_phi}
    \phi = (1 + \Omega c_R)^{-1} = J_f^{-1} \, ,
\end{equation}
resulting $0 \leq \phi \leq 1$. The dry state of the gel corresponds to $\phi = 1$, and $\phi < 1$ represents a swollen state. 

A generally adopted choice is to assume an isotropic swelling deformation ${\bf F}_f$, hence reading as:
\begin{equation} \label{eq:kinematic_constraint_lambdas}
   {\bf F}_f = \lambda^s \mathbf{I}\quad \text{with} \quad \lambda^s = J_f^{1/3} = \phi^{-1/3} = (1 + \Omega c_R)^{1/3} \, ,
\end{equation} 
where $\lambda^s$ represents the polymer network stretch due to swelling. It is noteworthy that, since $c_R>0$ by definition, it results in $J_f,\phi,\lambda^s >0$.

\subsubsection{Elastic deformation}

The total deformation of a hydrogel is obtained from the superimposition of the fluid-related deformation gradient ${\bf F}_f$  and the elastic one ${\bf F}_e$. The latter originates from the effects of mechanical actions to restore compatibility. Based on the previously introduced choices, we obtain:
\begin{equation}\label{eq:F-tot}
    {\bf F}=(1 + \Omega c_R)^{1/3} {\bf F}_e. \, 
\end{equation}

At this standpoint, depending on the volume change associated with the elastic deformation $J_e=\text{det}{\bf F}_e=J_f^{-1}J$, a general classification between two classes of models is introduced:
\begin{enumerate}
    \item \emph{elastic compressible models (non-constrained formulations)}. In this case, the elastic part of the deformation is assumed to be compressible. It is then allowed ${J}_e \neq 1$;
    
    \item \emph{elastic incompressible models (constrained formulations)}. In this case, the elastic part of the deformation is assumed to be perfectly incompressible, and the total volume change of the hydrogel is related only to fluid volume changes. In other words, it results in $J_e=1$ and the kinematic constraint,
    \begin{equation} \label{eq:J=Jf}
        J \stackrel{!}{=} J_f\, ,
    \end{equation}
    has to be enforced within the theoretical formulation. 
    
\end{enumerate}

\subsection{Governing partial differential equations}
The two governing partial differential equations (PDEs) for the vector-valued displacements $\mathbf{u}:\bar{\mathcal{B}}_R\to\mathbb{R}^d$ and scalar-valued, time-dependent, fluid content in terms of the concentration $c_R:\bar{\mathcal{B}}_R\times \bar{I}\to\mathbb{R}$ when expressed in the reference configuration, consist of:

\begin{enumerate}
    \item The local form of the \textbf{balance of linear momentum} reads: 
    \begin{equation}\label{eq:linear_momentum}
        \begin{cases}
            \text{Div} (\mathbf{P}) + \mathbf{b}_R = \bm{0}, & ~ \text{in} ~ \mathcal{B}_R, \\
            \mathbf{u} = \bar{\mathbf{u}}, & ~ \text{on} ~ \partial \mathcal{B}_{\mathbf{u}}, \\
            \mathbf{P}\mathbf{n}_R = \bar{\mathbf{t}}, & ~ \text{on} ~ \partial \mathcal{B}_{\bar{\mathbf{t}}}, \\
            \mathbf{P}\vert_{t = 0} = \mathbf{P}_0, & ~ \text{in} ~ \mathcal{B}_R\, .
        \end{cases}
    \end{equation}

    Here, $\mathbf{P}:\mathcal{B}_R\to\mathbb{R}^{d \times d}$ denotes the first Piola-Kirchhoff stress tensor, and $\mathbf{P}_0:\mathcal{B}_R\to\mathbb{R}^{d \times d}$ its initial value at $t = 0$. An alternative stress measure also commonly employed is the Cauchy stress tensor $\bm{\sigma} = J^{-1} \mathbf{P} \mathbf{F}^{T}$. External actions consist of body forces per unit deformed volume in the reference configuration $\mathbf{b}_R:\mathcal{B}_R\to\mathbb{R}^d$. Moreover, boundary conditions prescribe displacement $\bar{\mathbf{u}}:\mathcal{B}_R\to\mathbb{R}^d$ and traction $\bar{\mathbf{t}}:\partial\mathcal{B}_{\bar{\mathbf{t}}}\to\mathbb{R}^{d}$ on separate portions of the boundary. Notice that inertial effects have been neglected due to the considerably slow dynamics of the fluid diffusion evolution w.r.t. the time scale of the wave propagation. 
    
    \item The local form of the \textbf{mass balance of fluid content} inside the hydrogel reads:
    \begin{equation}\label{eq:fluid_balance}
        \begin{cases}
            \partial_t c_R + \text{Div} ( \mathbf{J}_R ) = 0, & ~ \text{in} ~ \mathcal{B}_R \times I,\\
            \mu = \bar{\mu}, & ~ \text{on} ~ \partial \mathcal{B}_{\varphi}\times I,\\
            -\mathbf{J}_R \cdot \mathbf{n}_R = \bar{J}_R, & ~ \text{on} ~ \partial \mathcal{B}_{\bar{J}_R}\times I,\\
            \mu\vert_{t = 0} = \mu_0, & ~ \text{in} ~ \mathcal{B}_R\times \{t=0\}\, .
        \end{cases}
    \end{equation}
    
    Here, $\mathbf{J}_R$ is the fluid flux vector in the reference configuration, related to the one ${\bf j}$ in the current configuration via $\mathbf{J}_R = J \mathbf{F}^{-1} {\bf j}$. Boundary conditions prescribe values of chemical potential $\bar{\mu}:\partial \mathcal{B}_{\varphi}\times I\to\mathbb{R}$ and fluid flux $\bar{J}_R: \partial \mathcal{B}_{\bar{J}_R} \times I\to\mathbb{R}^d$ in the reference configuration on the boundaries. Moreover, $\mu_0:\partial \mathcal{B}_{\varphi}\to\mathbb{R}$ refers to the initial value of the chemical potential inside the hydrogel. Notice that the mass balance of fluid content is written in terms of $c_R$ and ${\bf J}_R$, but its corresponding boundary and initial conditions involve a different variable, the chemical potential $\mu$. The connection of equation \eqref{eq:fluid_balance} with $\mu$ becomes clear by introducing a constitutive relation for ${\bf J}_R$, e.g., through Fick's laws of diffusion. 
\end{enumerate}

\subsection{Constitutive equations: stress and chemical potential}

This section introduces constitutive equations for the stress and chemical potential, addressing both compressible or perfectly incompressible formulations.

\subsubsection{Compressible formulations}

Following standard thermodynamic arguments \citep{wriggers2008nonlinear,gurtin2010mechanics,Chester2010DiffDeform,Chester2011thermo}, the local form of the second law of thermodynamics reads:
\begin{equation}\label{eq:secondlaw-0}
    {\bf P} : \dot{\bf F} + \mu \dot{c}_R - {\bf J}_R \cdot \text{Grad}(\mu) - \dot{\psi}_R \geq 0,
\end{equation}
where $\psi_R:\mathcal{B}_{R}\times I\to\mathbb{R}$ is the free energy density function (per unit reference volume). Guided by equation \eqref{eq:secondlaw-0}, the free energy density function can be regarded as a function of the total deformation ${\bf F}$ and fluid concentration $c_R$, that is: 
\begin{equation} \label{eq:PsiR}
\psi_R = \Psi_R({\bf F},c_R)\, . 
\end{equation}
Consequently, equation \eqref{eq:secondlaw-0} can be reformulated as:
\begin{equation}\label{eq:secondlaw}
    {\bf P} : \dot{\bf F} + \mu \dot{c}_R - {\bf J}_R \cdot \text{Grad}(\mu) - \frac{\partial \Psi_R}{\partial {\bf F}} : \dot{\bf F} - \frac{\partial \Psi_R}{\partial c_R} : \dot{c}_R \geq 0,
\end{equation}
and the following thermodynamically-consistent constitutive relations can be established for the \textbf{first Piola-Kirchhoff stress tensor} (PK1):
\begin{equation}\label{eq:gen_PK1}
    \mathbf{P} = \dfrac{\partial \Psi_R}{\partial \mathbf{F}},
\end{equation}
and for \textbf{the chemical potential}:
\begin{equation}\label{eq:gen_mu}
    \mu = \dfrac{\partial \Psi_R}{\partial c_R}\, .
\end{equation}
Alternative formulations can be found in the state-of-the-art for defining stresses and chemical potential, built upon elastic stress and active chemical potential concepts. These are reviewed and discussed in appendix \ref{sec:alternative-forms}, showing that both approaches lead to identical results.

\subsubsection{Incompressible formulations}

Incompressible models require a special treatment of the kinematic constraint in equation \eqref{eq:J=Jf} that has to be satisfied \emph{a priori} within the formulation. A first possibility is to extend the free energy $\Psi_R$ in equation \eqref{eq:PsiR} in the context of Lagrangian formulations by introducing a constrained free-energy $\Psi_R^{c}$ that reads:
\begin{equation} \label{eq:constrained-1}
    \Psi_R^{c}({\bf F},c_R,P) = \Psi_R({\bf F},c_R) + p ( J - J_f(c_R)),
\end{equation}
where $p$ represents a pressure-like Lagrange multiplier to enforce the kinematic constraint in equation \eqref{eq:J=Jf} through a variational framework. By replacing $\Psi_R$ with $\Psi_R^{c}$, equation \eqref{eq:secondlaw} reads:
\begin{equation}\label{eq:secondlaw-c}
    {\bf P} : \dot{\bf F} + \mu \dot{c}_R - {\bf J}_R \cdot \text{Grad}(\mu) - \left(\frac{\partial \Psi_R}{\partial {\bf F}}- p J {\bf F}^{-T} \right) : \dot{\bf F} - \left(\frac{\partial \Psi_R}{\partial c_R}  + \Omega p \right) \dot{c}_R \geq 0,
\end{equation}
since $\partial J/\partial {\bf F}=J {\bf F}^{-T}$ and $\partial J_f/\partial c_R=\Omega$. From equation \eqref{eq:secondlaw-c}, the following thermodynamically consistent choices can be introduced for the first Piola-Kirchhoff stress tensor:
\begin{equation}\label{eq:gen_PK1-c}
    \mathbf{P} = \dfrac{\partial \Psi_R}{\partial \mathbf{F}} - p J {\bf F}^{-T},
\end{equation}
and for the chemical potential:
\begin{equation}\label{eq:gen_mu-c}
    \mu = \dfrac{\partial \Psi_R}{\partial c_R} + \Omega p.  
\end{equation}
Such an approach has been followed, for instance, by \cite{Bouklas2012Linear-NL,Chester2010DiffDeform}.

Alternatively, the kinematic constraint in equation \eqref{eq:J=Jf} can be embedded directly within the free-energy function. For instance, this rationale is described and adopted by \cite{Liu2015GelsGenAnisotropic,Liu2016TransGels}. In this case, the fluid concentration $c_R$ is regarded as a dependent variable since it is related to the deformation gradient ${\bf F}$ (or displacements ${\bf u}$) through equation \eqref{eq:J=Jf}. Hence, the free energy can be reformulated in the form:
\begin{equation}\label{eq:constrained-2}
\psi_R=\tilde{\Psi}_R({\bf F})\, .
\end{equation}
Then, equation \eqref{eq:secondlaw} reads:
\begin{equation}\label{eq:secondlaw-Liu}
    {\bf P} : \dot{\bf F} + \frac{\mu}{\Omega} \dot{J} - {\bf J}_R \cdot \text{Grad}(\mu) - \left(\frac{\partial \tilde{\Psi}_R}{\partial {\bf F}} \right) : \dot{\bf F} \geq 0,
\end{equation}
since $\dot{c}_R=\dot{J}_f/\Omega=\dot{J}/\Omega$ from the incompressibility constraint. From equation \eqref{eq:secondlaw-Liu} and noting that $\dot{J}=J{\bf F}^{-T} : \dot{\bf F}$, the first Piola-Kirchhoff stress tensor can be expressed as:
\begin{equation}\label{eq:gen_PK1-Liu}
    \mathbf{P} = \frac{\partial \tilde{\Psi}_R}{\partial \mathbf{F}} - \frac{\mu J}{\Omega}  {\bf F}^{-T}.
\end{equation}
As previously noted, the fluid concentration $c_R$ can no longer be considered as an independent variable. Therefore, the chemical potential $\mu$ shall now be considered as the problem's primary variable. 
Consequently, a special treatment is required for the $\partial_t c_R$ term in the mass balance equation \eqref{eq:fluid_balance_mu}. As introduced by \cite{Liu2015GelsGenAnisotropic,Liu2016TransGels}, the following relationship holds true from equation \eqref{eq:J=Jf} between the time derivative of the fluid concentration and the determinant of the deformation gradient:
\begin{equation}
    \partial_t c_R = \dfrac{1}{\Omega} \partial_t J_f = \dfrac{J}{\Omega} \text{div}(\partial_t \mathbf{u}) = \dfrac{J}{\Omega} \text{Grad}(\partial_t \mathbf{u}):\mathbf{F}^{-T},
\end{equation}
which yields
\begin{equation}\label{eq:fluid_balance_mu_Liu}
    \dfrac{J}{\Omega} \text{Grad}(\partial_t \mathbf{u}):\mathbf{F}^{-T} + \text{Div} ( \mathbf{J}_R ) = 0, ~ \text{in} ~ \mathcal{B}_R \times I \, .
\end{equation}

\subsection{Constitutive equations: fluid flux}

From equation \eqref{eq:secondlaw}, a thermodynamically motivated choice for the \textbf{fluid flux} $\mathbf{j}$ is to assume that it is proportional to the gradient of the chemical potential in the current configuration, namely,
\begin{equation}\label{eq:flux_j}
    \mathbf{j} = - \dfrac{cD}{k_B T} \text{grad} (\mu),
\end{equation}
with $c$ the solvent concentration in the current configuration, related to the nominal concentration by $c_R = J c$, and $D$ is the solvent diffusivity assumed to be a constant. Notice that $\text{grad}(\mu)$ can be pulled back to the reference configuration by $\text{grad} (\mu) =  \mathbf{F}^{-T} \text{Grad} (\mu)$. Hence, the flux in the reference configuration reads:
\begin{equation}\label{eq:flux_J_R}
    \mathbf{J}_R = - J \mathbf{F}^{-1} \left( \dfrac{cD}{k_B T} \text{grad}(\mu) \right) = - \dfrac{c_R D}{k_B T} \mathbf{b}^{-1} \text{Grad} (\mu)\, .
\end{equation}
Inserting equation \eqref{eq:flux_J_R} into \eqref{eq:fluid_balance} yields for the mass balance equation:
\begin{equation}\label{eq:fluid_balance_mu}
    \partial_t c_R - \text{Div} \left( \dfrac{c_R D}{k_B T} \mathbf{b}^{-1} \text{Grad} (\mu) \right) = 0, ~ \text{in} ~ \mathcal{B}_R \times I\, ,
\end{equation}
with $\mu=\mu({\bf F},c_R)$ from equation \eqref{eq:gen_mu}. 

Notice that to make the units of equation \eqref{eq:fluid_balance_mu} consistent, $c_R$ must have units of [mol], $k_B$ units of [J K$^{-1}$], $T$ units of [K], $\mu$ units of [J mol$^{-1}$], $D$ units of [m$ ^2$ s$^{-1}$], and $\bm{b}$ is dimensionless. Some authors have defined constitutive equation \eqref{eq:flux_j} in terms of $RT$, where $R$ is the gasses constant, (see, e.g., \cite{Chester2011thermo} or \cite{Chester2015Abaqus}). In this case, $\mu$ units of [J]. 

\subsection{Specialization of constitutive theories}

Specific choices for the free energy function characterize different state-of-the-art models linking stress and chemical potential variations with diffusion-deformation mechanisms. We introduce constitutive models in accordance with the framework outlined in equation \eqref{eq:PsiR} for compressible formulations and equations \eqref{eq:constrained-1} or \eqref{eq:constrained-2} for incompressible formulations. 

Unless explicitly specified otherwise, we choose to characterize fluid content using the concentration variable $c_R$. As a result, we express fluid-related volume changes and polymer volume fractions as functions of $c_R$, denoting them as $J_f = J_f(c_R)$ and $\phi = \phi(c_R)$, respectively, based on equations \eqref{eq:kinematic_constraint} and \eqref{eq:kinematic_constraint_phi}, respectively. By inverting these relationships, we can reformulate the proposed theories using different primary variables to describe fluid content whenever needed.

The free energy function is, in general, written in a separable additive form:
\begin{equation}\label{eq:gen_psi}
   \psi_R = \psi_{R}^{\text{mix}} + \psi_{R}^{\text{mech}}\, .  
\end{equation}
Here, $\psi_{R}^{\text{mix}}$ describes the mixing of the solvent with the polymer network. Overall, there is a rather consensus agreement that this is well described by the Flory-Huggins/Flory-Rehner theory (see, e.g., \cite{Chester2010DiffDeform, Bouklas2012Linear-NL, Liu2016TransGels}), reading:
\begin{equation}\label{eq:psi_mix}
    \psi_{R}^{\text{mix}} = \Psi_R^{\text{mix}}(c_R) = \mu^0 c_R + \dfrac{k_B T }{\Omega} \left[ \Omega c_R \ln\left(\dfrac{\Omega c_R}{1 + \Omega c_R}\right) + \chi \left(\dfrac{\Omega c_R}{1 + \Omega c_R}\right)\right],
\end{equation}
where $\mu^0$ is the chemical potential of the unmixed pure solvent, $k_B$ refers to Boltzmann's constant, $T$ is the absolute temperature, and $\chi$ is a dimensionless parameter named Flory's interaction parameter. The latter represents the disaffinity between the polymer and the fluid. In particular, if $\chi$ is increased, the fluid molecules are expelled from the gel, and it shrinks, while if $\chi$ is decreased, the gel swells.

Furthermore, $\psi_{R}^{\text{mech}}$ is the contribution to the change in the free energy due to the deformation of the polymer network, for which elastic incompressible or compressible formulations differ by considering:
\begin{equation}\label{eq:psi_mech}
    \psi_{R}^{\text{mech}} = \begin{cases}
       \Psi_{R}^{\text{s}} (\mathbf{F},J) & \text{incompressible}, \\
       \Psi_{R}^{\text{s}} (\mathbf{F},J) + \Psi_{R}^{\text{en}} (J, c_R) & \text{compressible},
       \end{cases}
\end{equation}
where $\Psi_{R}^{\text{s}}$ is an entropic component and $\Psi_{R}^{\text{en}}$ an energetic contribution.

The entropic component is usually defined following the arguments of classical statistical mechanics models for rubber elasticity, \citep{Chester2011thermo}. For small to moderate values of stretching, Gaussian statistics provide an estimate of the entropy change due to mechanical stretching of the polymer network resulting in the form of a Neo-Hooke material that takes the form \citep{Huang2020GelsModels_Review}:

\begin{equation}\label{eq:psi_entropic}
    \Psi_{R}^{\text{s}} (\mathbf{F},J) = \dfrac{G_0}{2} \left[ I_{1}({\bf F}) - 3 - 2 \ln \left(J\right) \right] \, ,
\end{equation}
where $I_1$ is given in equation \eqref{eq:I1} and $G_0 \approx N k_B T$ represents the shear modulus, with $N$ being the number of polymer chains per unit reference volume, i.e., crosslink polymer network density. 

In contrast, different choices have been made by authors when it comes to defining the energetic component $\Psi_{R}^{\text{en}}$ of the free energy due to the deformation of the polymer network. Some available solutions will be discussed in Section \ref{sec:compressible-models}, and after that, some state-of-the-art perfectly incompressible models will be presented.

\subsubsection{Incompressible constitutive models}

Two incompressible constitutive models are presented here. 

\textit{Constitutive model I:} Following the ideas by \cite{Hong2008GelsTheory,Zhang2009Gels, Liu2016TransGels}, this model assumes a perfectly incompressible elastic material response, by introducing a constrained material response within the rationale presented in equation \eqref{eq:constrained-2}. Therefore, the constraint in equation \eqref{eq:J=Jf} is directly embedded in the free-energy function, leading to:
\begin{equation} \label{eq:Psi_Liu}
    \tilde{\Psi}_R({\bf F}) = \left.\left(\Psi_{R}^{\text{s}}({\bf F},J) + \tilde{\Psi}_{R}^{\text{mix}}(J)\right)\right|_{J=\text{det}({\bf F})}\, ,
\end{equation}
where the mixing part of the energy $\tilde{\Psi}_{R}^{\text{mix}}$ respects $\tilde{\Psi}_{R}^{\text{mix}}(J) =  \Psi_{R}^{\text{mix}}(c_{R}(J))$ with $c_{R}(J)=(J-1)/\Omega$ from equation \eqref{eq:kinematic_constraint} under the condition of equation \eqref{eq:J=Jf}.
Considering the entropic component in equation \eqref{eq:psi_entropic}, the first Piola-Kirchhoff stress tensor is derived from equation \eqref{eq:gen_PK1-Liu} as:
\begin{equation}\label{eq:stress_Liu}
    \mathbf{P} ({\bf F}, \mu) = \frac{\partial \Psi_{R}^{\text{s}}}{\partial {\bf F}} + \left(\frac{\partial \Psi_{R}^{\text{s}}}{\partial J}+ \frac{ \partial \tilde{\Psi}_{R}^{\text{mix}}}{\partial J} \right) J{\bf F}^{-T} - \frac{J \mu}{\Omega} {\bf F}^{-T} = \Big[G_0 \left( \mathbf{b} - \mathbf{I} \right)  + J p_{\mu} (J,\mu) \mathbf{I}\Big]\mathbf{F}^{-T}\, .
\end{equation}

Here, $p_{\mu}$ is the equivalent volumetric stress:
\begin{equation}\label{eq:p_mu_Liu}
    p_{\mu} (J,\mu) = -\dfrac{\mu}{\Omega} + \dfrac{k_B T}{\Omega} \left[ \ln \left( 1 - \dfrac{1}{J} \right) + \dfrac{1}{J} + \dfrac{\chi}{J^2} \right].
\end{equation}

\textit{Constitutive model II:} This model has been introduced by \cite{Chester2010DiffDeform}. Also, in this case, a perfectly incompressible elastic material response is assumed, introducing a constrained material response by following a Lagrangian approach as described in equation \eqref{eq:constrained-1}. By exploiting the kinematic constraints in equations \eqref{eq:kinematic_constraint} and \eqref{eq:J=Jf}, the entropic component in equation \eqref{eq:psi_entropic} is re-formulated as $\hat{\Psi}_R^{\text{e}}({\bf F},c_R)=\Psi_R^{\text{s}}({\bf F},J_f(c_R))$ and the total free-energy reads:
\begin{equation}\label{eq:Psi-constII}
    \Psi_R=\Psi_R^{\text{mix}}(c_R) + \hat{\Psi}_R^{\text{e}}({\bf F},c_R)\, .
\end{equation}

The first Piola-Kirchhoff stress tensor is derived from equation \eqref{eq:gen_PK1-c} yielding:
\begin{equation}\label{eq:stress_Chester2010}
    \mathbf{P} ({\bf F},P) = \frac{\partial \hat{\Psi}_R^{\text{s}}}{\partial {\bf F}} - p J {\bf F}^{-T} = G_0 \left( \mathbf{b} - P \mathbf{I} \right) \mathbf{F}^{-T}\, ,
\end{equation}
where $P = p J$ is the pressure term in the reference configuration. To be consistent with the original model, the fluid concentration $c_R$ is replaced in the formulation with the polymer volume fraction $\phi$ by means of equation \eqref{eq:kinematic_constraint_phi}. Accordingly, the chemical potential can be obtained from equations \eqref{eq:gen_mu-c}, \eqref{eq:psi_mix}, \eqref{eq:psi_entropic} and \eqref{eq:Psi-constII} as:
\begin{equation}\label{eq:mu_Chester2010}
    \mu(\phi,P) = \left.\left(\frac{\partial \hat{\Psi}_R^{\text{mix}}}{\partial c_R} + \frac{\partial \hat{\Psi}_R^{\text{s}}}{\partial c_R} + \Omega p\right)\right|_{c_R(\phi)} = \mu^0 + k_B T  \left[ \ln \left( 1 - \phi \right) + \phi + \chi \phi^2 \right] - \Omega G_0 \phi + \Omega P \phi \, ,
\end{equation}
where $P = p J = p J_f = p \phi^{-1}$ follows directly from equations \eqref{eq:kinematic_constraint}, \eqref{eq:kinematic_constraint_phi} and \eqref{eq:J=Jf}. 

\subsubsection{Compressible models} \label{sec:compressible-models}

Compressible models are characterized by the superposition of an entropic energetic component (given in equation \eqref{eq:psi_entropic}) and an energetic part of the free energy. The latter reflects changes in the internal energy associated with the volumetric mechanical deformation of the swollen elastomer. Three well-established constitutive models for $\Psi_{R}^{\text{en}}$ are here discussed: 
\begin{equation}\label{eq:psi_energetic}
    \Psi_{R}^{\text{en}} (J, c_R) = 
    \begin{cases}
       \Psi_{R,1}^{\text{en}} (J, c_R) = \dfrac{K}{2} \left[ J - J_f(c_R) \right]^2 & \text{\citep{Bouklas2015Nonlinear}} \\
        \Psi_{R,2}^{\text{en}} (J, c_R) =\dfrac{K}{2}  \ln (J\phi(c_R))^2 & \text{\citep{Chester2011thermo}} \\
        \Psi_{R,3}^{\text{en}} (J, c_R) =\phi(c_R)^{-1} \left[ \dfrac{K}{2}  \ln (J\phi(c_R))^2 \right] & \text{\citep{Chester2015Abaqus}}
    \end{cases}
\end{equation}
where $K>0$ represents the bulk modulus of the gel. 

\textit{Constitutive model III:} The following formulation is based on the ideas by \cite{Bouklas2015Nonlinear}. To be consistent with the original model, the swelling volume change $J_f$ is considered in place of $c_R$ within the formulation, but we highlight that this is straight linked through equation \eqref{eq:kinematic_constraint}.
Recalling that $J=\text{det}({\bf F})$, the first Piola-Kirchhoff stress tensor considers the entropic and energetic components in equations \eqref{eq:psi_entropic} and \eqref{eq:psi_energetic}$_1$, reading from equation \eqref{eq:gen_PK1}:

\begin{equation}\label{eq:stress_Bouklas2015}
    \mathbf{P} ({\bf F},J_f) = \frac{\partial \Psi_R^{\text{s}}}{\partial {\bf F}} + \left(\frac{\partial \Psi_R^{\text{s}}}{\partial J} + \frac{\partial \Psi_{R,1}^{\text{en}}}{\partial J}\right) J {\bf F}^{-T} = G_0 \left( \mathbf{F} + \alpha({\bf F},J_f) J {\bf F}^{-T} \right),
\end{equation}
with:
\begin{equation}\label{eq:alpha_H_Bouklas}
    \alpha ({\bf F},J_f) = -\dfrac{1}{J} + \dfrac{K}{G_0} \left(J - J_f \right).
\end{equation}

The chemical potential is obtained from equations \eqref{eq:gen_mu}, \eqref{eq:gen_psi}, \eqref{eq:psi_mix} and \eqref{eq:psi_energetic}$_1$ as:
\begin{equation}\label{eq:mu_Bouklas2015}
    \mu({\bf F},J_f) = \left.\frac{\partial \Psi_R^{\text{mix}}}{\partial c_R}\right|_{c_R(J_f)}  = \mu^0 + k_B T  \left[ \ln \left( 1 - \dfrac{1}{J_f} \right) + \dfrac{1}{J_f} + \dfrac{\chi}{J_f^2} \right] - \Omega K \left( J - J_f \right). 
\end{equation}
It is worth noting that the mass balance equation \eqref{eq:fluid_balance_mu} reads in this context as:
\begin{equation} \label{eq:fluid_balance_rewritten-1}
\dfrac{1}{\Omega}\partial_t J_f - \text{Div} \Bigl( {\bf M}({\bf F},J_f) \text{Grad} (\mu({\bf F},J_f)) \Bigr) = 0 \quad \text{in} ~ \mathcal{B}_R \times I,
\end{equation}
where ${\bf M}={\bf M}({\bf F},J_f)$ denotes the species mobility tensor:
\begin{equation}\label{eq:mobility_tensor}
    {\bf M}({\bf F},J_f) = \dfrac{D}{k_B T} \dfrac{J_f - 1}{\Omega} \mathbf{b}^{-1}\, .
\end{equation}

\textit{Constitutive models IV and V:} Starting from the original incompressible model introduced in \cite{Chester2010DiffDeform} (\emph{constitutive model II}), the same group of authors presented alternative compressible formulations by adding energetic components as given in equations \eqref{eq:psi_energetic}$_2$ (\cite{Chester2011thermo}, \emph{constitutive model IV}) and \eqref{eq:psi_energetic}$_3$ (\cite{Chester2015Abaqus}, \emph{constitutive model V}). To be consistent with the original models, the polymer volume fraction $\phi$ is considered in place of $c_R$ within the formulation, but we highlight that this is straight linked through equation \eqref{eq:kinematic_constraint_phi}. Hence, recalling that $J=\text{det}({\bf F})$, the first Piola-Kirchhoff stress tensor follows, for \emph{constitutive model IV}, from equation \eqref{eq:gen_PK1} with equations \eqref{eq:psi_entropic} and \eqref{eq:psi_energetic}$_2$ as:
\begin{subequations} \label{eq:PK1_Chester_energetic}
\begin{equation}\label{eq:PK1_Chester_energetic-IV}
    \mathbf{P} ({\bf F}, \phi) = 
      \dfrac{\partial \Psi_R^{\text{s}}}{\partial {\bf F}} + \left(\dfrac{\partial \Psi_R^{\text{s}}}{\partial J} + \dfrac{\partial \Psi_{R,2}^{\text{en}}}{\partial J}\right) J {\bf F}^{-T} = \left[ G_0 \left( \mathbf{b} - \mathbf{I} \right) + K \left( \ln (J \phi) \right) \mathbf{I} \right] \mathbf{F}^{-T} \, ,
\end{equation}
and, for \emph{constitutive model V} with \eqref{eq:psi_energetic}$_3$, as:
\begin{equation}\label{eq:PK1_Chester_energetic-V}
    \mathbf{P} ({\bf F}, \phi) = 
      \dfrac{\partial \Psi_R^{\text{s}}}{\partial {\bf F}} + \left(\dfrac{\partial \Psi_R^{\text{s}}}{\partial J} + \dfrac{\partial \Psi_{R,3}^{\text{en}}}{\partial J}\right) J {\bf F}^{-T} = \left[ G_0 \left( \mathbf{b} - \mathbf{I} \right) + \phi^{-1} K \left( \ln (J \phi) \right) \mathbf{I} \right] \mathbf{F}^{-T}\, .
\end{equation}
\end{subequations}

Equation \eqref{eq:gen_mu}, together with equation \eqref{eq:psi_mix}, \eqref{eq:psi_energetic}$_2$ and \eqref{eq:psi_energetic}$_3$, yields the chemical potential:
\begin{equation}\label{eq:mu_Chester_energetic}
    \mu({\bf F}, \phi) = \begin{cases}
     \left.\left(\dfrac{\partial \Psi_R^{\text{mix}}}{\partial c_R} + \dfrac{\partial \Psi_{R,2}^{\text{en}}}{\partial c_R}\right)\right|_{c_R(\phi)}  = \mu^0 + g({\bf F},\phi), & \text{\emph{model IV} }\\
      \left.\left(\dfrac{\partial \Psi_R^{\text{mix}}}{\partial c_R} + \dfrac{\partial \Psi_{R,3}^{\text{en}}}{\partial c_R}\right)\right|_{c_R(\phi)} = \mu^0 + g({\bf F},\phi) + \dfrac{1}{2} \Omega K \left( \ln (J \phi) \right)^2 & \text{\emph{model V}},
    \end{cases} 
\end{equation}
with:
\begin{equation}
g({\bf F},\phi) = R T  \left[ \ln \left( 1 - \phi \right) + \phi + \chi \phi^2  \right] + \begin{cases}
   -\Omega K \left( \ln (J \phi) \right) \phi & \text{\emph{model IV}}\\
   -\Omega K \left( \ln (J \phi) \right) & \text{\emph{model V}}
\end{cases}  \, . 
\end{equation}

It is noteworthy that, in the original papers, authors formulate the theories based on the elastic PK1 and the active chemical potential, leading, however, to identical results as proved in appendix \ref{sec:alternative-forms}. Moreover, in these contexts, the mass balance equation \eqref{eq:fluid_balance_mu} is conveniently reformulated in terms of $\phi$, instead of $c_R$, as:
\begin{equation} \label{eq:fluid_balance_rewritten-2}
    -\dfrac{1}{\Omega \phi^2} \partial_t \phi - \text{Div} \Bigl( {\bf M}({\bf F},\phi^{-1}) \text{Grad} (\mu({\bf F},\phi)) \Bigr) = 0 \quad \text{in} ~ \mathcal{B}_R \times I \, ,
\end{equation}
where ${\bf M}$ is given in equation \eqref{eq:mobility_tensor} and $J_f=\phi^{-1}$ from equation \eqref{eq:kinematic_constraint_phi}.

\subsubsection{Preliminary comparisons between models and final considerations}\label{sec:final-comments-theory}

In the context of compressible models, we can gain valuable insights by examining how different choices of the energy component $\Psi_R^{\text{en}}$ impact the stress constitutive relationships. To illustrate this, let us focus on equation \eqref{eq:stress_Bouklas2015} within \emph{constitutive model III}, which primarily penalizes substantial elastic deformations, that is, when $|J - J_f| = |J_e - 1 | J_f \gg 0$. In contrast, \emph{constitutive models IV and V} in equations \eqref{eq:PK1_Chester_energetic} incorporate penalties for both significant elastic deformations and the fully swollen state, as evidenced by the behavior of $|\ln (J \phi)| = |\ln(J_e)|$, which approaches $+\infty$ when $|J_e-1| \gg 0$, and inversely, it approaches $-\infty$ as the polymer volume fraction $\phi$ approaches $0$. However, it is worth noting that \emph{constitutive models IV and V} diverge from each other in treating large swelling deformations. Specifically, \emph{constitutive model V} penalizes these deformations, occurring when $\phi^{-1} = J_f \rightarrow + \infty$, whereas \emph{constitutive model IV} does not.

Furthermore, as previously highlighted, \emph{constitutive models IV and V} serve as the compressible counterparts to \emph{constitutive model II}. This connection becomes evident when we observe that the Lagrange multiplier $P$ in equation \eqref{eq:stress_Chester2010} is effectively replaced by a term related to bulk modulus in equations \eqref{eq:PK1_Chester_energetic}. However, from a numerical implementation standpoint, this substitution carries significant implications. In general, $P$ represents an additional primary variable that must be determined through the stationary conditions of the constrained functional \eqref{eq:constrained-1} with respect to the Lagrange multiplier. Only under specific circumstances, such as a traction-free condition in the presence of plane stresses, can $P$ be directly defined from equilibrium conditions. In these instances, \textit{constitutive model II} simplifies to having only two primary variables, namely, $\textbf{u}$ and $\phi$ (as demonstrated in, for example, equation \eqref{eq:P_slab} in the subsequent Section \ref{sec:res-II}). From this point forward, we will exclusively focus on these special cases.

\subsection{Weak formulations}

The solution of the coupled PDE system consists of a vector-valued field of displacements ($\mathbf{u}$) and, depending on the formulation, a scalar-valued field ($\varphi$) given either by the concentration ($c_R$), polymer volume fraction ($\phi$), or chemical potential ($\mu$). Hence, it results $\varphi \in \lbrace c_R, \phi, \mu \rbrace$.

Here we adopt standard notation for the usual Lebesgue and Sobolev spaces, e.g., \cite{Wlo87}. The functional space $H^1(\mathcal{B}_R)^d$ is a Sobolev space that consists of functions defined on a bounded domain $X \subset \mathbb{R}^d$, with square integrable partial derivatives up to the first order. 

Here, $X := \mathcal{B}_R$. Specifically, a function $w \in H^1(\mathcal{B}_R)^d$, if it satisfies the following conditions, namely $w$ is square integrable: $\int_{\mathcal{B}_R} |w(x)|^2 dx < \infty$
and the first-order partial derivatives of $w$ exist and are square-integrable such that $\int_{\mathcal{B}_R} |\nabla w(x)|^2 dx < \infty$.

The norm associated with this space is given by
\begin{equation}
    \vert \vert w \vert \vert_{H^1 (\mathcal{B}_R)^d} := \left( \int_{\mathcal{B}_R} \vert w(x) \vert^2 dx + \int_{\mathcal{B}_R} \vert \nabla w(x) \vert^2 dx \right)^{1/2}.
\end{equation}

This norm induces a complete metric space with respect to which the functions in $H^1(\mathcal{B}_R)^d$ can be well-defined and approximated.

The coupled system of equations is formulated in terms of a variational coupled system. 
To this end, we define the trial and test spaces as follows: 

\begin{align*}
    Q &:= H^1 \left( \mathcal{\mathcal{B}_R} \right), \quad Q_{\bar\varphi} : = \Big\lbrace \varphi \in Q \Big\vert ~ h(\varphi) = \bar{h} ~~ \text{on} ~~ \partial \mathcal{B}_{\varphi} \Big\rbrace, \quad 
    Q_{0} : = \Big\lbrace \varphi \in Q \Big\vert ~ \varphi = 0 ~~ \text{on} ~~ \partial \mathcal{B}_{\varphi} \Big\rbrace,\\
    V &:= H^1 \left( \mathcal{\mathcal{B}_R} \right)^d, \quad V_{\bar{\mathbf{u} }} : = \Big\lbrace \mathbf{u} \in V \Big\vert ~ \mathbf{u}  = \bar{\mathbf{u} } ~~ \text{on} ~~ \partial \mathcal{B}_{\mathbf{u}}\Big\rbrace, 
    \quad V_{0} : = \Big\lbrace \mathbf{u} \in V \Big\vert ~ \mathbf{u}  = 0 ~~ \text{on} ~~ \partial \mathcal{B}_{\mathbf{u}}\Big\rbrace.
\end{align*}

We notice that in $Q_{\bar\varphi}$, the boundary conditions may be given explicitly
or implicitly through the relation $h(\varphi) = \bar{h}$. This is seen in the set of equations \eqref{eq:fluid_balance}.
To elucidate this, let us consider two distinct scenarios: 
\textbf{\textit{i.}} for \textit{constitute models II, IV, and V}, we seek $\phi$, and the boundary condition is $\phi = \bar{\phi}$, i.e., $h(\varphi) := \phi$ and $\bar{h}:=\bar{\phi}$. On the other hand, \textbf{\textit{ii.}} in \textit{constitutive model III}, we have $\varphi := J_f$ and $\bar{h}:= \bar{\mu}$. Then, we solve the coupled problem for $J_f$ and $\mu$, where $J_f$ is implicitly obtained from equation \eqref{eq:mu_Bouklas2015}. Once we know $J_f$ or $\phi$, then $c_R$ can be recovered from \eqref{eq:kinematic_constraint}. 

To formulate both problem statements in an abstract fashion, we introduce 
for the displacement system the semi-linear form $a((\mathbf{u},\varphi))(\mathbf{v})$, which is nonlinear in the first argument (trial function) and linear in the second argument (test function). Furthermore, let $b(\mathbf{v})$ be the given right-hand side data. Next, for the balance of fluid concentration, 
we use $c((\mathbf{u},\varphi))(q)$ and $d(q)$.
Then, the weak formulation can be written as:
\begin{formulation} (Diffusion-deformation of gels in $\mathcal{B}_R$). 
\label{form_1}
    Find $(\mathbf{u}, \varphi) \in V_{\bar{\mathbf{u} }}\times Q_{\bar\varphi}$, with $\varphi(0) = \varphi_0$, such that for $t\in I$
    it holds
    \begin{equation}
        \begin{aligned}
            a((\mathbf{u},\varphi))(\mathbf{v}) + b(\mathbf{v}) & = \mathbf{0}, ~~ \forall \mathbf{v} 
            \in V_0, \\
            c((\mathbf{u},\varphi))(q) + d(q) & = 0, ~~ \forall q \in Q_0,
        \end{aligned}
    \end{equation}
    where
    \begin{align}
        a((\mathbf{u},\varphi))(\mathbf{v}) & = \int_{\mathcal{B}_R} \mathbf{P} ( \mathbf{u}, \varphi):\nabla \mathbf{v} dV, \\
        b(\mathbf{v}) & = - \int_{\mathcal{B}_R} \mathbf{b}_R \cdot \mathbf{v} dV, \\
        c((\mathbf{u},\varphi))(q) & = \int_{\mathcal{B}_R} \partial_t c_R(\varphi) \cdot q dV + \int_{\mathcal{B}_R} {\bf M}({\bf u}, \varphi) \nabla \mu \left( \mathbf{u}, \varphi \right) \cdot \nabla q dV, \\
        d(q) & = 0,
    \end{align}
    with $``~:~"$ denoting the double contraction of the second-order tensors $\mathbf{P}$ and $\nabla \mathbf{v}$,
    where $\mathbf{P} ( \mathbf{u}, \varphi)$ is defined by either equation \eqref{eq:stress_Liu}, \eqref{eq:stress_Chester2010}, \eqref{eq:stress_Bouklas2015}, or \eqref{eq:PK1_Chester_energetic}, depending of the constitutive model adopted. Whereas ${\bf M}({\bf u}, \varphi)$ is given by equation \eqref{eq:mobility_tensor}.
    
    Notice that the two balance equations are fully coupled through the constitutive equations of the stress $\mathbf{P} (\mathbf{u}, \varphi)$ and the species mobility tensor ${\bf M}({\bf u}, \varphi)$.
\end{formulation}

\section{Classifications, discretization, and numerical solution}
\label{sec:numerics}
In this section, based on Formulation \ref{form_1}, we explain numerical coupling strategies,
provide mathematical classifications, and introduce spatial and temporal discretizations.
These derivations serve as a starting point for the implementation in FEniCS. 
The reader is referred to the introduction to understand the importance of this section, particularly when testing novel algorithms, comparing them, and pursuing our own numerical developments, including code debugging. This section provides the link between the strong form problem statements in Section \ref{sec:gels_theory} and the numerical simulations carried out in Section \ref{sec:Sim_results} by following the road map outlined in \cite{Wi23_st}[Section 12.3].

\subsection{Coupling strategies}
\label{sec_coupling_strategies}

There exist several ways for realizing numerically the coupling of several PDEs. Here, we discuss two fundamental strategies to be implemented later, namely, monolithic and partitioned approaches. 
In the former, Formulation \ref{form_1}, it treated all-at-once, while in a partitioned 
approach, the two subproblems are decoupled and solved in an iterative fashion.
Here, we closely follow the concepts and notation introduced by 
\cite{Wick2020multiphysicsPFF}[Chapter 3].

\textbf{Variational-monolithic coupling:} 
In the variational-monolithic setting, the coupling conditions are realized in an exact fashion in the weak (i.e., variational) formulation. Formulation \ref{form_1} is given in such a variational-monolithic fashion and, more specifically, the coupling conditions 
are of volume-coupling type \cite{Wick2020multiphysicsPFF}[Section 3.3.3].

In the monolithic approach, the entire PDE system can be either solved all-at-once, which usually requires physics-based preconditioners. Either the system is decoupled on the solver level within an outer monolithic iteration, e.g., GMRES (generalized minimal residual) or multigrid, and the preconditioner is constructed based on decoupled subproblems. In general, monolithic solutions can be computationally demanding depending on the complexity of the problem at hand.

The monolithic scheme can be naturally extended to account for time-dependent problems. In this case, we need to introduce a suitable time discretization scheme and solve the previous problem at each time step. As an alternative, a global space-time formulation can be formulated, discretized, and solved accordingly; see e.g., \cite{GaNeu16}
for a specific space-time multigrid realization and analysis on the linear level.

\textbf{Partitioned (staggered) approach:}
Conversely, in the partitioned approach, the system of PDEs is broken down into smaller subsystems, and each subsystem is solved independently using its own numerical method. The solutions of these subsystems are then coupled to obtain the solution of the entire system. The partitioned approach typically involves the following steps:

\begin{enumerate}
    \item \textbf{Initialization:} provide initial guesses $\hat{\bf u}^0$ and $\hat{\varphi}^0$ for the unknown fields $\mathbf{u}$ and $\varphi$.
    
    \item \textbf{Iteration:} Let $(\hat{\bf u}^j,\hat{\varphi}^j)$ and $(\hat{\bf u}^{j-1},\hat{\varphi}^{j-1})$ be the trial values of $\mathbf{u}$ and $\varphi$ at the current and previous iterations, respectively.
    
    \begin{algor}
    For $j = 1, 2, 3, \dots$, iterate:
    
    \begin{itemize}
        \item Given $\hat{\varphi}^{j-1}$, find $\hat{\bf u}^j$: 
        \begin{equation}
            a((\hat{\bf u}^j, \hat{\varphi}^{j-1}))(\mathbf{v}) + b(\mathbf{v}) = \mathbf{0}
        \end{equation}
        
        \item Given $\hat{\bf u}^{j}$, find $\hat{\varphi}^j$:
        \begin{equation}
            c((\hat{\bf u}^{j},\hat{\varphi}^{j}))(q) + d(q)  = 0
        \end{equation}
    \end{itemize}
    \end{algor}
    
    \item \textbf{Check for convergence:} compare the updated and previous trial values. The iteration is considered converged if the difference is below a specified tolerance. That is, check whether
    \begin{equation}\label{eq:staggered_conver}
        \max \left( \vert\vert \hat{\bf u}^j - \hat{\bf u}^{j-1} \vert\vert, \vert\vert \hat{\varphi}^j - \hat{\varphi}^{j-1} \vert\vert \right) < TOL.
    \end{equation}

    If the criterion is fulfilled, stop and assign ${\bf u} = \hat{\bf u}^j$ and $\varphi = \hat{\varphi}^j$. If not, increment $j \rightarrow j + 1$ and return to \textbf{step 2}.
    
\end{enumerate}

The previous procedure extends to discrete formulations of time-dependent problems, for which a sequence of unknown fields $\mathbf{u}^{n}$ and $\varphi^{n}$ at time points $t_n$ with $n = 1, 2, 3, ..., N$ is sought for. Here, the previously introduced algorithm remains identical and performed at each time point $t_n$. In this context, trial quantities at iteration step $j$ and time $t_n$ can be denoted as $\hat{\bf u}^{n,j}$ and $\hat{\varphi}^{n,j}$.

\subsection{Mathematical classification}

We follow the ideas presented by \cite{Wick2020multiphysicsPFF} and classify the diffusion-deformation prototype problem in Formulation \ref{form_1}. This is the starting point to design appropriate algorithms, which are of interest in this work and 
have been introduced before. Moreover, such classifications are required for mathematical 
and numerical analysis, which both exceed the focus of this work.
To this end, we can formally analyze the problem statement as follows:

\begin{enumerate}
    \item \textbf{Orders in time and space:} Equation \eqref{eq:linear_momentum} represents a quasi-static problem with no time derivatives and second order in space. In the thermodynamic context, this quasi-static problem is in a thermodynamic equilibrium at each instance. Equation \eqref{eq:fluid_balance} is a nonlinear, time-dependent, and of advection-diffusion type when solved for $c_R$ or $\lambda^s$.
    Equation \eqref{eq:fluid_balance} is first order in time and second order in space, i.e., 
    a nonlinear parabolic PDE.
    On the other hand, when it is solved for $\phi$, 
    a minus sign appears in front of the time derivative term (see equation \eqref{eq:fluid_balance_rewritten-2}), which is rather unusual.
    
    \item \textbf{Nonlinearities:} They appear due to two reasons in Formulation \ref{form_1}.
    First, the constitutive equations for $\mathbf{P}$ and $\mu$ are nonlinear. The constitutive theories are formulated in a large deformation setting with compressibility constraints. Second, the coupling terms enter in a nonlinear fashion in the respective other problem.
    
    \item \textbf{Type of coupling:} The coupling is of domain-type and occurs via coefficients and solution variables. In a partitioned approach (see Section \ref{sec_coupling_strategies}), assuming that one variable is given, the displacement PDE displays only the geometric nonlinearity coming from the Neo-Hookean type of constitutive equation for the deformation. On the other hand, the fluid balance concentration PDE can become quasi-linear if the problem is solved for $c_R$ or $\lambda^s$ because a lower-order term of the solution variable is multiplied with the highest derivative. It can also remain fully nonlinear if the problem is solved for $\phi$. 
\end{enumerate}

\subsection{Temporal and spatial discretization}

In this section, we discuss the discretization in time and space. A classical finite difference scheme is employed for the temporal discretization of the fluid balance concentration PDE, resulting in a quasi-stationary solution in space at each time point. The spatial discretization is based on a Galerkin FEM formulation \cite{Cia87}. Here, due to the structure of Formulation \ref{form_1} of saddle-point type, the inf-sup stability, i.e., LBB (Ladyzhenskaya-Babuska-Brezzi), must be guaranteed, which requires on the discrete level using Taylor-Hood elements, also known as $\mathcal{P}_2 / \mathcal{P}_1$ elements. The Taylor-Hood elements consist of quadratic basis functions ($\mathcal{P}_2$) for the displacement field and linear basis functions ($\mathcal{P}_1$) for the fluid concentration-related field. As just mentioned, the main reason for using Taylor-Hood elements lies in their ability to satisfy the LBB  stability condition, which helps avoid numerical instabilities like spurious chemical potential oscillations \citep{Bouklas2015Nonlinear}. 
The notation in this section mainly matches the notation adopted by \cite{Logg2012FenicsBook} and \cite{Wick2020multiphysicsPFF}.

Assume the computational domain $\mathcal{B}_R$ is partitioned into open elements $K$ that depend on the spatial dimension $d$. A mesh consists of quadrilateral, triangular, or hexahedron cells $K$, all of them available in FEniCS. Here, we employ hexahedron cells. 
They perform a non-overlapping cover of the computational domain $\mathcal{B}_R \subset \mathbb{R}^d$. 
Let $\mathcal{T}_h = \lbrace K \rbrace$ be a conforming mesh of the bounded domain $\mathcal{B}_R \subset \mathbb{R}^d$, with mesh size $h$.

We employ Taylor-Hood elements, i.e., a pair of finite element spaces 
$(V_h^{TH},Q_h^{TH})$, 
where:
\begin{itemize}
    \item $Q_h^{TH} \subseteq Q$ is the space of continuous, piecewise linear functions on $\mathcal{T}_h$, i.e.,
    \begin{equation}
        Q_h^{TH}(\mathcal{T}_h) : = \Big\lbrace q_h \in \left[ C \left( \mathcal{B}_h \right) \right] \Big\vert ~ q_h\vert_K \in \left[\mathcal{P}_1 \left( K \right)\right] ~~ \forall K \in \mathcal{T}_h, ~ q_h \vert_{\partial \mathcal{B}_{q_h}} = 0 \Big\rbrace.
    \end{equation}
    
    \item $V_h^{TH} \subseteq V$ is the space of continuous, piecewise quadratic functions on $\mathcal{T}_h$, i.e.,
    \begin{equation}
        V_h^{TH}(\mathcal{T}_h) : = \Big\lbrace \mathbf{v}_h \in \left[ C \left( \mathcal{B}_h \right) \right]^d \Big\vert ~ \mathbf{v}_h\vert_K \in  \left[\mathcal{P}_2 \left( K \right)\right]^d ~~ \forall K \in \mathcal{T}_h, ~ \mathbf{v}_h \vert_{\partial \mathcal{B}_{\mathbf{v}_h}} = 0 \Big\rbrace. 
    \end{equation}
    
\end{itemize}
Here, $\mathcal{P}_1(K)$ denotes the space of linear polynomials over the element $K$ in $\mathbb{R}$ and $\mathcal{P}_2(K)^d$ denotes the space of quadratic polynomials over the element $K$ in $\mathbb{R}^d$.

Thus, the discrete variational monolithic formulation for the diffusion-deformation model reads: 

\begin{formulation} (Semi-discrete in space variational monolithic diffusion-deformation of gels in $\mathcal{T}_h$). 
\label{form_2}
    Find $(\mathbf{u}_h, \varphi_h) \in \{\bar{\mathbf{u}}_h|_{\partial \mathcal{B}_{\bar u}} + V_h^{TH}\} \times \{\bar\varphi_h|_{\partial \mathcal{B}_{\bar\varphi}} + Q_h^{TH}\}$ , with $\varphi_h(0) = \varphi_0$, such that for $t 
    \in I$ it holds
    \begin{equation}\label{eq:discrete_var_mono}
        \begin{aligned}
        a((\mathbf{u}_h, \varphi_h))(\mathbf{v}_h) + b(\mathbf{v}_h) & = \mathbf{0}, ~~ \forall \mathbf{v}_h \in V_h^{TH}, \\
        c((\mathbf{u}_h, \varphi_h))(q_h) + d(q_h) & = 0, ~~ \forall q_h \in Q_h^{TH}.
        \end{aligned}
    \end{equation}
\end{formulation}

Moreover, the discrete variational formulation using the staggered approach reads:

\begin{formulation} (Semi-discrete in space variational diffusion-deformation of gels in $\mathcal{T}_h$). 
\label{form_3}
    Find $\mathbf{u}_h \in \{\bar{\mathbf{u}}_h|_{\partial \mathcal{B}_{\bar u}} + V_h^{TH}\}$ and $\varphi_h \in \{\bar\varphi_h|_{\partial \mathcal{B}_{\bar\varphi}} + Q_h^{TH}\} $, with $\varphi_h(0) = \varphi_0$, such that, for the nonlinear iterations $j = 1, 2, \dots$  and  $t \in I$ it holds
    \begin{equation}\label{eq:discrete_var_stagg}
        \begin{aligned}
            a((\hat{\bf u}_h^{j}, \hat{\varphi}_h^{j-1}))(\mathbf{v}_h) + b(\mathbf{v}_h) & = \mathbf{0}, ~~ \forall \mathbf{v}_h \in V_h^{TH}, \\
            c((\hat{\bf u}_h^{j}, \hat{\varphi}_h^{j}))(q_h) + d(q_h) & = 0, ~~ \forall q_h \in Q_h^{TH},
        \end{aligned}
    \end{equation}
    until the iteration converges, i.e., equation \eqref{eq:staggered_conver} is fulfilled.
\end{formulation}

The time-dependent term in the balance of fluid concentration is approximated using the first-order implicit Euler discretization for $n = 1, 2, \dots, N_f$, with $N_f$ the final simulation time index at the final time $T$, as
\begin{equation}
    \partial_t c_R(\varphi) \approx \frac{c_R(\varphi^n) - c_R(\varphi^{n-1})}{\Delta t},
\end{equation}
where $\varphi^n$ and $\varphi^{n-1}$ are the value of $\varphi$ at the current and previous time step, respectively, and $\Delta t$ is the time step increment. Following \cite{Wick2020multiphysicsPFF}[Chapter 5, Definition 52, p. 90],
we split the semi-linear form $c((\mathbf{u},\varphi))(q)$ into time derivative and non-derivative
terms. To this end, we have
\begin{align}
c((\mathbf{u},\varphi))(q) &:= c_T((\varphi))(q) + c_E((\mathbf{u},\varphi))(q) \nonumber \\ 
&:= \int_{\mathcal{B}_R} \partial_t c_R (\varphi) \cdot q\, dV + \int_{\mathcal{B}_R} {\bf M}({\bf u}, \varphi) \nabla \mu \left( \mathbf{u}, \varphi \right) \cdot \nabla q dV\, .
\end{align}
Then, the difference approximation of the time derivative with time step increment $\Delta t$ yields:
\begin{equation}
c_T((\varphi))(q) \approx c_T^{\Delta t}((\varphi^{n}))(q) := 
\int_{\mathcal{B}_R}  \frac{c_R(\varphi^n) - c_R(\varphi^{n-1})}{\Delta t}\cdot q dV\, .
\end{equation}

Then, the fully discrete variational monolithic formulation for the diffusion-deformation model reads: 

\begin{formulation} (Fully-discrete in space variational monolithic diffusion-deformation of gels in $\mathcal{T}_h$). 
 \label{form_4}
    Let $h$ be the spatial discretization parameter and $n$ the current time point index.
    Find $( \mathbf{u}_h^n, \varphi_h^n) \in \{\bar{\mathbf{u}}_h|_{\partial \mathcal{B}_{\bar u}} + V_h^{TH}\} \times \{\bar\varphi_h|_{\partial \mathcal{B}_{\bar\varphi}} + Q_h^{TH}\}$, with $\varphi(0) = \varphi_0$, such that for $n=1,\ldots, N_f$ it holds
    \begin{equation}\label{eq:fully_discrete_var_mono}
        \begin{aligned}
        a(( \mathbf{u}_h^n, \varphi_h^n))(\mathbf{v}_h) + b(\mathbf{v}_h) & = \mathbf{0}, ~~ \forall \mathbf{v}_h \in V_h^{TH}, \\
        c_T^{\Delta t}((\varphi_h^{n}))(q_h) + c_E(( \mathbf{u}_h^n, \varphi_h^n ))(q_h) + d(q_h) & = 0, ~~ \forall q_h \in Q_h^{TH}.
        \end{aligned}
    \end{equation}
\end{formulation}
We notice that the abstract cycle from monolithic problem statements until the final linear solution is outlined in \cite{Wick2020multiphysicsPFF}[Section 7.8.4]. Using the same abstract concept, but replacing the monolithic nonlinear solution with some iteration (Formulation \ref{form_3}), we obtain the fully discrete variational formulation using the staggered approach:

\begin{formulation} (Fully-discrete in space variational diffusion-deformation of gels in $\mathcal{T}_h$). 
 \label{form_5}
    Let $h$ be the spatial discretization parameter and $n$ the current time point index.
    Find $\mathbf{u}_h^n \in \{\bar{\mathbf{u}}_h|_{\partial \mathcal{B}_{\bar u}} + V_h^{TH}\}$ and $\varphi_h^n \in \{\bar\varphi_h|_{\partial \mathcal{B}_{\bar\varphi}} + Q_h^{TH}\}$, with $\varphi(0) = \varphi_0$, such that, for the nonlinear iterations $j = 1, 2, \dots$  and  $n=1,\dots, N_f$ it holds
    \begin{equation}\label{eq:fully_discrete_var_stagg}
        \begin{aligned}
            a((\hat{\bf u}_h^{n,j}, \hat{\varphi}_h^{n,j-1}))(\mathbf{v}_h) + b(\mathbf{v}_h) & = \mathbf{0}, ~~ \forall \mathbf{v}_h \in V_h^{TH}, \\
            c_T^{\Delta t}((\hat{\varphi}_h^{n} ))(q_h) + c_E(( \mathbf{u}_h^n, \varphi_h^n ))(q_h)
            + d(q_h) & = 0, ~~ \forall q_h \in Q_h^{TH},
        \end{aligned}
    \end{equation}
    until the iteration converges at time point $t_n$, i.e., equation \eqref{eq:staggered_conver} is fulfilled, and then proceeds to $t_{n+1}$. Then, we set as initial guesses of the nonlinear iterative scheme at the next time step $( \hat{\bf u}_h^{n+1,0}, \hat{\varphi}_h^{n+1,0}):= ( \mathbf{u}_h^{n}, \varphi_h^{n})$.
\end{formulation}

\subsection{Numerical solution}
\label{sec_numerical_solution}
The numerical consequences of our previous classification (specifically the type of 
nonlinearities) are that we will always have to solve at least one fully nonlinear PDE, independently of whether the problem is solved in a monolithic, Formulation \ref{form_4}, or staggered way, Formulation \ref{form_5}. Here, we utilize a Newton-type solver
\citep{Deuflhard2011} and the consistent linearization of the system of PDEs at each nonlinear iteration step. 
Specifically, for Formulation \ref{form_4}, a convenient way (see \cite{Wick2020multiphysicsPFF}[Section 3.3.3.2 and Section 3.3.3.3]) is to formulate a common semi-linear form
\begin{subequations} \label{eq:final-solving-system}
\begin{equation}
    A(\mathbf{U}_h^n)(\mathbf{\Psi}_h) := a(( \mathbf{u}_h^n, \varphi_h^n))(\mathbf{v}_h) 
+ c_T^{\Delta t}(( \varphi_h^{n}))(q_h)+ c_E(( \mathbf{u}_h^n, \varphi_h^n ))(q_h)\, ,
\end{equation}
and corresponding right-hand side:
\begin{equation}
    F(\mathbf{\Psi}_h) := b(\mathbf{v}_h) + d(q_h)
\end{equation}
\end{subequations}
with the joint unknown, trial function, $\mathbf{U}_h^n := (\mathbf{u}_h^n,\varphi_h^n)$ and the joint test function 
$\mathbf{\Psi}_h := (\mathbf{v}_h,q_h)$. This corresponds to Step 4 \cite{Wick2020multiphysicsPFF}[Section 7.8.4.1]. 
Then, we can proceed with Step 5 (Newton's method) in \cite{Wick2020multiphysicsPFF}[Section 7.8.4.1] for the nonlinear solution. The resulting solving system is analogous to equations \eqref{eq:final-solving-system} for Formulation \ref{form_5}, but with $\mathbf{U}_h^n$ replaced by $\hat{\mathbf{U}}_h^{n,j} := (\hat{\bf u}_h^{n,j},\hat{\varphi}_h^{n,j})$ that collects trial values of unknown fields at step $j$ within the iterative solution scheme.

In this work, automatic differentiation offered by FEniCS was employed rather than calculating the Jacobian by hand.
Sparse LU decomposition (Gaussian elimination) is used inside each Newton step to solve the arising linear equation systems. As we have the specific derivations in Formulation \ref{form_4} and Formulation \ref{form_5} at hand, a future extension is to employ iterative methods, like GMRES - generalized minimal residuals \citep{Saad2003}, or multigrid solvers \citep{Hack1985}, for which however, preconditioners need to be developed.

Finally, it is well known that Dirichlet boundary conditions on the chemical potential, see equation \eqref{eq:fluid_balance}$_4$, might be the source of spurious numerical oscillations due to large sudden pressures within the hydrogel at the start of the simulation. Hence, whenever needed, such boundary condition is incrementally applied during the simulation, a strategy known as \textit{time-ramping boundary condition}. In each case, the boundary condition is increased as fast as possible to meet a good compromise between numerical stability and physical reality. This is achieved by introducing a time-dependent exponential term that multiplies the Dirichlet boundary condition, i.e., $h(\varphi,t) = \bar{h}(\varphi)\left( 1 - \text{exp}(-\alpha_r t) \right)$, where $\alpha_r$ is a positive constant that determines the rate of ramping. 

\section{General settings for the numerical simulation campaign}\label{sec:general-settings}

The coupled diffusion-deformation problem is faced by solving equations \eqref{eq:linear_momentum} with null body forces ${\bf b}_R={\bf 0}$ and \eqref{eq:fluid_balance} and discretized either as in equation \eqref{eq:fully_discrete_var_mono} (Formulation \ref{form_4}, monolithic) or equation \eqref{eq:fully_discrete_var_stagg} (Formulation \ref{form_5}, staggered) together with the specific choices of \textit{Constitutive models I} to \textit{V} introduced in Section \ref{sec:gels_theory}. Table \ref{tab:modelsFeatures} summarizes the main features of the models considered in this work and refers to their respective equations.

Two campaigns of numerical simulations will be presented. Section \ref{sec:Sim_results} addresses different representative prototype problems by adopting parameter settings presented in the original papers where each constitutive model has been originally presented. The aim is to explore the robustness of the developed numerical implementations. Section \ref{res:benchmark} presents a unified benchmark problem to have a unique reference simulation example that shows the differences in the response for each constitutive equation.

The solution strategy outlined in Section \ref{sec_coupling_strategies} is adopted in all simulation cases. The algorithms are implemented in FEniCS, and the code is provided so the reader can reproduce and verify the results (online repository link: \url{https://doi.org/10.25835/5v49yfk0}).
The following material parameters remain constant in all cases: $k_B = 1.38065 \times 10^{-23} ~ [J mol^{-1}]$, $T = 298 ~ [K]$, $\Omega = 1.7 \times 10^{-28} ~ [mol^{-1}]$ and $N\Omega = 0.001 ~ [m^3 mol^{-1}]$. The remaining material parameters vary depending on the specific constitutive model and the adopted simulation setup. They can be found in the caption of the figures associated with each numerical result. For the sake of notation, let ${\bf E}_1, {\bf E}_2, {\bf E}_3$ be introduced as a Cartesian coordinate system in the reference configuration (resp., ${\bf e}_1, {\bf e}_2, {\bf e}_3$ in the current one), parametrized in $X$, $Y$ and $Z$ (resp., $x$, $y$ and $z$). Moreover, let the following stress components be introduced:
\begin{equation}
    P_X = {\bf P}:({\bf e}_1 \otimes {\bf E}_1)\, , \quad P_Y = {\bf P}:({\bf e}_2 \otimes {\bf E}_2)\, , \quad  \sigma_x = \boldsymbol{\sigma}:({\bf e}_1 \otimes {\bf e}_1)\, .
\end{equation}

Numerical results will be reported in terms of displacements $\mathbf{u}$, stretch $\lambda$, and the evolution of $\mu$ since they are the most relevant to analyze from a physical viewpoint.

\begin{sidewaystable}
\centering
\caption{Models' main features summary.}
\label{tab:modelsFeatures}
\begin{tabularx}{\textwidth}{cllccccc}
\toprule
\multirow{2}{*}{\makecell{\textbf{Constitutive} \\ \textbf{model}}} &
\multirow{2}{*}{\textbf{Reference}} &
\multirow{2}{*}{\textbf{Equations}} &
\multirow{2}{*}{\makecell{\textbf{Primary} \\ \textbf{variables}}} &
\multicolumn{2}{c}{\textbf{Solution strategy}} &
\multicolumn{2}{c}{\makecell{\textbf{Incompressible}}} \\ 
\cmidrule(lr){5-6} \cmidrule(lr){7-8}
& & & & \textbf{monolithic} & \textbf{staggered} & \textbf{yes} & \textbf{no} \\ 
\midrule
I &
  \cite{Liu2016TransGels} &
  \eqref{eq:linear_momentum}, \eqref{eq:fluid_balance_mu_Liu}, and \eqref{eq:stress_Liu} &
  $\mu$ and $\mathbf{u}$ &
  x &
   &
  x &
   \\
II &
  \cite{Chester2010DiffDeform}\footnotemark[1] &
  \eqref{eq:linear_momentum}, \eqref{eq:stress_Chester2010}, \eqref{eq:mu_Chester2010}, and \eqref{eq:fluid_balance_rewritten-2}&
  $\phi$ and $\mathbf{u}$&
  x &
  x &
  x &
   \\
III &
  \cite{Bouklas2015Nonlinear} &
  \eqref{eq:linear_momentum}, \eqref{eq:stress_Bouklas2015}, \eqref{eq:mu_Bouklas2015}, and \eqref{eq:fluid_balance_rewritten-1} &
  $J_f$, $\mu$ and $\mathbf{u}$ &
   &
  x &
   &
  x \\
IV &
  \cite{Chester2011thermo} &
  \eqref{eq:linear_momentum}, \eqref{eq:PK1_Chester_energetic-IV}, \eqref{eq:mu_Chester_energetic}$_1$, and \eqref{eq:fluid_balance_rewritten-2} &
  $\phi$ and $\mathbf{u}$ &
  x &
  x &
   &
  x \\
V &
  \cite{Chester2015Abaqus} &
  \eqref{eq:linear_momentum}, \eqref{eq:PK1_Chester_energetic-V}, \eqref{eq:mu_Chester_energetic}$_2$, and \eqref{eq:fluid_balance_rewritten-2} &
  $\phi$ and $\mathbf{u}$ &
  x &
  x &
   &
  x \\
\bottomrule
\end{tabularx}
\footnotetext[1]{\textit{Constitutive model II} originally has $P$ as a primary variable but, as discussed in Section \ref{sec:final-comments-theory}, a functional dependency $P=P(\phi)$ can be analytically determined in some special cases, like those addressed in this work. \\
\textbf{Note:} The rationale behind the selected solution strategies is detailed in the text of Section \ref{sec:Sim_results} for each constitutive model.}
\end{sidewaystable}

\section{Representative prototype problems}
\label{sec:Sim_results}

\begin{figure}
\centering
\includegraphics[width=0.75\textwidth]{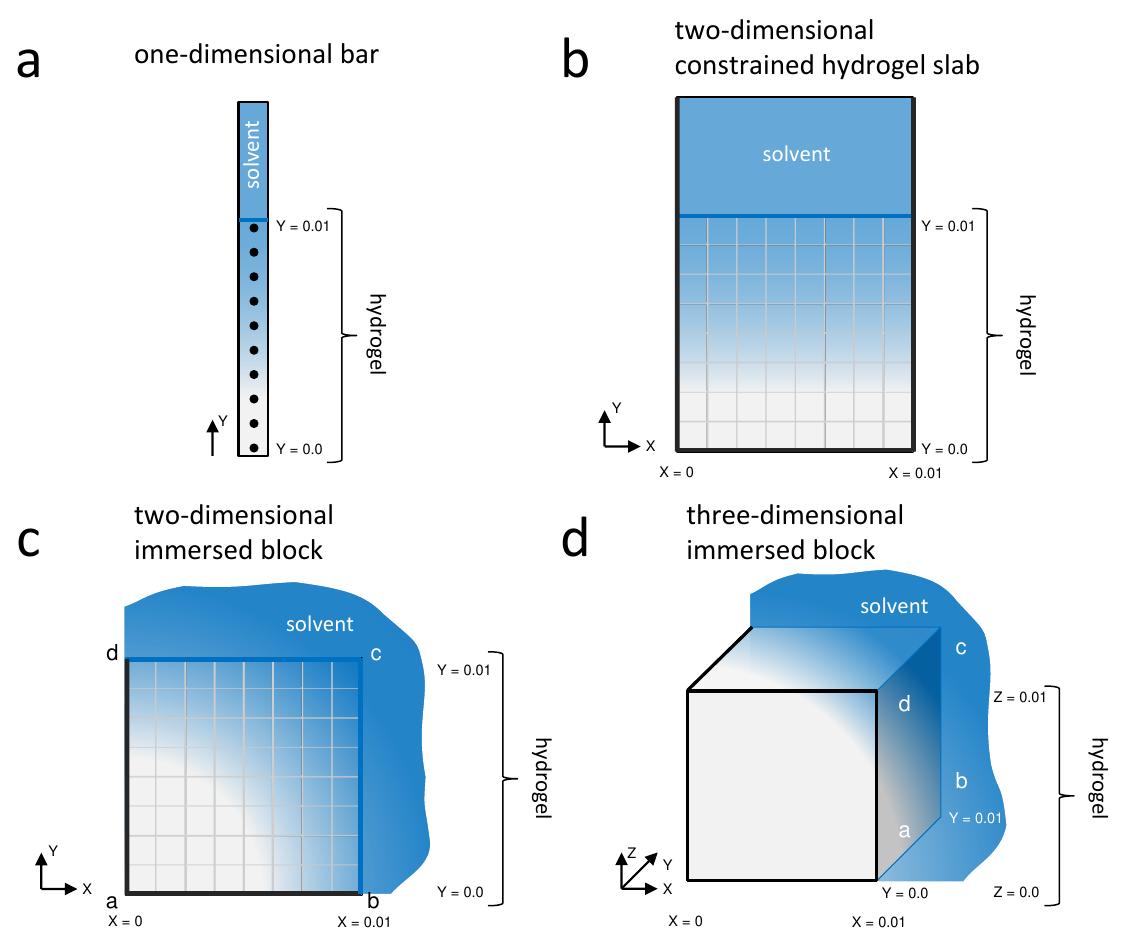}
\caption{\textbf{Representative examples setup}. \textbf{a.} One-dimensional transient swelling of a hydrogel bar. The bar is fixed at $Y = 0.0~[m]$, while the opposite end, $Y = 0.01~[m]$, is exposed to a non-reactive solvent. \textbf{b.} Two-dimensional transient swelling of a constrained hydrogel slab. In this example, the hydrogel block is placed in a rigid container with frictionless walls and the deformation in the $X$ direction is constrained. The top surface at $Y = 0.01~[m]$ keeps traction-free and is in contact with the solvent during deformation. At the bottom surface $Y = 0.0~[m]$, the gel is fixed to the container wall and no fluid is allowed to diffuse through it. Due to the solvent absorption, the hydrogel can only swell along the $Y$ direction. \textbf{c.} Two-dimensional hydrogel block is immersed in a non-reactive solvent with a reference chemical potential $\mu^0 = 0$. Only a quarter of the whole model is considered because of the symmetry of the block. For the mechanical boundary conditions, the nodes along edge \textbf{ab} are prescribed to have displacement component $u_y = 0$, while the nodes along edge \textbf{ad} are prescribed to have $u_x = 0$. The edges \textbf{bc} and \textbf{cd} are taken to be traction-free. For the solvent concentration boundary conditions, the edges \textbf{ab} and \textbf{ad} (the symmetry edges) are prescribed a zero fluid flux, and on the edges \textbf{bc} and \textbf{cd}, the chemical potential is prescribed as $\mu = 0$ on $\partial \mathcal{B}_{\mu},~t=\lbrace 0, T \rbrace$. \textbf{d.} Three-dimensional cube immersed in a non-reactive solvent. Only a quarter of the whole model is considered because of the symmetry of the 3D cube. The mechanical boundary conditions are specified such that the $u_y = 0$ in the front face, $u_x = 0$ in the left face, and $u_z = 0$ in the face in the bottom part. For the solvent concentration boundary conditions, the front, left, and bottom faces (the symmetry faces) are prescribed a zero fluid flux, and on the back, right, and top faces, the chemical potential is prescribed as $\mu = 0$ on $\partial \mathcal{B}_{\mu},~t=\lbrace 0, T \rbrace$. \textbf{Note:} the remaining boundary conditions, together with the initial conditions, are defined depending on the specific constitutive theory adopted to study the diffusion-deformation process.}
\label{fig:case_studies_setup}
\end{figure}

Four representative prototype problems are introduced. First, we consider a \textbf{one-dimensional transient swelling} of a hydrogel bar along the $Y$-direction. The bar is fixed at $Y=0$ and free at $Y=0.01~[m]$ where $P_Y = 0$. At this latter end, the bar is exposed to a non-reactive solvent (see Figure \ref{fig:case_studies_setup}\textbf{a}). The deformation gradient takes the form
\begin{equation}\label{eq:kinematics-1D}
   \text{1D:} \quad \mathbf{F} = \text{diag}(1, \lambda , 1) \, , 
\end{equation}
occurring along the $Y$ direction. For perfectly incompressible models (i.e., \emph{constitutive models I} and \emph{II}), it then results in $\text{det} (\mathbf{F}) = \lambda = J_f = \phi^{-1}$, that is the total stretch is equal to the swelling volume change and inversely proportional to polymer volume fraction. Furthermore, from the equilibrium condition \eqref{eq:linear_momentum}, it follows that $\partial \mathbf{P}/\partial Y = 0$, which, after considering traction free boundary conditions, leads to $P_Y = 0$ and, in consequence, $\sigma_y = 0$. This problem has been studied previously by both linear and nonlinear theories, e.g., \cite{Bouklas2012Linear-NL,Bouklas2015Nonlinear}. Due to the simplicity of its numerical settings, this case study will serve as a reference, providing an estimate of the correct order of magnitude of the quantities of interest, i.e., $\lambda$, $\mu$, $P_X$ or $\sigma_x$, in more complex case studies.

As a second example, we investigate the \textbf{transient swelling of a constrained hydrogel slab in a two-dimensional setting in-plane strain (in the $(X,Y)$ plane)}. In this example, the hydrogel block is placed in a rigid container with frictionless walls, and the deformation in the $X$ direction is constrained. Only the upper part of the hydrogel is exposed to a non-reactive solvent (see Figure \ref{fig:case_studies_setup}\textbf{b}).
This example has been previously considered by, e.g., \cite{Chester2010DiffDeform} and \cite{Liu2016TransGels}, and represents the 2D counterpart of the previously introduced 1D example. However, it is noteworthy that the one-dimensional problem is numerically solved for a single scalar field (representing either $\lambda$ or $\phi$ depending on the constitutive model), and the other quantities of interest are computed in the post-processing stage. On the other hand, the numerical solution of the two-dimensional problem is obtained by considering the complete sets of unknowns, that is, a scalar field describing the fluid content and the vector displacement field $\mathbf{u}$. Then, quantities of interest (e.g., the chemical potential $\mu$) are computed in the post-processing stage.

As a third example, we consider the \textbf{transient free-swelling in a two-dimensional setting (in the $(X,Y)$ plane)} of a polymer gel with an initially square cross-section. A free hydrogel block is immersed in a non-reactive solvent. Due to the symmetry of the deformation, only a quarter of the whole model needs to be considered (see Figure \ref{fig:case_studies_setup}\textbf{c}). A similar setup to this example can be found in, e,g., \cite{Chester2011thermo} and \cite{Liu2016TransGels}. At the steady state, the deformation gradient is uniform within the domain, reading in the Cartesian representation as:
\begin{equation}\label{eq:kinematics-2D}
\text{Free swelling:} \qquad \mathbf{F}_{\infty}^{2D} = \text{diag}\left(\lambda_{\infty}^{2D}, \lambda_{\infty}^{2D}, 1 \right)\, , 
\end{equation}
with $\lambda_{\infty}^{2D}$ referring to the (constant) 2D final stretch at the steady state.

The fourth and last example corresponds to \textbf{the extension of the two-dimensional block example into three-dimensions}, namely, a cube is immersed in a non-reactive solvent to swell due to the solvent absorption freely (see Figure \ref{fig:case_studies_setup}\textbf{d}). In the steady state, the deformation gradient results:
\begin{equation}
    \text{Free swelling:} \qquad \mathbf{F}_{\infty}^{3D} = \lambda_{\infty}^{3D} \mathbf{I}
\end{equation}
with $\lambda_{\infty}^{3D}$ referring to the (constant) 3D final stretch at the steady state.
This simulation resembles that presented, e.g., by \cite{Lucantonio2013transientGels} and will be only performed considering the \textit{constitutive model I}. 

We estimate the convergence order with a well-known heuristic formula.
Let us denote the errors by $E_h$, $E_{h/2}$, and $ E_{h/4}$, where $h$ is the 
mesh size parameter as before. Under the assumption that our discretized problem has a convergence order of $p$, then each error should be roughly $(1/2)^p$ times the previous error. Therefore, we can estimate $p$ by taking the logarithm base $2$ of the error ratios as:
\begin{equation}
    p = \dfrac{1}{\log (2)} \log \left( \left\vert \dfrac{E_h - E_{h/2}}{ E_{h/2} - E_{h/4}} \right\vert \right).
\end{equation}

\subsection{Constitutive model I}

\textbf{One-dimensional transient swelling.} Here, we follow the ideas presented by \cite{Liu2016TransGels}. 
Recalling equations \eqref{eq:stress_Liu} and \eqref{eq:p_mu_Liu} with equation \eqref{eq:kinematics-1D}, the balance of linear momentum \eqref{eq:linear_momentum} reduces to:
\begin{equation}\label{eq:mu_1D_Liu}
    P_Y = -\dfrac{\mu}{\Omega} + \dfrac{k_B T}{\Omega} \left[ \ln \left( 1 - \dfrac{1}{\lambda} \right) + \dfrac{1}{\lambda} + \dfrac{\chi}{(\lambda)^2} \right] + G_0 \left(\lambda - \dfrac{1}{\lambda} \right) = 0.
\end{equation}

Additionally, the fluid balance equation \eqref{eq:fluid_balance} in a one-dimensional setting yields
\begin{equation}\label{eq:fluid_balance_1D_Liu}
    \begin{cases}
       \dfrac{\partial_t \lambda}{\Omega} = \dfrac{\partial}{\partial Y} \left( \dfrac{D}{\Omega k_B T} (\lambda)^{-1} \dfrac{\partial \mu}{\partial Y} \right), & ~ \text{in} ~ \mathcal{B}_{R}, \\
        \lambda = \bar{\lambda}, & ~ \text{at} ~ Y = 0.01, \\
        \dfrac{\partial \lambda}{\partial Y} = 0, & ~ \text{at} ~ Y = 0.0, \\
        \lambda(t = 0) = \lambda_0, & ~ \text{in} ~ \mathcal{B}_R.
    \end{cases}
\end{equation}

Differentiating equation \eqref{eq:mu_1D_Liu} with respect to $Y$ yields
\begin{equation}\label{eq:du_dY_Liu}
    \dfrac{\partial \mu}{\partial Y} = k_B T \left[  \dfrac{1}{\lambda(\lambda - 1)} - \dfrac{1}{(\lambda)^2} - \dfrac{2\chi}{(\lambda)^3} \right]\dfrac{\partial \lambda}{\partial Y} + \Omega G_0 \left( 1 + \dfrac{1}{(\lambda)^2} \right)\dfrac{\partial \lambda}{\partial Y} \, . 
\end{equation}

Substituting equation \eqref{eq:du_dY_Liu} into equation \eqref{eq:fluid_balance_1D_Liu} leads to a nonlinear partial differential equation with respect to $\lambda(Y, t)$. 

The weak discretized form of equation \eqref{eq:fluid_balance_1D_Liu} reads
\begin{equation}\label{eq:weak_form_mu_1D_Liu}
    \int_{\mathcal{B}_R} \left( \dfrac{\lambda_h^n - \lambda_h^{n-1}}{\Delta t} q_h + \dfrac{D}{\Omega k_B T} (\lambda_h^n)^{-1}  \dfrac{\partial \mu_h^n}{\partial Y} \dfrac{\partial q_h}{\partial Y} \right)dY = 0, ~ \forall q_h \in Q_h \, , 
\end{equation}
where $\partial \mu_h^n/\partial Y$ is given by equation \eqref{eq:du_dY_Liu} evaluated at $\lambda_h^n$. 
After solving equation \eqref{eq:weak_form_mu_1D_Liu}, the chemical potential can be obtained from \eqref{eq:mu_1D_Liu} and the stress component $\sigma_x$ in the direction orthogonal to swelling as $\sigma_x = p_{\mu}(\lambda,\mu)$, with $p_{\mu}$ from equation \eqref{eq:p_mu_Liu}.

\begin{figure}
\centering
\includegraphics[width=0.75\textwidth]{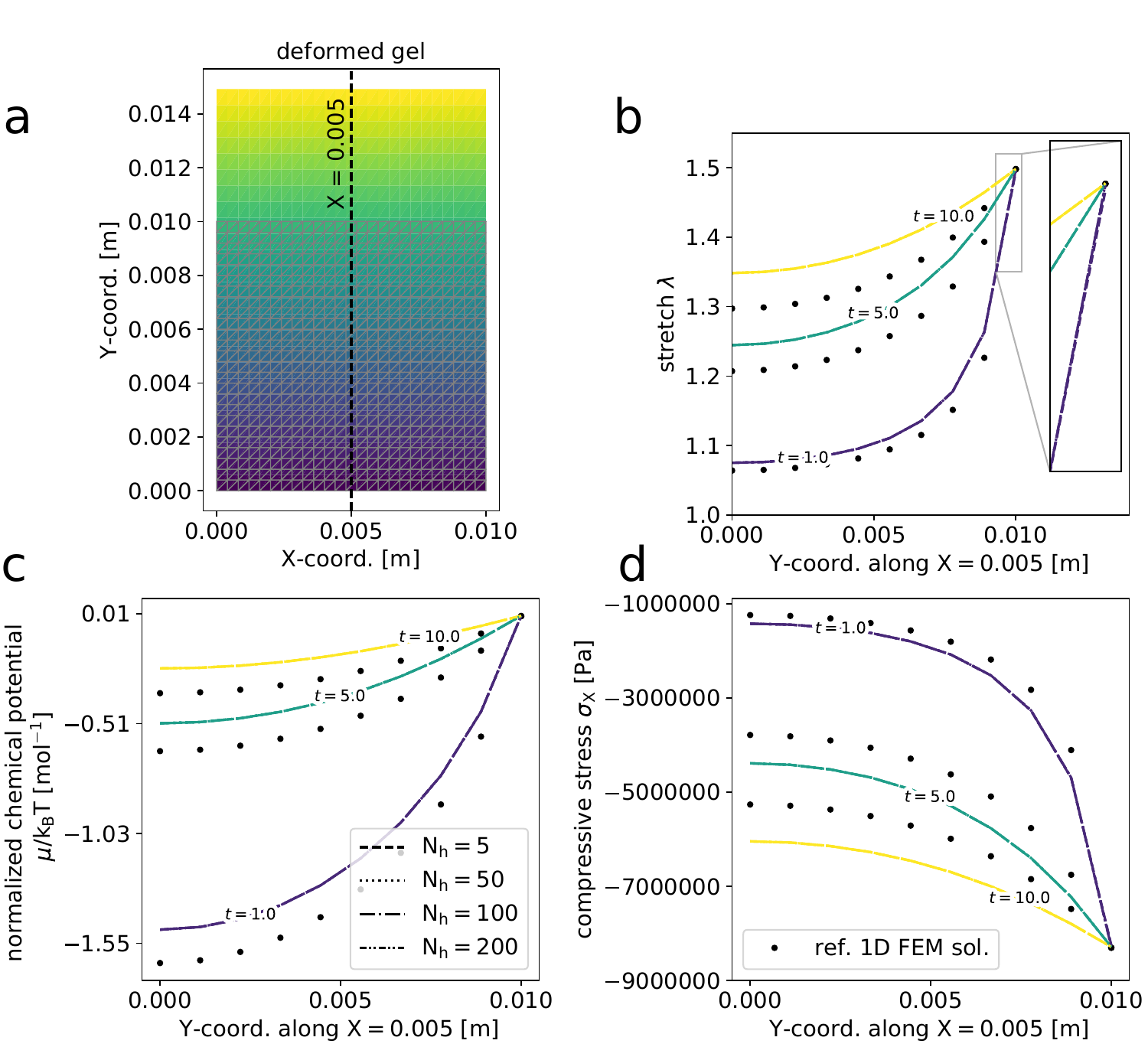}
\caption{\textbf{\emph{Constitutive model I}:} \textbf{one-dimensional bar (black dots) and two-dimensional hydrogel constrained slab (colored lines) numerical solution comparison for different mesh densities $N_h$ and at different simulation times.} \textbf{a.} Deformed two-dimensional constrained slab at $t = 10.0~[s]$. 
\textbf{b.} Stretch due to swelling $\lambda$. \textbf{c.} Chemical potential $\mu$ normalized by $k_B T$. \textbf{d.} Cauchy compressive stress $\bm{\sigma}_X$. \textbf{Simulation parameters:} $G_0 = 10~[MPa], ~~ \chi = 0.2~[--], ~~ D = 2.0\times10^{-5}~[m^2 s^{-1}]$.}
\label{fig:slab_Liu_conv}
\end{figure}

The numerical results after solving equation \eqref{eq:weak_form_mu_1D_Liu} with the FEM are presented in Figure \ref{fig:slab_Liu_conv} (black dots). A one-dimensional mesh is created to discretize the hydrogel bar, with the number of elements and time steps defined from a convergence study on $\lambda(Y,t)$ (results not shown). In the end, the reference solution is obtained with $200$ elements and $50$ time steps. The boundary condition at $Y = 0.01~[m]$ is obtained after solving equation \eqref{eq:mu_1D_Liu} for $\lambda$ with $\mu = 0$, which yields $\lambda(Y=0.01, t) = \bar{\lambda} = 1.498$. The initial condition is defined as $\lambda_0 = 1.0$.
It is worth mentioning that the obtained results accurately capture the ones presented by \cite{Liu2016TransGels} in Figure 3 therein for the same problem setup.   

\textbf{Two-dimensional constrained hydrogel slab.} \textit{Constitutive model I} is now solved for the constrained hydrogel slab considered in Figure \ref{fig:case_studies_setup}\textbf{b}. 

Figure \ref{fig:slab_Liu_conv} presents the comparison between the solution previously obtained for the one-dimensional bar (black dots) and the two-dimensional constrained slab examples (different colored lines). 
For the two-dimensional case, we show results from different simulations with increasing number of elements and time steps. The deformed hydrogel at $t = 10.0~[s]$ is displayed in Figure \ref{fig:slab_Liu_conv}\textbf{a}, and a time step equal to $0.2$ ($50$ time steps) was used to get Figure \ref{fig:slab_Liu_conv}\textbf{a}. It is observed from Figures \ref{fig:slab_Liu_conv}\textbf{b} - \textbf{d} that differences in numerical solution are rather small for mesh densities higher than $50$. A zoom-in is included in Figure \ref{fig:slab_Liu_conv}\textbf{b} to better distinguish between the different mesh densities. The discrepancy between the one- and two-dimensional cases can be assessed from Figure \ref{fig:slab_Liu_conv}, which shows the effect of approximating the time derivative in the fluid concentration balance through the displacement and increasing the problem's dimension. 

From Figure \ref{fig:slab_Liu_conv}\textbf{c}, it is observed that the chemical potential's rate increases fast in the beginning when the solvent starts entering the gel and becomes slower when it tends to the steady state. Stress $\sigma_x$ follows a similar pattern as observed in Figure \ref{fig:slab_Liu_conv}\textbf{d}. This behavior can be understood as follows. The gradient of $\mu$ is relatively large close to the top surface, at $Y = 0.01~[m]$, and the compressive stress in the interior part of the hydrogel is the smallest, which are both helpful for the diffusion of the solvent content. The hydrogel's network is relaxed and allows easy solvent absorption. The gradient of $\mu$ becomes smaller and $\sigma_x$ larger as the solvent's concentration increases in the hydrogel. This prevents the solvent from penetrating the hydrogel further and slows the diffusion process. There is less space within the hydrogel network, and it starts to saturate. Therefore, each quantity approaches the corresponding steady solution at a decreasing rate.

Next, a computational convergence analysis is performed to investigate the robustness and computational cost of the monolithic approach. We focus on investigating the effect of mesh density and the time step size on the behavior of the implemented numerical algorithm. We aim to understand how these parameters influence the performance of the algorithm in solving the nonlinear system of equations associated with the FEM discretization.

\begin{figure}[htbp]
\centering
\includegraphics[width=0.75\textwidth]{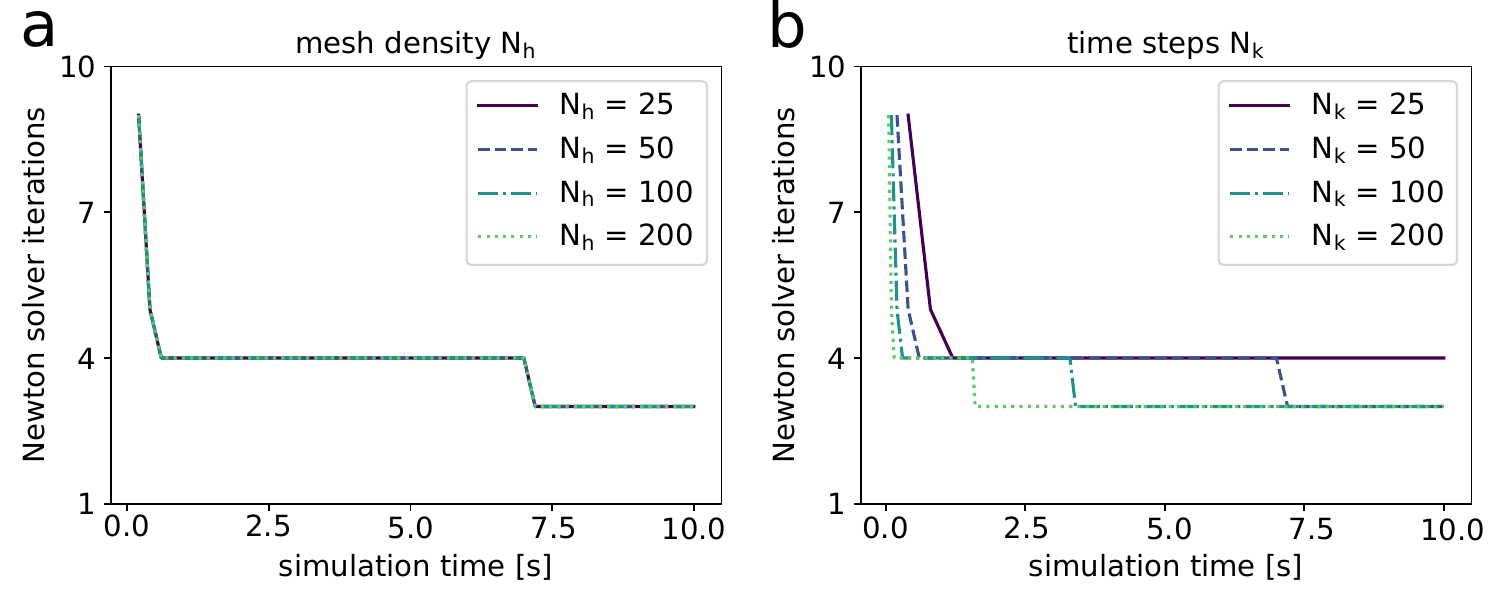}
\caption{\textbf{\emph{Constitutive model I}:} \textbf{newton iterations along the time steps.} \textbf{a}. For different mesh densities ($N_h$). \textbf{b}. For different time step sizes ($N_k$).}
\label{fig:slab_Liu_conv_Newton_iteration}
\end{figure}

First, we conduct performance studies of the Newton solver for each time step to assess its efficiency in solving the nonlinear equations. We observe whether the mesh density and time step size influence the number of Newton iterations, determining if it remains consistent throughout the simulation. Figure \ref{fig:slab_Liu_conv_Newton_iteration} depicts time versus the number of Newton iterations, where Figure \ref{fig:slab_Liu_conv_Newton_iteration}\textbf{a} showcases variations in mesh density and Figure \ref{fig:slab_Liu_conv_Newton_iteration}\textbf{b} displays variations in time step size. From these findings, it is clear that at the beginning of the simulation, the number of Newton iterations is highest but subsequently decreases until it reaches a steady value of three iterations.

\begin{figure}[htbp]
\centering
\includegraphics[width=0.75\textwidth]{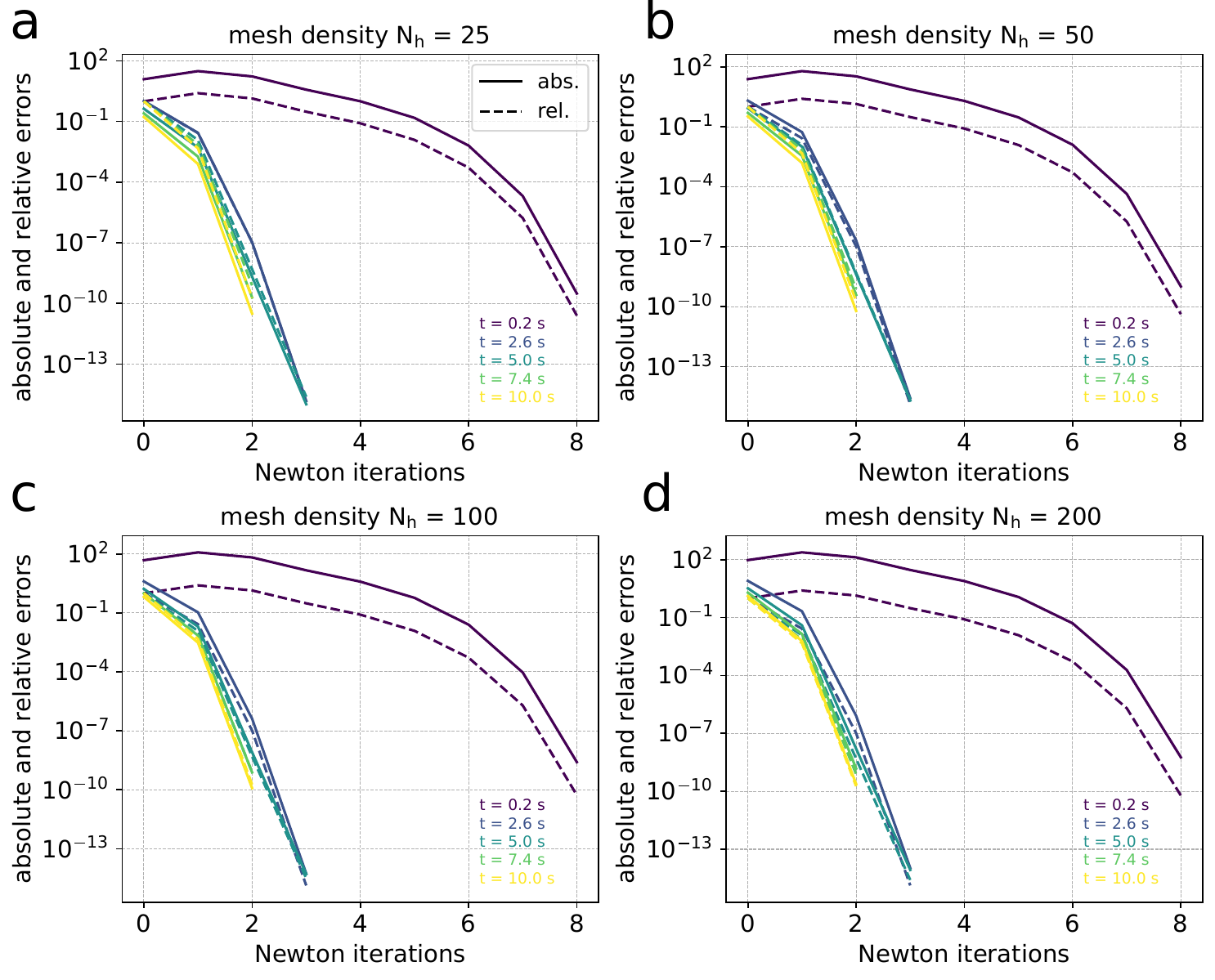}
\caption{\textbf{\emph{Constitutive model I}:} \textbf{convergence analysis of the Newton solver for different mesh densities ($N_h$).} \textbf{a}. $N_h = 25$. \textbf{b}. $N_h = 50$. \textbf{c}. $N_h = 100$. \textbf{d}. $N_h = 200$.}
\label{fig:slab_Liu_conv_Newton_error_mesh}
\end{figure}

\begin{figure}[htbp]
\centering
\includegraphics[width=0.75\textwidth]{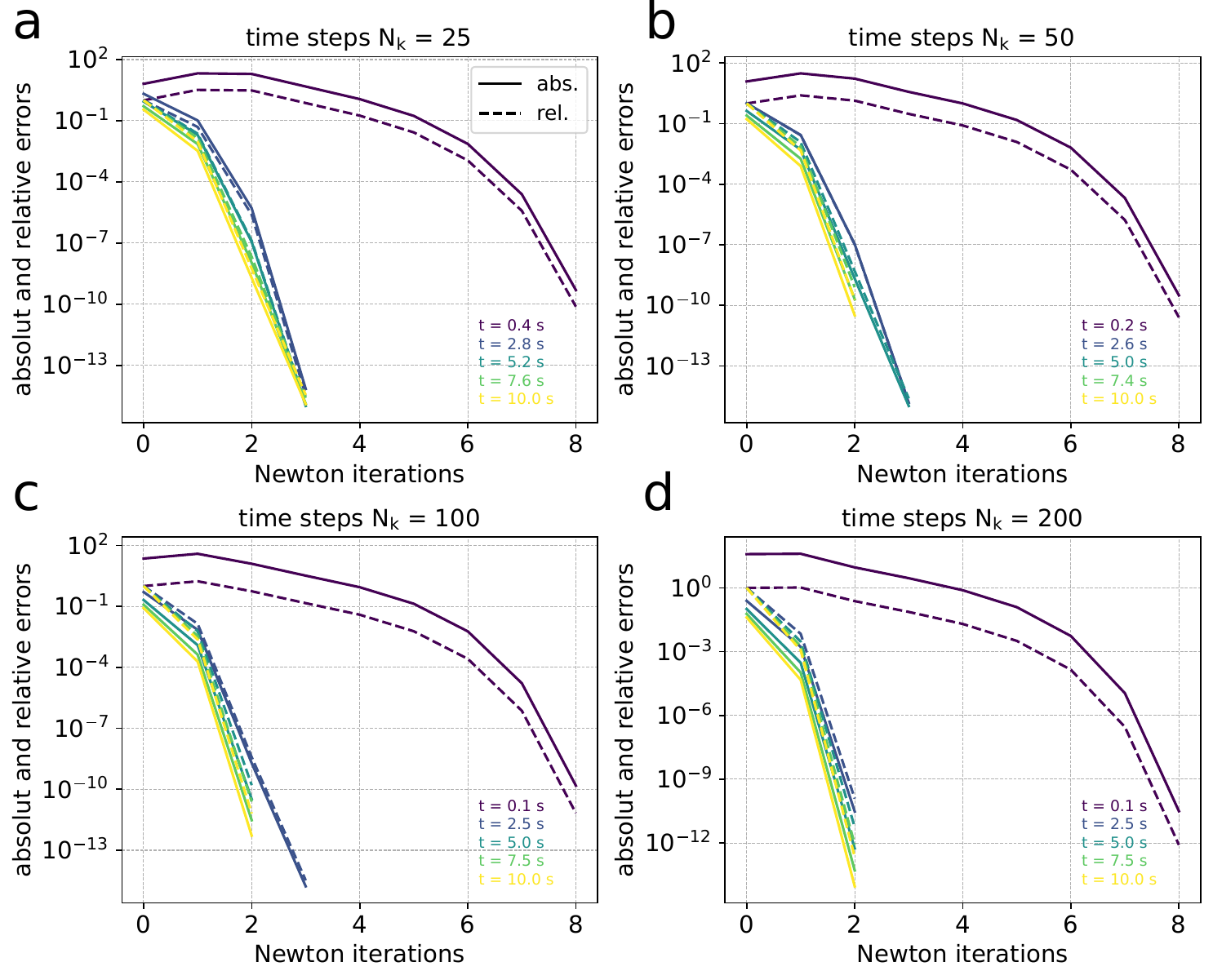}
\caption{\textbf{\emph{Constitutive model I}:} \textbf{convergence analysis of the Newton solver for different time steps size ($N_k$).} \textbf{a}. $N_k = 25$. \textbf{b}. $N_k = 50$. \textbf{c}. $N_k = 100$. \textbf{d}. $N_k = 200$.}
\label{fig:slab_Liu_conv_Newton_error_time}
\end{figure}

Additionally, we examine the Newton algorithm's convergence behavior at five different time points during the simulation. We aim to confirm whether the Newton iteration has quadratic convergence in error.
We also investigate whether changes in mesh density or time steps impact the Newton iteration's convergence. Figures \ref{fig:slab_Liu_conv_Newton_error_mesh} and \ref{fig:slab_Liu_conv_Newton_error_time} demonstrate the convergence behavior, considering both absolute and relative errors. Figure \ref{fig:slab_Liu_conv_Newton_error_mesh} presents the results for various mesh density values, while Figure \ref{fig:slab_Liu_conv_Newton_error_time} displays the results for different time step sizes. Figures \ref{fig:slab_Liu_conv_Newton_error_mesh} and \ref{fig:slab_Liu_conv_Newton_error_time} reveal that the error decays very quickly for all the cases, except for the first iteration. But, the Newton iteration also converges in only $8$ iterations in this case. These results support the efficiency and robustness of the Newton algorithm. 
These findings are relevant as they assure the reliability of the numerical solution for the coupled problem solved using a monolithic approach.

After establishing that the Newton solver is reliable, we test whether using Taylor-Hood elements and the Euler method leads to the expected convergence in space and time. Specifically, we measure the $L_2$ error for various mesh densities and time step sizes with respect to the highest fidelity solution and the values of both displacement and chemical potential at the center of the top face of the two-dimensional hydrogel slab.

The convergence analysis is detailed in Tables \ref{tab:conv_analysis_constI_mesh} and \ref{tab:conv_analysis_constI_time}, where the results demonstrate a second-order convergence for spatial discretization and a first-order convergence for temporal discretization. These findings are in accordance with the theoretically predicted orders from the FEM and Euler discretization scheme. 

\begin{sidewaystable}
\centering
\caption{Spatial discretization convergence analysis for the two-dimensional constrained hydrogel slab example considering \textit{constitutive model I}.}
\label{tab:conv_analysis_constI_mesh}
\begin{tabular}{cccccccccc}
\hline
\textbf{Level} &
  \textbf{\begin{tabular}[c]{@{}c@{}}Mesh density\\ $N_h$\end{tabular}} &
  \textbf{\begin{tabular}[c]{@{}c@{}}Time steps \\ $N_k$\end{tabular}} &
  \bm{$h$} &
  \textbf{Elements} &
  \textbf{DoFs} &
  \textbf{$\mu$(0.5.1.0)} &
  \textbf{L2 error for $\mu$} &
  \textbf{$\mathbf{u}$(1.0.1.0)} &
  \textbf{L2 error for $\mathbf{u}$} \\ \hline
1 &	25 &	50 &	0.04 &	1250 &	5878 &	-0.44017195 &	1.3896e-4 &	0.33459041 &	3.5884e-5 \\
2 &	50 &	50 &	0.02 &	5000 &	23003 &	-0.43999694 &	3.4425e-5 &	0.33461516 & 	9.0131e-6 \\
3 &	100 &	50 &	0.01 &	20000 &	91003 &	-0.43994541 &	7.1092e-6 &	0.33462156 &	1.9465e-6 \\
4 &	200 &	50 &	0.005 &	80000 &	362003 &	-0.43993062 &	-- &	0.33462318 &	--           \\ \hline
\textbf{conv. order} & \textbf{} & \textbf{} & \textbf{} & \textbf{} & \textbf{} & \textbf{1.80} & \textbf{1.93} & \textbf{1.98} & \textbf{1.93} \\ \hline
\end{tabular}%
\end{sidewaystable}



\begin{sidewaystable}
\centering
\caption{Time discretization convergence analysis for the two-dimensional constrained hydrogel slab example considering \textit{constitutive model I}.}
\label{tab:conv_analysis_constI_time}
\begin{tabular}{cccccccc}
\hline
\textbf{Level} &
  \textbf{\begin{tabular}[c]{@{}c@{}}Mesh density\\ $N_h$\end{tabular}} &
  \textbf{\begin{tabular}[c]{@{}c@{}}Time steps \\ $N_k$\end{tabular}} &
  \bm{$k$} &
  \textbf{$\mu$(0.5.1.0)} &
  \textbf{L2 error for $\mu$} &
  \textbf{$\mathbf{u}$(1.0.1.0)} &
  \textbf{L2 error for $\mathbf{u}$} \\ \hline
1 &	25 &	25 &	0.4 &	-0.44915384 &	1.2353e-2 & 0.33195688 &	3.1447e-3 \\
2 &	25 &	50 &	0.2 &	-0.44017195 &	5.5713e-3 &	0.33459041 &	1.4267e-3 \\
3 &	25 &	100 &	0.1 &	-0.43534375 &	1.9333e-3 &	0.33601343 &	4.9668e-4 \\
4 &	25 &	200 &	0.05 &	-0.43277477 &	-- &	0.33677258 &	--                      \\ \hline
\textbf{conv. order} & \textbf{} & \textbf{} & \textbf{} & \textbf{0.91} & \textbf{0.90} & \textbf{0.91} & \textbf{0.88} \\ \hline
\end{tabular}
\end{sidewaystable}


Finally, it is noteworthy that we tested both monolithic and partitioned approaches for the solution of the \emph{constitutive model I} by \cite{Liu2016TransGels}. However, only the monolithic implementation produced physically consistent results. The reason seems to be due to the reformulation of the concentration-time derivative. The time-dependent concentration term is expressed in terms of $\text{Grad}(\partial_t \mathbf{u}):\mathbf{F}^{-T}$ in equation \eqref{eq:fluid_balance_mu_Liu}. Hence, such a term is highly nonlinear in terms of displacements, and its accuracy is highly affected by the adopted spatial discretization. Moreover, if the deformation process is very slow, i.e., $\partial_t \mathbf{u} \approx 0$, which is the case here, the time-dependent term in the balance of fluid concentration may become negligible, and this equation becomes quasi-static. These issues cause some numerical difficulties in the staggered approach to capture the transient behavior of the coupled problem. 

\textbf{Two-dimensional free swelling of a square block.} A square block is immersed in a solvent with a reference chemical potential $\mu^0 = 0$, as illustrated in Figure \ref{fig:case_studies_setup}\textbf{c}. 
Recalling equations \eqref{eq:stress_Liu} and \eqref{eq:p_mu_Liu}, and following the same procedure as in \cite{Liu2016TransGels}, the theoretical value of the steady state stretching results $\lambda_{\infty}^{2D} = 1.35$ (see equation \eqref{eq:kinematics-2D}). 

\begin{figure}
\centering
\includegraphics[width=1.0\textwidth]{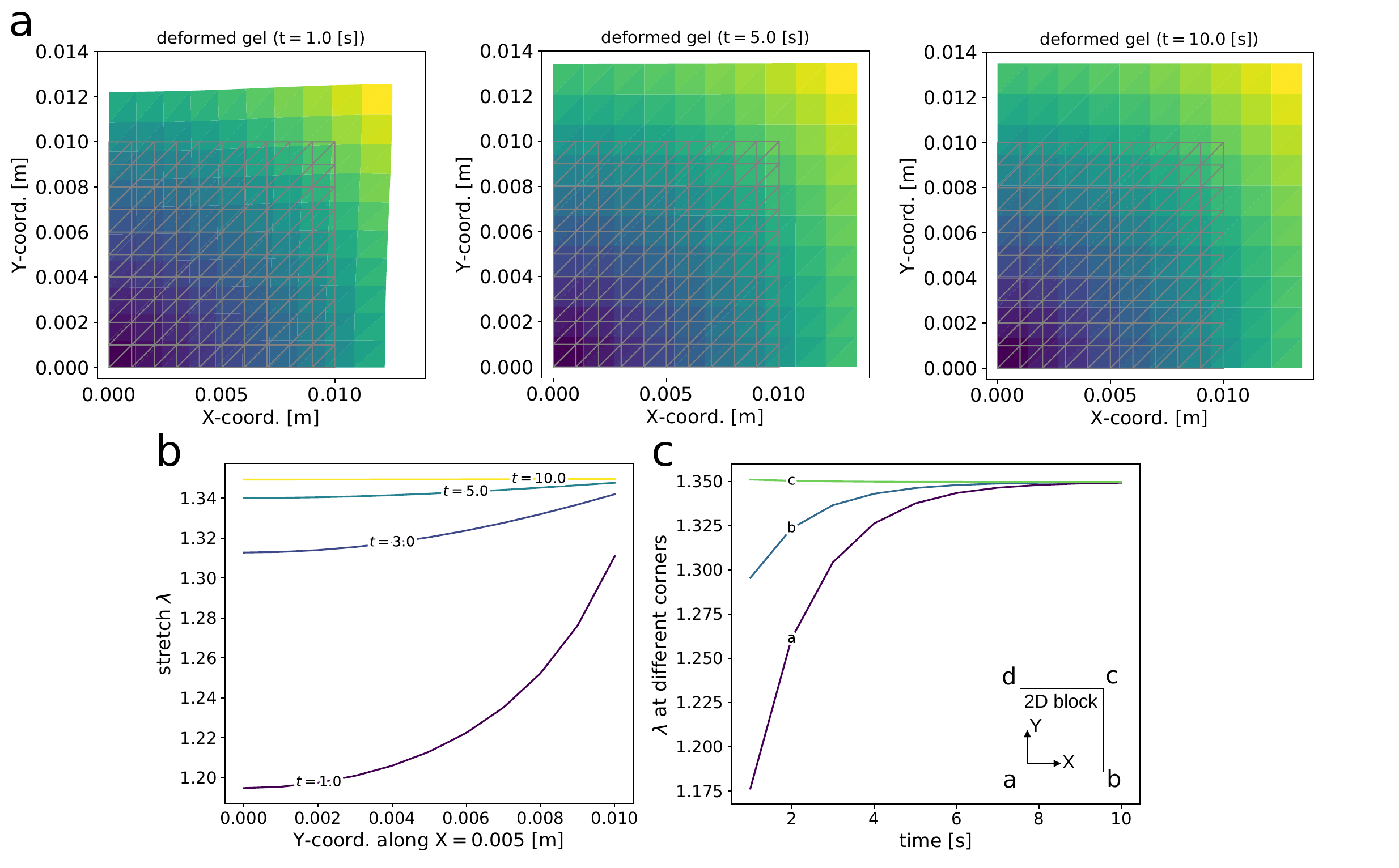}
\caption{\textbf{\emph{Constitutive model I}:} \textbf{two-dimensional hydrogel block of an initially square cross-section immersed in a non-reactive solvent at different simulation times.} \textbf{a.} Deformed two-dimensional block at three different time steps $t = 1.0~[s]$, $t = 5.0~[s]$, and $t = 10.0~[s]$. 
\textbf{b.} Stretch due to swelling $\lambda$ at different times across $X = 0.005$. \textbf{c.} Transient evolution of $\lambda$ measured at different corners of the two-dimensional block. 
\textbf{Simulation parameters:} $G_0 = 10~[MPa], ~~ \chi = 0.2~[--], ~~ D = 5.0\times10^{-5}~[m^2 s^{-1}]$.}
\label{fig:2D_block_Liu}
\end{figure}

Figure \ref{fig:2D_block_Liu} displays the simulation results of the transient diffusion-deformation process for the two-dimensional square block. The thick gray line in each subfigure in Figure \ref{fig:2D_block_Liu}\textbf{a} indicates the reference body. From Figure \ref{fig:2D_block_Liu}\textbf{a}, it is evidenced that the initially square block gets distorted at the beginning of the deformation process while swelling. The origin is the pronounced $\lambda$ gradient in the early stage of the transient behavior as evidenced in Figure \ref{fig:2D_block_Liu}\textbf{b}. This distortion vanishes as time progresses and all corners reach a similar stretching value as Figure \ref{fig:2D_block_Liu}\textbf{c} shows. 
The two-dimensional blocks exhibit a diffusion-deformation process that aligns with the one observed in the two-dimensional slab. 
The final stretch reaches the steady state at the previously computed theoretical value $\lambda_{\infty}^{2D} = 1.35$. 

\textbf{Three-dimensional free swelling cubic block.} The last example corresponds to the extension of the previous example from two to three dimensions as illustrated in Figure \ref{fig:case_studies_setup}\textbf{d}. 
The steady-state stretching is estimated to be $\lambda_{\infty}^{3D} = 1.45$, which can be used to verify the simulation results at the steady state.

\begin{figure}
\centering
\includegraphics[width=1.0\textwidth]{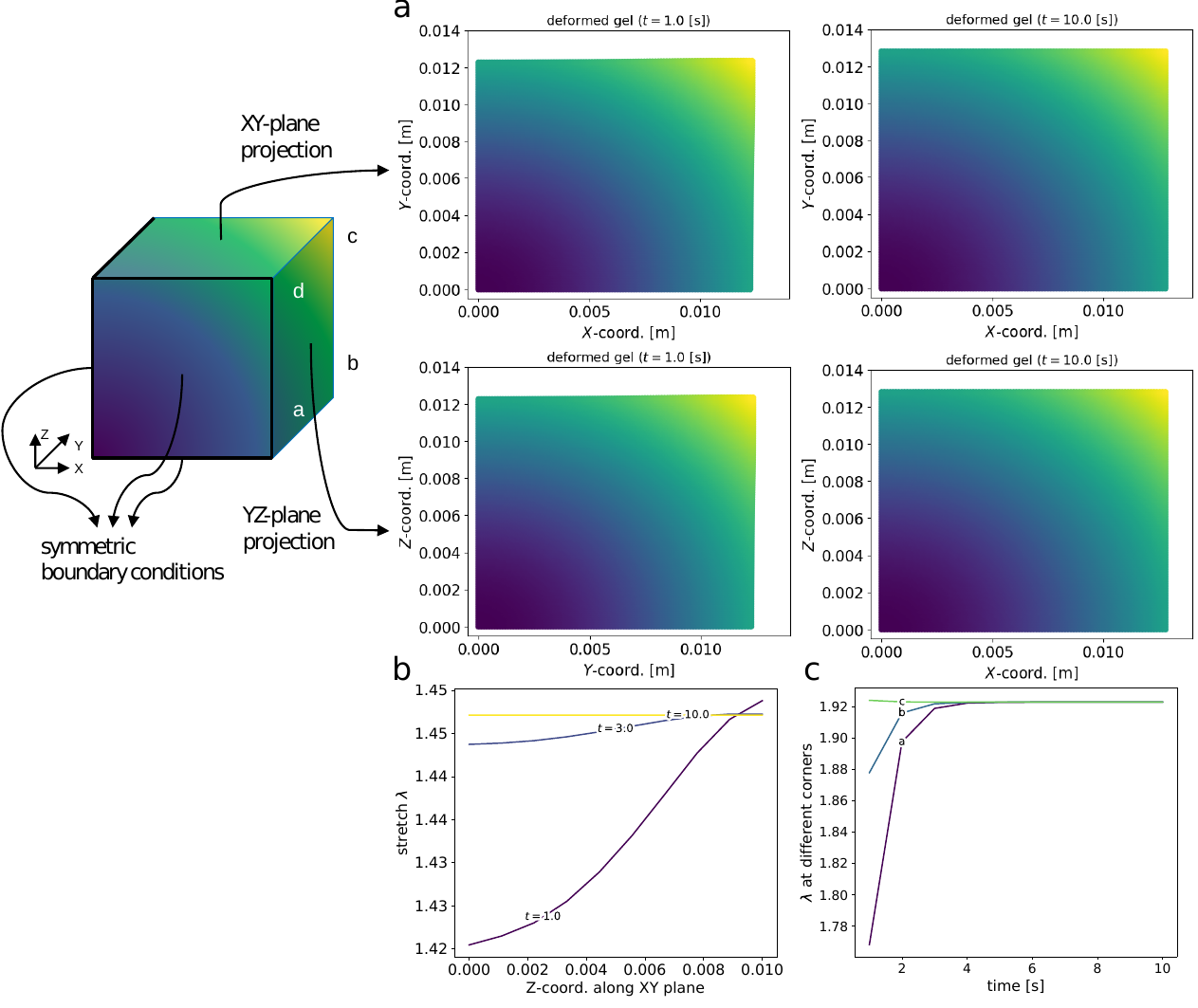}
\caption{\textbf{\emph{Constitutive model I}:} \textbf{three-dimensional hydrogel cubic block immersed in a non-reactive solvent at different simulation times.} \textbf{a.} Deformed three-dimensional cube projected to planes $XY$ and $YZ$ at two different time steps $t = 0.1$ and $t = 1.0$. 
\textbf{b.} Stretch due to swelling $\lambda$ along $Y$ at different times across the plane $XZ = 0.01~[m]$. \textbf{c.} Transient evolution of $\lambda$ measured at different corners of the three-dimensional cubic block. 
\textbf{Simulation parameters:} $G_0 = 10~[MPa], ~~ \chi = 0.2~[--], ~~ D = 7.5\times10^{-5}~[m^2 s^{-1}]$.}
\label{fig:3D_block_Liu}
\end{figure}

Figure \ref{fig:3D_block_Liu} presents the simulation results of the transient diffusion-deformation process for the three-dimensional block. Compared to the two-dimensional block, the cube is less distorted, as seen in the $XY$ and $YZ$ plane projections in Figure \ref{fig:3D_block_Liu}\textbf{a}. This can be explained by the increment in the diffusivity coefficient $D$ from $2.5\times10^{-5}$ to $5.0\times10^{-5}~[m^2 s^{-1}]$, where the latter is the minimum value that leads to the Newton algorithm iteration convergence. The diffusivity effect is observed in Figure \ref{fig:3D_block_Liu}\textbf{b} when compared to Figure \ref{fig:2D_block_Liu}\textbf{b}. The former displays a faster evolution of the stretch over time. Figure \ref{fig:3D_block_Liu}\textbf{c} shows the stretch evolution at different corners. It is observed that each observed corner presents a different stretch at the beginning of the simulation, but it fades over time. Consequently, as illustrated in Figure \ref{fig:3D_block_Liu}\textbf{a}, the three-dimensional block recovers its cube shape. 

It is noted that the two-dimensional and three-dimensional problems are addressed without altering the diffusion-deformation model or its constitutive equations. Nonetheless, adjusting the diffusivity coefficient value was necessary. This is because a pronounced stretch gradient results in high-stress values and excessive distortion of the elements used for domain discretization as more degrees of freedom are added to the problem. This stability issue is commonly encountered in diffusion-deformation studies of hydrogels with non-reactive solvent absorption (refer to \cite{Bouklas2015Nonlinear} for an in-depth discussion). Although beyond the scope of this study, stabilization methods can be employed to overcome this issue (see \cite{Krischok2016MixedFEMGels,Boger2017minimizationHydrogels} for more information). Our results provide a foundation for testing some of the approaches.

\subsection{Constitutive model II} \label{sec:res-II}

The second model under consideration is composed of the balance equations \eqref{eq:linear_momentum} and \eqref{eq:fluid_balance} together with constitutive equations \eqref{eq:stress_Chester2010} and \eqref{eq:mu_Chester2010}. 

\textbf{One-dimensional transient swelling.} Here, we follow the ideas presented by \cite{Chester2010DiffDeform}. 
By considering the 1D deformation gradient in equation \eqref{eq:kinematics-1D} under the incompressibility condition, the stress equation \eqref{eq:stress_Chester2010} yields:
\begin{equation}
    P_{Y} = G_0\phi^{-1} - P\phi = 0 \, .
\end{equation}

Hence, it results:
\begin{equation}\label{eq:P_slab}
    P = G_0 \phi^{-2} \, .
\end{equation}

Notice that this expression for $P$ is case specific and equation \eqref{eq:P_slab} holds true exactly only in the present 1D case.
After some manipulations, the balance of fluid concentration given in equation \eqref{eq:fluid_balance_rewritten-2} can be rewritten as
\begin{equation}\label{eq:fluid_balance_1D_Chester2010}
    \begin{cases}
        \partial_t \phi = \dfrac{D}{k_B T} \left[ \dfrac{\partial \phi}{\partial Y} \dfrac{\partial \mu}{\partial Y} - \phi^2 \left( \dfrac{1 - \phi}{\phi} \right) \dfrac{\partial^2 \mu}{\partial Y^2} \right], & ~ \text{in} ~ \mathcal{B}_{R}, \\
        \phi = \bar{\phi}, & ~ \text{at} ~ Y = 0.01, \\
        \dfrac{\partial \phi}{\partial Y} = 0, & ~ \text{at} ~ Y = 0.0, \\
        \phi(t = 0) = \phi_0, & ~ \text{in} ~ \mathcal{B}_R,
    \end{cases}
\end{equation}
with
\begin{equation}\label{eq:du_dY_Chester2010}
    \dfrac{\partial \mu}{\partial Y} = k_B T \left[- \dfrac{1}{1 - \phi} + 1 + 2\chi\phi \right]\dfrac{\partial \phi}{\partial Y} - \Omega G_0 \left( 1 + \dfrac{1}{\phi^2} \right)\dfrac{\partial \phi}{\partial Y},
\end{equation}
obtained after differentiating equation \eqref{eq:mu_Chester2010} with respect to $Y$.

The discretized weak form of equation \eqref{eq:fluid_balance_1D_Chester2010} reads
\begin{equation}\label{eq:weak_form_mu_1D_Chester2010}
    \int_{\mathcal{B}_R} \left( \dfrac{\phi_h^n - \phi_h^{n-1}}{\Delta t} q_h - \dfrac{D}{k_B T} \left[ \dfrac{\partial \phi_h^n}{\partial Y} \dfrac{\partial \mu_h^n}{\partial Y} q_h - (\phi_h^n)^2 \left( \dfrac{1 - \phi_h^n}{\phi_h^n} \right) \dfrac{\partial \mu_h^n}{\partial Y} \dfrac{\partial q_h}{\partial Y} \right] \right)dY = 0, ~ \forall q_h \in Q_h,
\end{equation}
where $\partial \mu_h^n/\partial Y$ is given by equation \eqref{eq:du_dY_Chester2010} evaluated at $\phi_h^n$. Again, equation \eqref{eq:weak_form_mu_1D_Chester2010} can be directly implemented in FEniCs with the corresponding boundary and initial conditions and solved using the FEM. Solution of equation \eqref{eq:weak_form_mu_1D_Chester2010} serves as a reference solution for the diffusion-deformation problem adopting \textit{constitutive model II}. The stress can be computed replacing $P$ into equation \eqref{eq:stress_Chester2010}, yielding:
\begin{equation}
    \sigma_x = J^{-1} P_X (\phi, \mathbf{u}) = G_0 \phi \left[ 1 - \phi^{-2} \right],
\end{equation}
which for $\phi \in [0, 1]$ results in a compressive stress state. Whereas the chemical potential in equation \eqref{eq:mu_Chester2010} becomes
\begin{equation}
    \mu(\phi) = \mu^0 + k_B T  \left[ \ln \left( 1 - \phi \right) + \phi + \chi \phi^2 \right] + \Omega G_0 \left[ \phi^{-1} - \phi \right],
\end{equation} 
after the definition of $P$.

\begin{figure}
\centering
\includegraphics[width=0.75\textwidth]{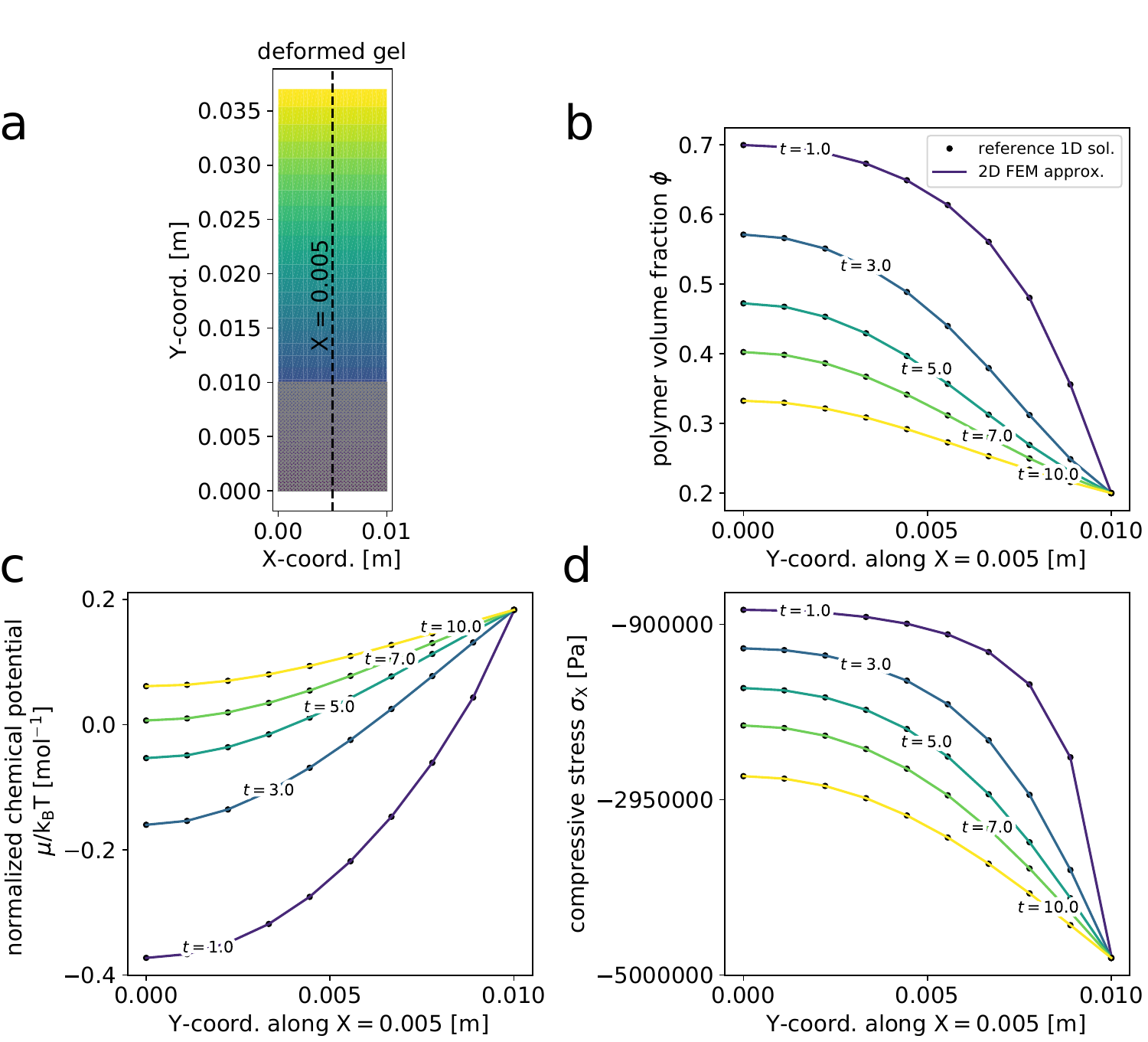}
\caption{\textbf{\emph{Constitutive model II}:} \textbf{one-dimensional bar (black dots) and two-dimensional hydrogel constrained slab (colored lines) numerical solution comparison for different mesh densities $N_h$ and at different simulation times.} \textbf{a.} Deformed two-dimensional constrained slab at $t = 1.0$. 
\textbf{b.} Polymer volume fraction $\phi$. \textbf{c.} Chemical potential $\mu$ normalized by $k_B T$. \textbf{d.} Cauchy compressive stress $\bm{\sigma}_X$. \textbf{Simulation parameters:} $G_0 = 1~[MPa], ~~ \chi = 0.2~[--], ~~ D = 2.5\times10^{-5}~[m^2 s^{-1}]$.}
\label{fig:slab_Chester_comparison}
\end{figure}

The numerical solution of equation \eqref{eq:weak_form_mu_1D_Chester2010} is presented in Figure \ref{fig:slab_Chester_comparison} (black dots). A one-dimensional mesh is created to discretize the hydrogel bar. As a result of a convergence study, the final number of elements equals $200$, and $50$ time steps are chosen ($\Delta t = 0.02$). The boundary condition at $Y = 0.01~[m]$ is determined from equation \eqref{eq:mu_Chester2010}, which gives $\phi(Y=0.01, t) = 0.2$. The initial condition is defined as $\phi_0 = 0.75$, which corresponds to a pre-swollen state conveniently chosen to alleviate the numerics and allow for a comparison between the one and two-dimensional models. In the original work by \cite{Chester2010DiffDeform}, this one-dimensional problem is numerically approximated using a finite difference method in space. The simulation results here obtained with the FEM can be quantitatively compared to those reported by \cite{Chester2010DiffDeform} for a similar example (see Figures 3, 5, and 7 in \cite{Chester2010DiffDeform}).

\textbf{Two-dimensional constrained hydrogel slab.} \textit{Constitutive model II} is solved for the constrained hydrogel slab considered in Figure \ref{fig:case_studies_setup}\textbf{b}. 
Initial and boundary conditions for $\phi$ are kept analogous to the one-dimensional case. Hence, since we have a dominant diffusion along the $Y$-direction also in this case and an analogous stress state, we approximate the Lagrange multiplier $P$ by employing the one obtained in the 1D case, i.e. with equation \eqref{eq:P_slab}. Under the limitations of such approximation, the stress for the two-dimensional case hence reads:
\begin{equation}
    \sigma_x = J^{-1} P_{X} (\phi, \mathbf{u}) = J^{-1} G_0 \left[ \mathbf{b} - \phi^{-2} \mathbf{I} \right],
\end{equation}
after replacing equation \eqref{eq:P_slab} into equation \eqref{eq:stress_Chester2010}. Since the same approximation for $P$ would be inaccurate in a free swelling condition, we will not face these case studies for \textit{constitutive model II}. 

Figure \ref{fig:slab_Chester_comparison} presents the comparison between the one-dimensional bar (black dots) and two-dimensional constrained slab examples (different colored lines). The one-dimensional simulation is taken as the reference solution to the problem. The two-dimensional slab problem is solved for different mesh densities and different number of time steps. It was concluded that a mesh density of $N_h = 30$ and $25$ time steps are enough to produce accurate results, so the following simulation results are produced using this simulation setup.
The deformed hydrogel at $t = 10.0 [s]$ is displayed in Figure \ref{fig:slab_Chester_comparison}\textbf{a}.
Figures \ref{fig:slab_Chester_comparison}\textbf{b} - \textbf{d} show the time evolution of $\phi$, $\mu$ and $\sigma_x$. It is observed that the numerical solutions for the one and two-dimensional problems coincide. This is not surprising since the large deformation displayed by the hydrogel slab makes it behave like a one-dimensional bar.

In Figure \ref{fig:slab_Chester_comparison}\textbf{c}, it is observed that the chemical potential's rate increases fast in the beginning when the solvent starts entering the gel and becomes slower when it tends to the steady state. 
The pronounced gradient of $\mu$ along the spatial dimension in the beginning and a relaxed polymer network facilitates solvent absorption. As the gradient of $\mu$ flattens out and the compressive stress increases, solvent diffusion slows down.

\cite{Bouklas2012Linear-NL} and \cite{Chester2015Abaqus} proposed to solve an additional nonlinear equation either for $J_f$ or $\phi$ at the Gauss integration point to fully determine the time derivative term and then solve the balance of fluid concentration for the chemical potential. If equations \eqref{eq:fluid_balance_rewritten-1} and \eqref{eq:fluid_balance_rewritten-2} are solved for $\mu$, either equation \eqref{eq:mu_Bouklas2015}, \eqref{eq:mu_Chester2010}, or \eqref{eq:mu_Chester_energetic} must be solved as an implicit nonlinear equation at each Gauss integration point for either $J_f$ or $\phi$, depending of the adopted constitutive model, as suggested by \cite{Bouklas2015Nonlinear} and \cite{Chester2010DiffDeform, Chester2015Abaqus}.

In \cite{Chester2015Abaqus}, the coupled problem was solved using the commercial software Abaqus, via an UMAT routine. The coupled problem was solved for $\phi$, $\mu$, and $\mathbf{u}$. In particular, $\phi$ was defined as a local variable, and the nonlinear equation \eqref{eq:mu_Chester2010} was solved at each Gauss integration point. This approach allowed to fully determine the time derivative in equation \eqref{eq:fluid_balance_rewritten-2}$_2$ at each time instance. Thus, a static problem for $\mu$ and $\mathbf{u}$ is solved also at each time step. A similar idea was adopted by \cite{Bouklas2015Nonlinear}. 

We decided to test an alternative approach. That is, to find a suitable expression for $\text{Grad} (\mu)$ in equation \eqref{eq:fluid_balance_rewritten-2} from the constitutive equation \eqref{eq:mu_Chester2010}. Then, solve the couple problem only for $\phi$ and $\mathbf{u}$. Such an approach allows us to avoid the computation of $\phi$ at each integration point.
By following this approach, we noticed that the solution becomes less prompt to numerical instabilities, and monolithic and staggered formulations can be adopted to solve the resulting coupled problem. 
It is worth remarking that without the automatic differentiation capabilities of FEniCs, the computation of $\text{Grad} (\mu)$ can be a very tedious task, and it is not difficult to imagine that was the reason why \cite{Chester2015Abaqus} opted for their approach.

We conduct an investigation to assess the staggered approach's behavior. We verify the number of staggered iterations between the displacement and polymer volume fraction sub-systems and, as for the monolithic case, we examine the effects of mesh density and time step size on the algorithm's performance, focusing particularly on the Newton solver's efficiency and convergence.

For the staggered solution, we modify the strategy slightly. Instead of computing the total number of Newton iterations required for convergence, we compute the average number of Newton iterations over the inner iteration loop for each time increment for the displacement and polymer volume fraction sub-systems.
This approach provides us with insights into the staggered solution's convergence behavior and whether the number of staggered and Newton iterations remains stable throughout the simulation time. It allows us to gauge the efficiency and stability of the staggered approach, and compare it to the monolithic approach.

\begin{figure}[htbp]
\centering
\includegraphics[width=1.0\textwidth]{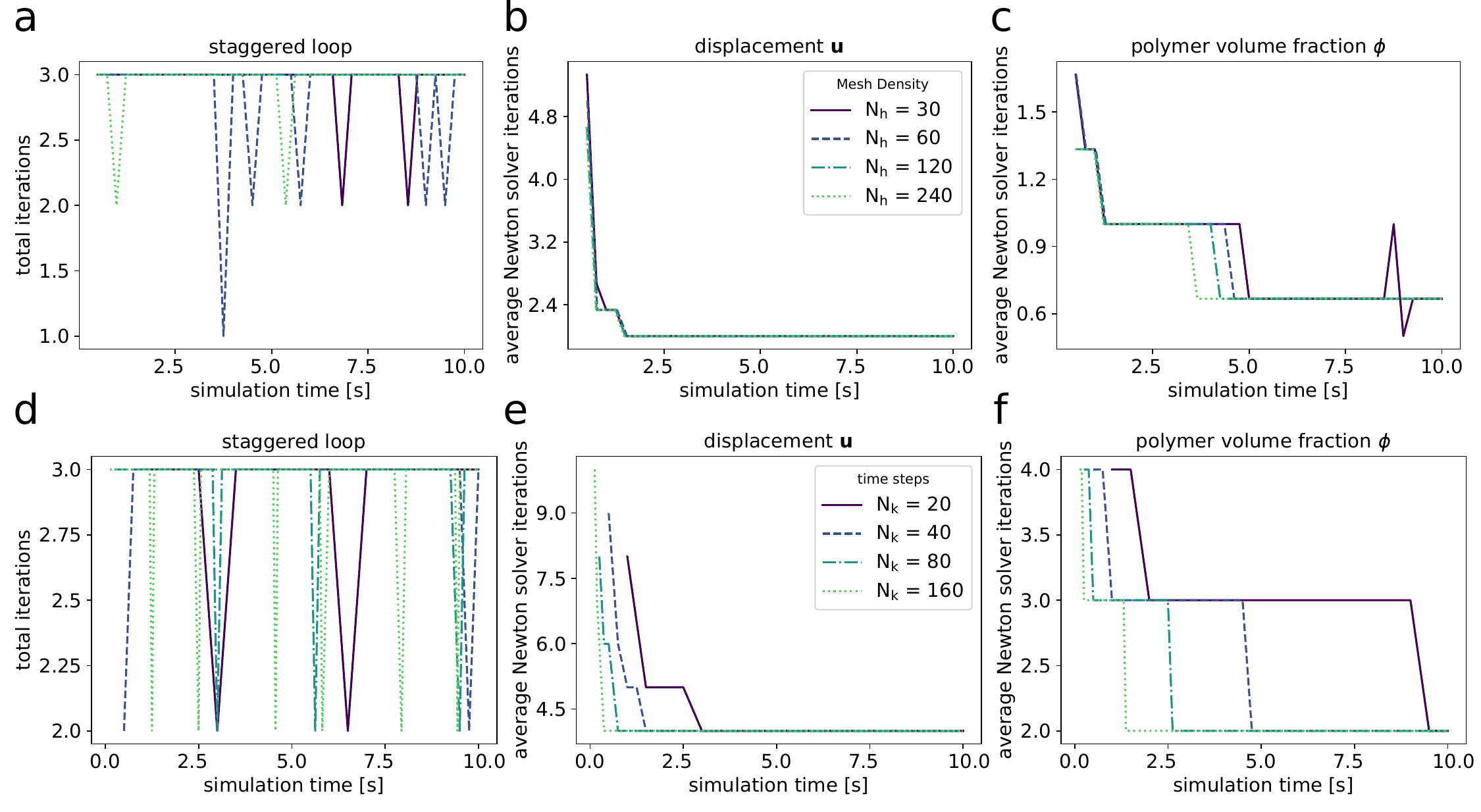}
\caption{\textbf{\emph{Constitutive model II}:} \textbf{staggered algorithm and Newton iterations along the time steps.} For different mesh densities ($N_h$): \textbf{a.} for the staggered scheme, \textbf{b}. for displacement, and \textbf{c.} for the polymer volume fraction. For different time step sizes ($N_k$): \textbf{d.} for the staggered scheme, \textbf{e.} for displacement, and \textbf{f.} for the polymer volume fraction.}
\label{fig:slab_Chester_conv_Newton_iteration}
\end{figure}

Figure \ref{fig:slab_Chester_conv_Newton_iteration} illustrates the time versus the staggered and Newton iterations for different mesh densities and time step sizes. Figures \ref{fig:slab_Chester_conv_Newton_iteration}\textbf{a} and \ref{fig:slab_Chester_conv_Newton_iteration}\textbf{d} show how many staggered iterations are necessary at each time step to reach an $L_2$ error lower than $10^{-10}$ both for $\mathbf{u}$ and $\phi$, respectively. It is observed that the number of iterations is never higher than three, independent of the mesh density and time step size. This highlights the effectiveness of the staggered approach.

The convergence patterns for the Newton iterations and the impact of variations in mesh density and time step size were consistent with those observed in the monolithic approach.
Figures \ref{fig:slab_Chester_conv_Newton_iteration}\textbf{b, e} focuses on the Newton iteration for the displacement sub-system, whereas Figure \ref{fig:slab_Chester_conv_Newton_iteration}\textbf{c,f} reports the Newton iterations along time for the polymer volume fraction sub-system, for different mesh densities and time step sizes, respectively. 
Results in Figure \ref{fig:slab_Chester_conv_Newton_iteration} reveal a similar trend to the monolithic case regarding the Newton iterations at each time increment.

\begin{figure}[htbp!]
\centering
\includegraphics[width=0.75\textwidth]{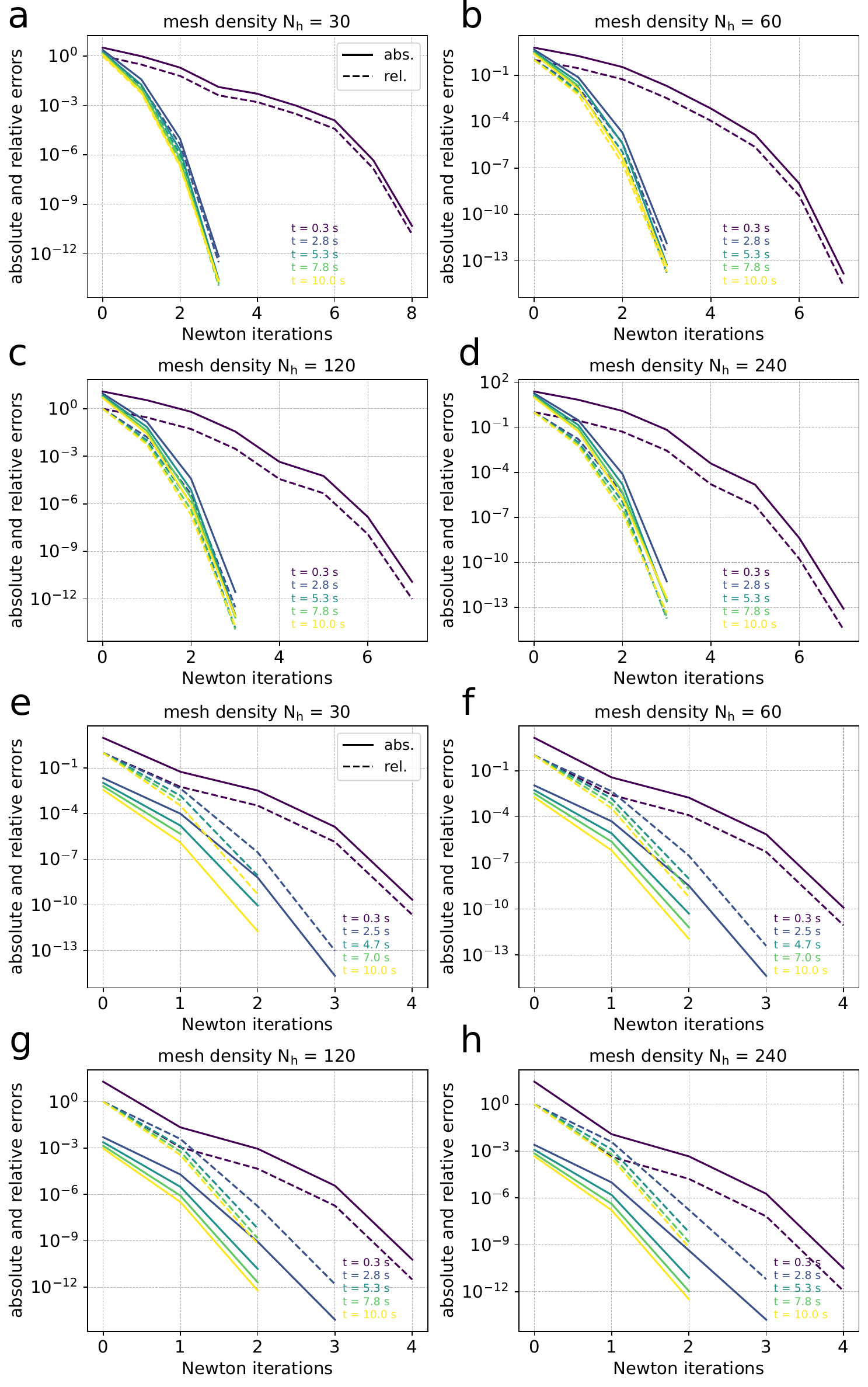}
\caption{\textbf{\emph{Constitutive model II}:} \textbf{convergence analysis of the Newton solver for different mesh densities ($N_h$).} For the displacement: \textbf{a}. $N_h = 30$. \textbf{b}. $N_h = 60$. \textbf{c}. $N_h = 120$. \textbf{d}. $N_h = 240$. For the polymer volume fraction: \textbf{e}. $N_k = 25$. \textbf{f}. $N_h = 50$. \textbf{g}. $N_h = 100$. \textbf{h}. $N_h = 200$.}
\label{fig:slab_Chester_conv_Newton_error_mesh}
\end{figure}

Additionally, we explored the Newton algorithm's convergence behavior at five distinct simulation moments, examining the quadratic convergence in error and the influence of mesh density and time step size. Figure \ref{fig:slab_Chester_conv_Newton_error_mesh} shows the error decay with respect to the Newton iterations. Sub-figures \ref{fig:slab_Chester_conv_Newton_error_mesh}\textbf{a, b, c, d} correspond to the displacement, while sub-figures \ref{fig:slab_Chester_conv_Newton_error_mesh}\textbf{e, f, g, h} refer to the polymer volume fraction, both for different mesh densities. The results are analogous to the monolithic approach, where the absolute and relative errors decay fast as the Newton iteration increases, demonstrating the staggered approach's robustness.

The observed similarities in the convergence behaviors between the monolithic and staggered approaches further underline the robustness and reliability of the numerical solutions, confirming the applicability of both methods to the coupled problem at hand.

Concerning convergence analyses for the discretization itself, for the Taylor-Hood elements and Euler method, our findings for spatial and temporal discretization yielded a second-order and first-order convergence, respectively. These outcomes are summarized in Tables \ref{tab:conv_analysis_constII_mesh} and \ref{tab:conv_analysis_constII_time}, reinforcing the staggered approach's validity in solving the coupled diffusion-deformation problem.

\begin{sidewaystable}
\centering
\caption{Spatial discretization convergence analysis for the two-dimensional constrained hydrogel slab example considering \textit{constitutive model II}.}
\label{tab:conv_analysis_constII_mesh}
\begin{tabular}{cccccccccc}
\hline
\textbf{Level} &
  \textbf{\begin{tabular}[c]{@{}c@{}}Mesh density\\ $N_h$\end{tabular}} &
  \textbf{\begin{tabular}[c]{@{}c@{}}Time steps \\ $N_k$\end{tabular}} &
  \bm{$h$} &
  \textbf{Elements} &
  \textbf{DoFs} &
  \textbf{$\phi$(0.5.1.0)} &
  \textbf{L2 error for $\phi$} &
  \textbf{$\mathbf{u}$(1.0.1.0)} &
  \textbf{L2 error for $\mathbf{u}$} \\ \hline
1 &	30 &	40 &	0.033 &	1800 &	8403 &	0.31318141 &	2.2590e-5 &	2.83246074 &	1.4800e-4 \\
2 &	60 &	40 &	0.016 &	7200 &	33003 &	0.31309735 &	5.4560e-6 &	2.83288465 &	3.4709e-5 \\
3 &	120 &	40 &	0.008 &	28800 &	130803 &	0.31307316 &	1.2288e-6 &	2.83299197 &	9.0328e-6 \\
4 &	240 &	40 &	0.004 &	115200 &	520803 &	0.31306639 &	-- &	2.83301882 &	--          \\ \hline
\textbf{conv. order} & \textbf{} & \textbf{} & \textbf{} & \textbf{} & \textbf{} & \textbf{1.83} & \textbf{2.02} & \textbf{2.00} & \textbf{2.14} \\ \hline
\end{tabular}
\end{sidewaystable}



\begin{sidewaystable}
\centering
\caption{Time discretization convergence analysis for the two-dimensional constrained hydrogel slab example considering \textit{constitutive model II}.}
\label{tab:conv_analysis_constII_time}
\begin{tabular}{cccccccc}
\hline
\textbf{Level} &
  \textbf{\begin{tabular}[c]{@{}c@{}}Mesh density\\ $N_h$\end{tabular}} &
  \textbf{\begin{tabular}[c]{@{}c@{}}Time steps \\ $N_k$\end{tabular}} &
  \bm{$k$} &
  \textbf{$\phi$(0.5.1.0)} &
  \textbf{L2 error for $\phi$} &
  \textbf{$\mathbf{u}$(1.0.1.0)} &
  \textbf{L2 error for $\mathbf{u}$} \\ \hline
1 &	30 &	20 &	0.5 &	0.31996593 &	8.2017e-3 &	2.72595105 &	9.6975e-02 \\
2 &	30 &	40 &	0.25 &	0.31309911 &	3.7086e-3 &	2.80367321 &	4.3200e-02 \\
3 &	30 &	80 &	0.125 &	0.30939811 &	1.2936e-3 &	2.84445820 &	1.4871e-02 \\
4 &	30 &	160 &	0.0625 &	0.30741375 &	-- &	2.86582984 &	--             \\ \hline
\textbf{conv. order} & \textbf{} & \textbf{} & \textbf{} & \textbf{0.90} & \textbf{0.90} & \textbf{0.93} & \textbf{0.92} \\ \hline
\end{tabular}
\end{sidewaystable}


\subsection{Constitutive model III}

The third model under consideration comprises the balance equations \eqref{eq:linear_momentum} and \eqref{eq:fluid_balance} together with constitutive equations \eqref{eq:stress_Bouklas2015} and \eqref{eq:mu_Bouklas2015}.

\textbf{One-dimensional transient swelling.} The basic ingredients to describe the hydrogel deformation considering an energetic constraint were presented by \cite{Bouklas2012Linear-NL,Bouklas2015Nonlinear}. 
By considering the 1D deformation gradient in equation \eqref{eq:kinematics-1D}, the balance of fluid concentration reads
\begin{equation}\label{eq:fluid_balance_1D_Bouklas2010}
    \begin{cases}
        \gamma(\lambda) \partial_t \lambda = D \dfrac{\partial}{\partial Y} \left[ \xi(\lambda) \dfrac{\partial \lambda}{\partial Y} \right], & ~ \text{in} ~ \mathcal{B}_{R}, \\
        \lambda = \bar{\lambda}, & ~ \text{at} ~ Y = 0.01, \\
        \dfrac{\partial \lambda}{\partial Y} = 0, & ~ \text{at} ~ Y = 0.0, \\
        \lambda(t = 0) = \lambda_0, & ~ \text{in} ~ \mathcal{B}_R,
    \end{cases}
\end{equation}
with 
\begin{equation}\label{eq:complementary_eq_1D_Bouklas_gamma}
    \gamma(\lambda) = \lambda_0 + \dfrac{Nk_B T}{K \lambda_0}\left( 1 + \dfrac{1}{\lambda^2} \right), 
\end{equation}
and
\begin{equation}\label{eq:complementary_eq_1D_Bouklas_xi}
    \xi(\lambda) = \left( \dfrac{1}{(\lambda_0^2\lambda)^2} - \dfrac{2\chi (\lambda_0^2\lambda - 1)}{(\lambda_0^2\lambda)^3}  \right) \left( \frac{Nk_B T (1 + \lambda^2)}{K(\lambda_0\lambda^2)^2} + \frac{\lambda_0^2}{\lambda^2}\right) + \dfrac{N\Omega(\lambda_0^2\lambda - 1)(1 + \lambda^2)}{(\lambda_0\lambda^2)^2},
\end{equation}
where $\lambda_0$ refers to the initial swelling ratio of the gel. 

The discretized weak form of equation \eqref{eq:fluid_balance_1D_Bouklas2010} reads
\begin{equation}\label{eq:weak_form_mu_1D_Bouklas2015}
    \int_{\mathcal{B}_R} \left( \gamma(\lambda_h^n) \dfrac{\lambda_h^n - \lambda_h^{n-1}}{\Delta t} q_h + D \xi(\lambda_h^n) \dfrac{\partial \lambda_h^n}{\partial Y} \dfrac{\partial q_h}{\partial Y} \right)dY = 0, ~ \forall q_h \in Q_h,
\end{equation}
where $\gamma(\lambda_h^n)$ and $\xi(\lambda_h^n)$ are given by equations \eqref{eq:complementary_eq_1D_Bouklas_gamma} and \eqref{eq:complementary_eq_1D_Bouklas_xi} evaluated at $\lambda_h^n$, respectively. Equation \eqref{eq:weak_form_mu_1D_Bouklas2015} can be solved using the FEM with FEniCS. 
The postprocessing quantities $\mu$ and $\sigma_x$ results:
\begin{equation}\label{eq:mu_Bouklas_1D}
    \mu(\lambda_h) = k_B T \left[ \ln \left( \dfrac{\lambda_0^2\lambda_h - 1}{\lambda_0^2\lambda_h} \right) +  \dfrac{\lambda_0^2\lambda_h + \chi}{(\lambda_0^2\lambda_h)^2} + \dfrac{N\Omega}{\lambda_0^2} \left( \lambda_h - \dfrac{1}{\lambda_h} \right)  \right], 
\end{equation}
and
\begin{equation}\label{eq:sigma_Bouklas_1D}
    \sigma_x(\lambda_h) = Nk_B T \left[ \left( \lambda_0 - \dfrac{1}{\lambda_0} \right) - K \left( \dfrac{\lambda_h}{\lambda_0} \right) \left( \lambda_h - \dfrac{1}{\lambda_h} \right) \right],
\end{equation}
respectively. Notice that $J_f$ in equation \eqref{eq:mu_Bouklas2015} was replaced in by $\lambda_0^2\lambda_h$ to get equation \eqref{eq:mu_Bouklas_1D}. More details about the derivation of equations \eqref{eq:mu_Bouklas_1D} and \eqref{eq:sigma_Bouklas_1D} can be found in \textbf{appendix A} of \cite{Bouklas2015Nonlinear}.

\textbf{Two-dimensional constrained hydrogel slab.} Example's geometry corresponds to that illustrated in Figure \ref{fig:case_studies_setup}\textbf{b}. 
Initial and boundary conditions correspond to those defined for the one-dimensional case. 

\begin{figure}
\centering
\includegraphics[width=0.75\textwidth]{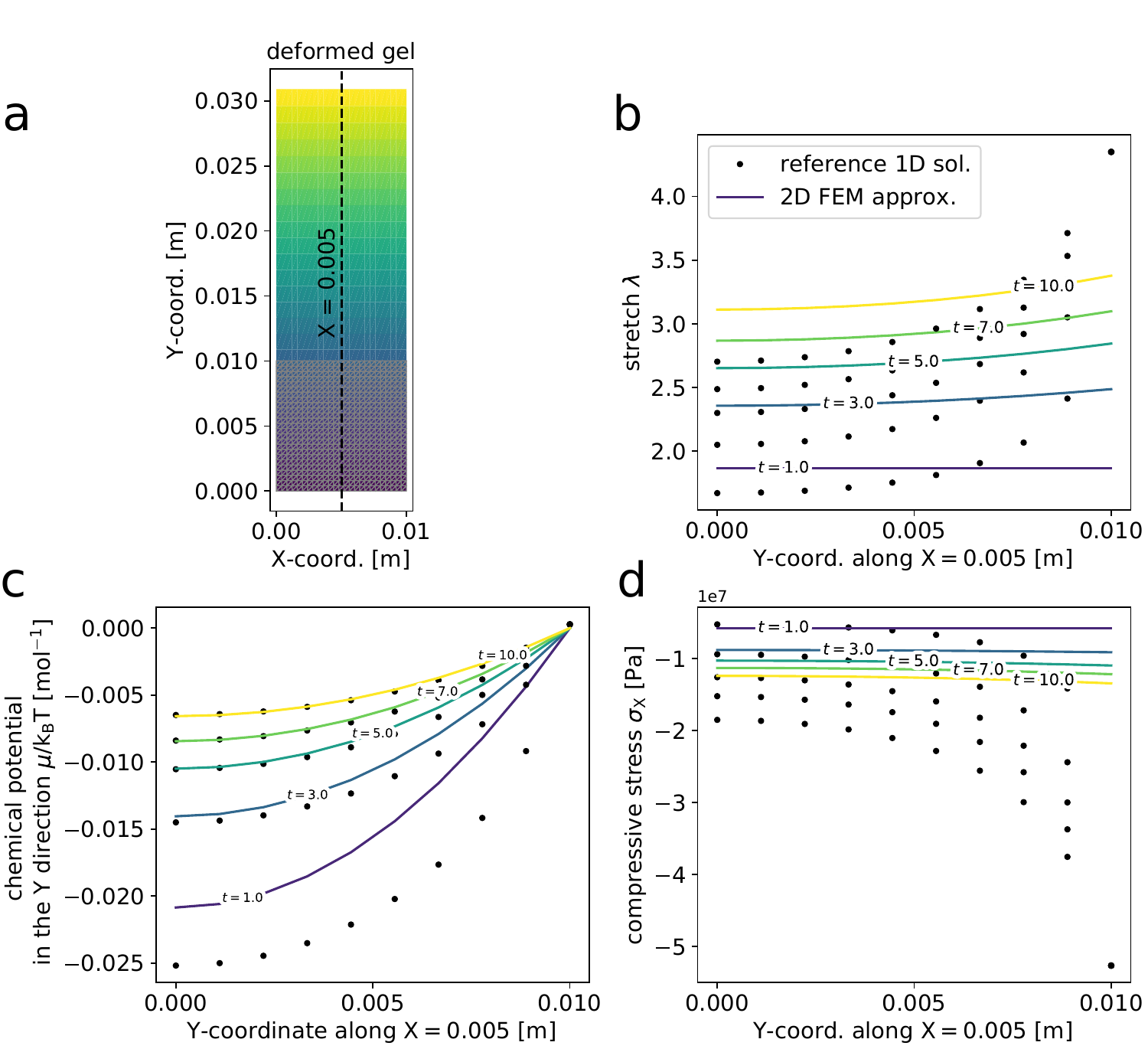}
\caption{\textbf{\emph{Constitutive model III}:} \textbf{one-dimensional bar (black dots) and two-dimensional hydrogel constrained slab (colored lines) numerical solution comparison at different simulation times.} \textbf{a.} Deformed two-dimensional constrained slab at $t = 10.0~[m]$. 
\textbf{b.} Stretch due to swelling $\lambda$. \textbf{c.} Chemical potential $\mu$ normalized by $k_B T$. \textbf{d.} Cauchy compressive stress $\bm{\sigma}_X$. \textbf{Simulation parameters:} $G_0 = Nk_B T \approx 41~[Pa], K = 100~Nk_B T ~ [Pa], ~~ \chi = 0.4~[--], ~~ D = 2.5\times10^{-6}~[m^2 s^{-1}]$.}
\label{fig:slab_Bouklas}
\end{figure}

Figure \ref{fig:slab_Bouklas} presents the comparison between the one-dimensional bar (black dots) and two-dimensional constrained slab examples (different colored lines). The two-dimensional slab problem is solved for a mesh density $N_h = 25$ and $10$ time steps ($\Delta t = 1.0~[s]$). This setup yields an acceptable numerical accuracy for the two-dimensional configuration.
The deformed hydrogel at $t = 10.0~[s]$ is displayed in Figure \ref{fig:slab_Bouklas}\textbf{a}.
Figures \ref{fig:slab_Bouklas}\textbf{b} - \textbf{d} show the time evolution of $\lambda = J_f$, $\mu$ and $\sigma_x$. It is observed that the numerical solutions for the one and two-dimensional problems are on the same order of magnitude and get closer as the two-dimensional slab becomes larger and progressively resembles a one-dimensional domain. The same pattern was observed for the \textit{constitutive model I}. However, the hydrogel experiences a lower level of stretch when considering the two-dimensional setup. This is the effect of determining first $\lambda$ through the nonlinear equation and then $\mu$ from the static PDE. The boundary condition can only be imposed on $\mu$, not $\lambda$ as in the one-dimensional case.

The transient behavior of the hydrogel follows the already observed evolution for \textit{constitutive models I and II}. At the beginning of the deformation, the solvent enters the gel faster because of the high $\mu$ gradient and low compressible stress. From Figure \ref{fig:slab_Bouklas}, it is noticed that the hydrogel slab undergoes a rather large deformation, i.e., the final size is about $3$ times the original one, despite the presence of the energetic constrain in the free energy function. The small shear modulus $G_0$ value can justify this large deformation.  

\textit{Constitutive model III} as considered in this work was formally presented by \cite{Bouklas2012Linear-NL} and solved using the FEM by \cite{Bouklas2015Nonlinear}. The coupled problem was solved for $c_R$, $\mu$, and $\mathbf{u}$. In particular, $c_R$ was defined as a local variable, and the nonlinear equation \eqref{eq:mu_Bouklas2015} was solved at each Gauss integration point to determine the time derivative in equation \eqref{eq:fluid_balance_rewritten-1} at each time instant. Next, $\mu$ and $\mathbf{u}$ are computed by the solution of a static problem defined by equations \eqref{eq:linear_momentum} and \eqref{eq:fluid_balance} at each time step.

Here we tried to find a suitable expression for $\text{Grad} (\mu) $ in equation \eqref{eq:fluid_balance_rewritten-2} from the constitutive equation \eqref{eq:mu_Bouklas2015} using FEniCS automatic differentiation tools and avoid the solution of the nonlinear equation at the Gauss points for the two-dimensional example. Although we got some numerical results, these appear highly inaccurate compared to the reference 1D outcomes. Consequently, we only report the results obtained following \cite{Bouklas2015Nonlinear} original formulation. 

It should be highlighted that among the constitutive models accounted for in this work, \textit{constitutive model III} numerical solution is the most fragile. 
There is not much room to explore different values for the model parameters. Small changes in their values would lead to the divergence of the Newton-Raphson algorithm. 
It is also not possible to arbitrarily increase the mesh density or make the time step smaller. This is not surprising since \cite{Bouklas2015Nonlinear} presented a detailed analysis of the induced instabilities in hydrogels in the presence of geometrical constraints. 
These instabilities arise because the exposure of a gel to a fluid not only leads to a large increase in volume but also to a wave-like buckling pattern. Free surfaces expand due to the species influx and are bonded to unswollen inner parts of the gel simultaneously. For high osmotic pressure, this mechanism leads to buckling patterns, that have extensively been analyzed within experimental setups (see, e.g., \cite{Guvendiren2010wrinklesGels,Dervaux2012instabilitiesGels,Liaw2019wrinklingHydrogels} for more details on hydrogels instabilities).

Due to this numerical fragility, it was not possible to perform a convergence analysis in the same manner as for the two previous constitutive models. However, for the chosen simulation setup presented in this sub-section, both the absolute and relative errors decay faster as the Newton iterations increase, in a manner similar to that observed for \textit{constitutive model II} (see Figures \ref{fig:slab_Liu_conv_Newton_error_mesh} and \ref{fig:slab_Liu_conv_Newton_error_time}).

\textbf{Two-dimensional free swelling of a square block.}
Figure \ref{fig:case_studies_setup}\textbf{c} illustrates the considered example. 
For the solvent concentration boundary conditions, the edges \textbf{ab} and \textbf{ad} (the symmetry edges) are prescribed a zero fluid flux, and on the edges \textbf{bc} and \textbf{cd}, the chemical potential is set equal to zero. The initial condition for $J_f$ is defined as $J_{f_0} = 1.4$.
In this case, it was not necessary to apply a time-ramping strategy.

\begin{figure}
\centering
\includegraphics[width=1.0\textwidth]{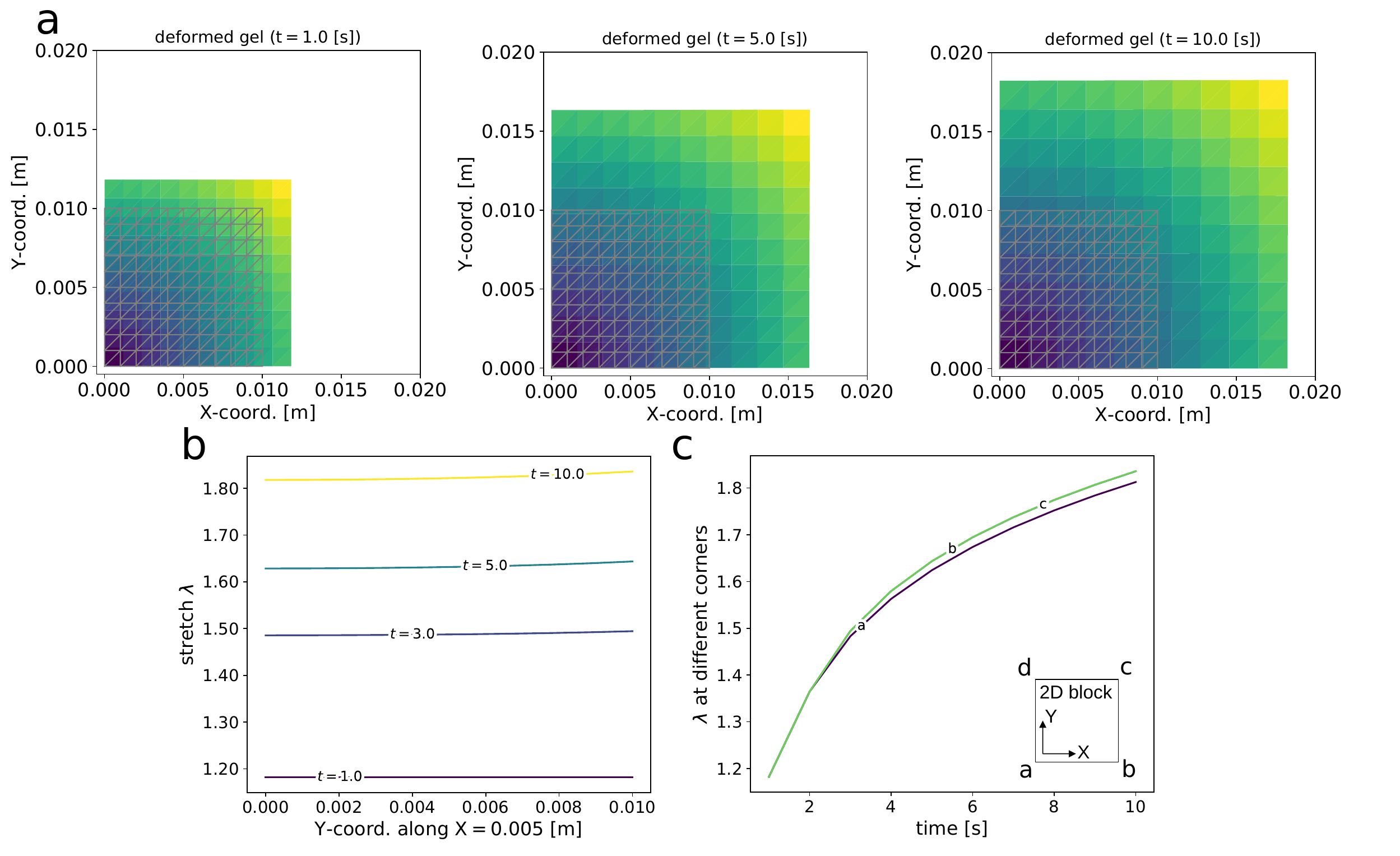}
\caption{\textbf{\emph{Constitutive model III}:} \textbf{two-dimensional hydrogel block of an initially square cross-section immersed in a non-reactive solvent at different simulation times.} \textbf{a.} Deformed two-dimensional block at three different time steps $t = 1.0~[s]$, $t = 5.0~[s]$, and $t = 10.0~[s]$.
\textbf{b.} Polymer volume fraction $\phi$ at different times across $X = 0.005~[m]$. \textbf{c.} Transient evolution of $\phi$ measured at different corners of the two-dimensional block.
\textbf{Simulation parameters:} $G_0 = Nk_B T \approx 0.41~[MPa], K = 100~Nk_B T ~ [MPa], ~~ \chi = 0.4~[--], ~~ D = 2.5\times10^{-6}~[m^2 s^{-1}]$.}
\label{fig:2D_block_Bouklas}
\end{figure}

The simulation results for the two-dimensional square block are presented in Figure \ref{fig:2D_block_Bouklas}. The box plotted with a thick gray line in Figure \ref{fig:2D_block_Bouklas}\textbf{a} corresponds to the reference body before undergoing deformation. In contrast to the \textit{constitutive model I}, the block tends to keep its square shape along the simulation time. This can be attributed to the value of the diffusivity that prevents high $\mu$ gradients as observed in Figure \ref{fig:2D_block_Bouklas}\textbf{b}. However, we chose this simulation setup because it leads to numerical convergence. For simulation setup producing more pronounced $\mu$ gradients, bigger distortion, or larger deformations, suitable stabilization techniques are required (see, e.g., \cite{Krischok2016MixedFEMGels,Boger2017minimizationHydrogels} for some approaches), which are out of the scope of the current research. An important difference between \textit{constitutive model III} and the previous constitutive models is that $\lambda$ does not tend to a steady state value as time simulation progresses. This can be appreciated from Figure \ref{fig:2D_block_Bouklas}\textbf{c}.

\subsection{Constitutive models IV and V}

The fourth and fifth models under consideration are composed of the balance equations \eqref{eq:linear_momentum} and \eqref{eq:fluid_balance} together with constitutive equations \eqref{eq:PK1_Chester_energetic} and \eqref{eq:mu_Chester_energetic} for the stress and chemical potential, respectively.

To avoid redundancies, we only present results for the two-dimensional free swelling of a square block with \textit{constitutive models IV and V}. 
It is worth recalling that the main difference between \textit{constitutive model II and IV - V} is the addition of an energetic component in the free energy function, thus eliminating the need to introduce a Lagrange multiplier as in \textit{constitutive model II}. 

\begin{figure}
\centering
\includegraphics[width=1.0\textwidth]{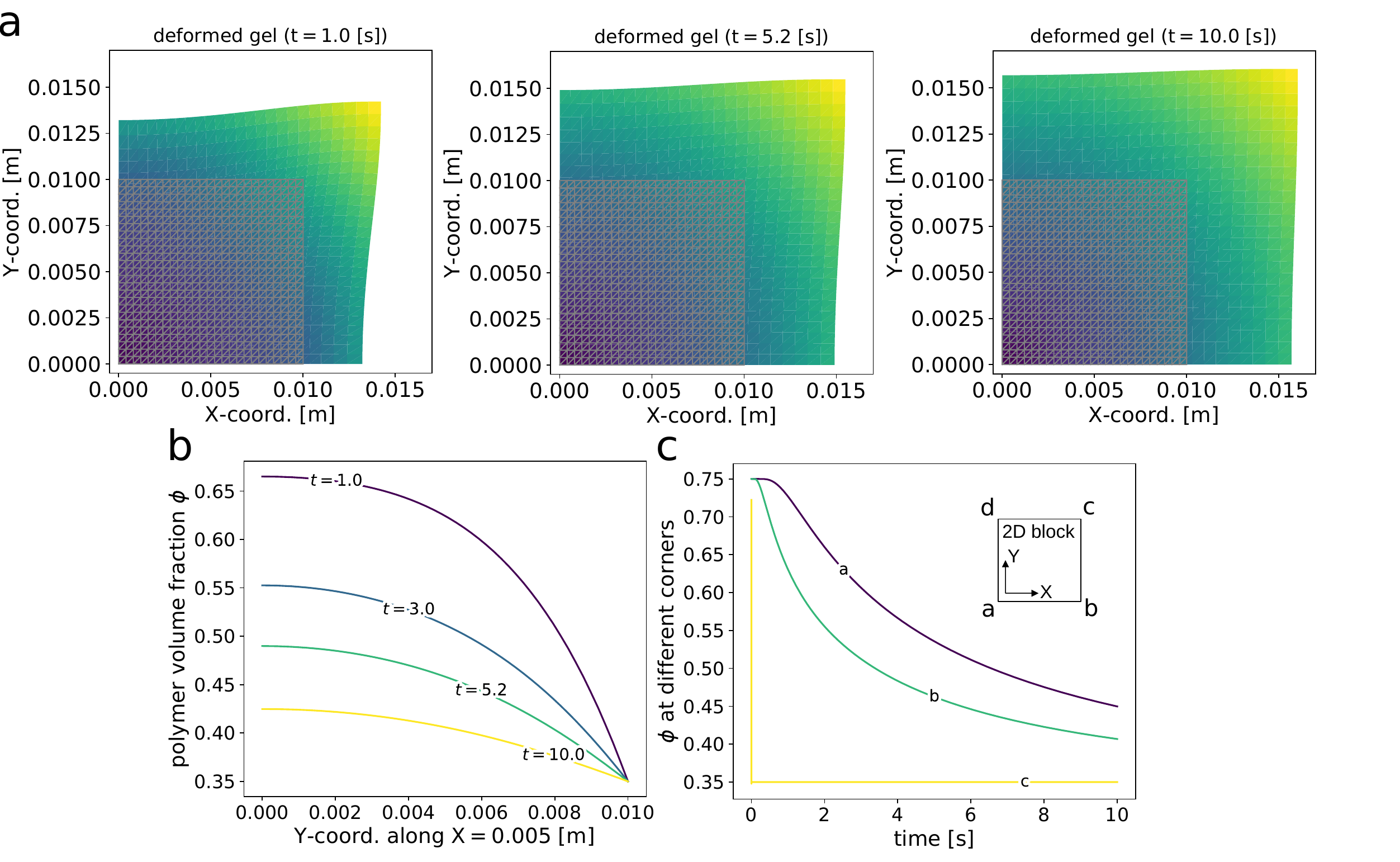}
\caption{\textbf{\emph{Constitutive model IV}:} \textbf{Two-dimensional hydrogel block of an initially square cross-section immersed in a non-reactive solvent at different simulation times.} \textbf{a.} Deformed two-dimensional block at three different time steps $t = 1.0~[s]$, $t = 5.0~[s]$, and $t = 10.0~[s]$.
\textbf{b.} Polymer volume fraction $\phi$ at different times across $X = 0.005~[m]$. \textbf{c.} Transient evolution of $\phi$ measured at different corners of the two-dimensional block. 
\textbf{Simulation parameters:} $G_0 = 1~[MPa], ~~ K = 100~[MPa], ~~ \chi = 0.2~[--], ~~ D = 7.5 \times 10^{-9}~[m^2 s^{-1}], ~~ \alpha_r = 10.0~[--]$.}
\label{fig:2D_block_Chester_2011}
\end{figure}

\begin{figure}
\centering
\includegraphics[width=1.0\textwidth]{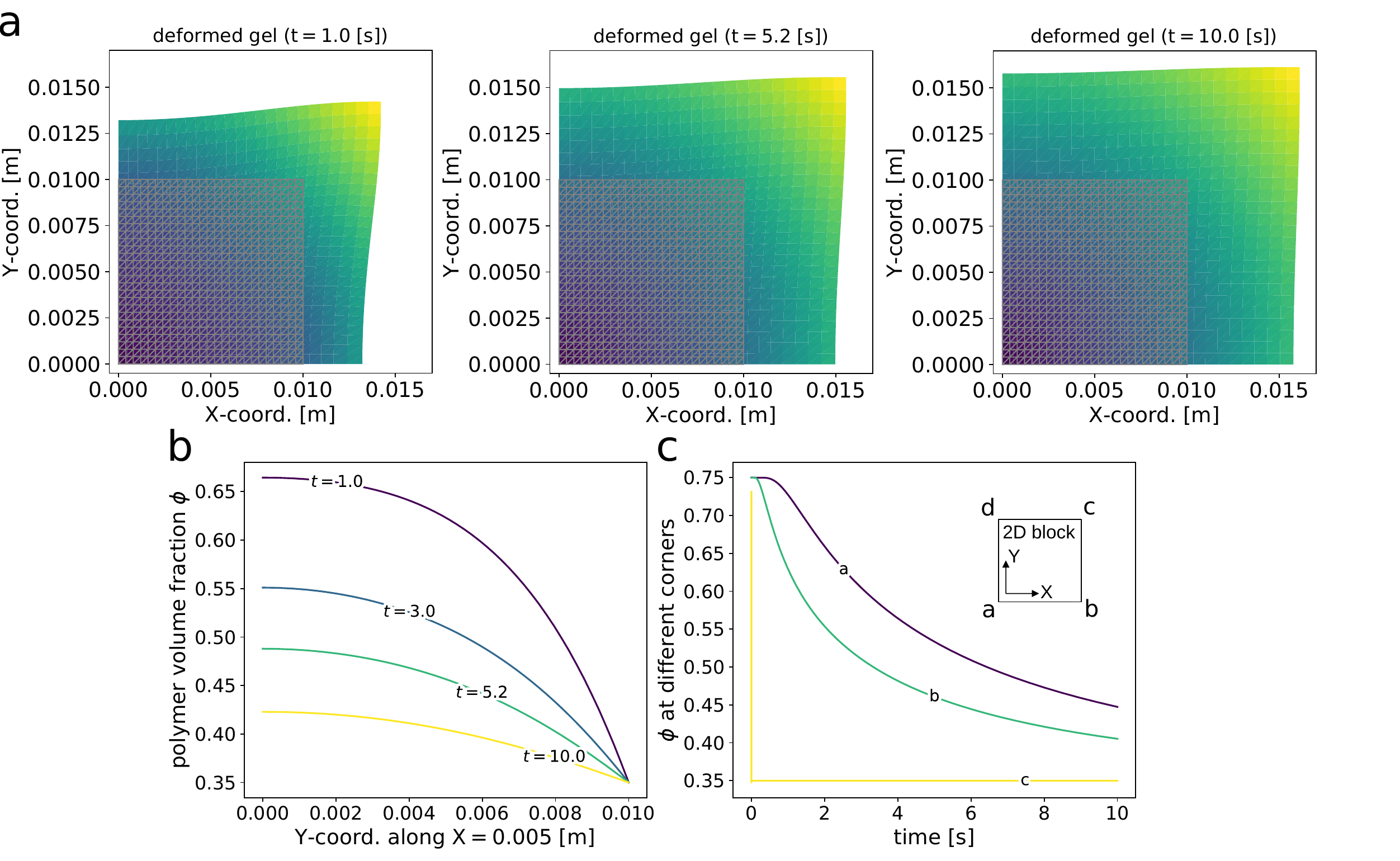}
\caption{\textbf{\emph{Constitutive model V}:} \textbf{Two-dimensional hydrogel block of an initially square cross-section immersed in a non-reactive solvent at different simulation times.} \textbf{a.} Deformed two-dimensional block at three different time steps $t = 1.0~[s]$, $t = 5.0~[s]$, and $t = 10.0~[s]$.
\textbf{b.} Polymer volume fraction $\phi$ at different times across $X = 0.005~[m]$. \textbf{c.} Transient evolution of $\phi$ measured at different corners of the two-dimensional block. 
\textbf{Simulation parameters:} $G_0 = 1~[MPa], ~~ K = 100~[MPa], ~~ \chi = 0.2~[--], ~~ D = 7.5 \times 10^{-9}~[m^2 s^{-1}], ~~ \alpha_r = 10.0~[--]$.}
\label{fig:2D_block_Chester_2015}
\end{figure}

The simulation results obtained for \textit{constitutive models IV and V} are displayed in Figures \ref{fig:2D_block_Chester_2011} and \ref{fig:2D_block_Chester_2015}, respectively. From these results, it becomes evident that both constitutive models lead to a similar level of deformation and time evolution of $\phi$. 

\section{A reference benchmark problem}\label{sec:benchmark}

In this section, the two-dimensional square block problem is studied again as a unified benchmark problem for the diffusion-deformation of hydrogels. As a matter of fact, previous results cannot be directly compared due to the disparate value of the material parameters selected by authors in the original papers (and adopted for the prototype problems presented in the previous section). Our goal here is to have a unique reference simulation example and show the differences in the response for each constitutive equation. We are interested in testing how much the diffusion-deformation behavior is affected by the different constitutive choices under the same simulation setup.

Hydrogels are a type of material that can vary significantly in their physical properties due to factors such as the particular polymer used, the degree of crosslinking, and the presence of any additives \citep{Caccavo2018hydrogels}. As such, the shear modulus ($G_0$) and bulk modulus ($K$) can have a wide range of values. Typically, the shear modulus of hydrogels can range from $1 ~ [Pa]$ to $1 ~ [MPa]$, with softer, more water-rich gels being at the lower end of the scale and more crosslinked or polymer-rich gels being at the higher end. Among the studies considered in this research, we found very different values for the shear modulus ranging from $0.1 ~ [MPa]$ in \cite{Chester2010DiffDeform,Chester2011thermo} to $10 ~ [MPa]$ as reported by \cite{Liu2010buclingGels}. 

Similarly, the bulk modulus, which is a measure of a material's resistance to uniform compression, can also vary widely for hydrogels. The bulk modulus is typically larger than the shear modulus. For many hydrogels, the bulk modulus can range from around $10 ~ [kPa]$ to several $[MPa]$. However, these values should only be used as a rough guide, as the specific values for any particular hydrogel can vary significantly depending on its formulation and preparation details. One needs to refer to specific experimental data for the particular hydrogel of interest for a precise value. The most common value for $K$ reported in the reviewed papers was $K = 100~G_0$.

In the present study, motivated by state-of-the-art values, we choose $G_0 = 1 ~ [MPa]$ and $K = 100 ~ [MPa]$. The diffusion coefficient is used as the free parameter to tune the simulation such that the results become comparable and the numerical stability is guaranteed. 
All other parameters remain the same as for the previously reported simulation results. Notice that this simulation setup is similar to the one used to solve \textit{constitutive models II, and IV - V} and it was inspired by \cite{Chester2015Abaqus}.

\begin{figure}{H}
\centering
\includegraphics[width=1.0\textwidth]{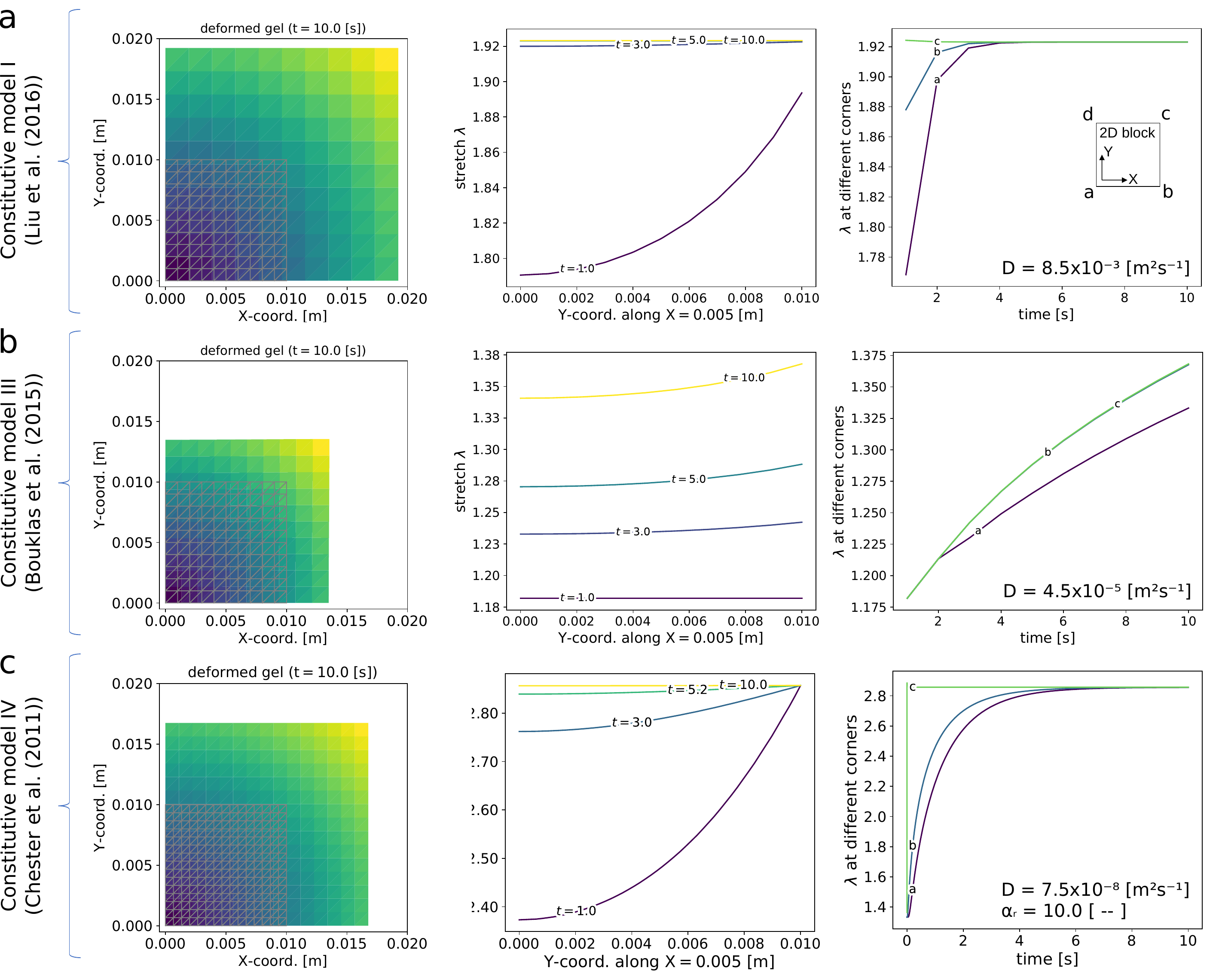}
\caption{\textbf{\emph{Reference benchmark}:} \textbf{two-dimensional hydrogel block of an initially square cross-section immersed in a non-reactive solvent at different simulation times.} \textbf{a.} \textit{constitutive model I}, \textbf{b.} \textit{constitutive model III}, and \textbf{c.} \textit{constitutive model IV}.
\textbf{Common simulation parameters:} $G_0 = 1~[MPa], ~~ K = 100~[MPa], ~~ \chi = 0.2~[--]$.}
\label{fig:2D_block_Models_comparison}
\end{figure}

Figure \ref{fig:2D_block_Models_comparison} shows the simulation results for the two-dimensional square block problem. Results are reported for \textit{constitutive models I, III, and IV}. The reason for this selection is twofold: \textbf{i.} \textit{constitutive model II} introduces an additional field represented by the Lagrange multiplier, thus requiring ad hoc numerical treatments outside of the scopes of present work;
\textbf{ii.} the same example was solved for \textit{constitutive model V} with the same parameters and results in Figure \ref{fig:2D_block_Models_comparison} can be directly compared with those plotted in Figure \ref{fig:2D_block_Chester_2015}.

From Figure \ref{fig:2D_block_Models_comparison}, first column, it is possible to observe that the level of deformation achieved in each case by the end of the simulation is different despite the similarity of the parameters. However, it is not on the deformation or stretch where the main difference between the different models becomes evident. It is on the transient evolution of the solvent as appreciated in the second and third columns of the figure.
\textit{Constitutive models I and IV} reach their steady state in about $4.0~[s]$ (Figure \ref{fig:2D_block_Models_comparison}\textbf{a}, third column) and $6.0~[s]$ (Figure \ref{fig:2D_block_Models_comparison}\textbf{c}, third column), respectively, whereas \textit{constitutive model III} is far from reaching the steady state (Figure \ref{fig:2D_block_Models_comparison}\textbf{b}, third column). 

It is important to note that the diffusivity values used in the benchmark problem for \textit{constitutive models I} and \textit{III} are the lowest possible before encountering numerical issues. For \textit{constitutive models III - IV}, it is feasible to set the diffusivity to $8.5\times10^{-3} [m^2s^{-1}]$, matching \textit{constitutive model I}. However, given our simulation setup, this is an exceptionally high value. Consequently, the system will reach steady in both scenarios almost immediately, eliminating any transient behavior. In summary, the mentioned diffusivity values differ significantly from each other. They belong to different time scales. This difference can be attributed to the distinct solution strategies adopted for solving the diffusion equation in each case. We delved into the specifics of this when introducing each \textit{constitutive model}.

One takeaway drawn from this comparison study is that further refinement may be required for the constitutive models such that closer predictions are achieved.
In particular, each model predicts a different level of deformation and requires a fine-tuning of some parameters to ensure numerical convergence. This can complicate the model's experimental validation because it could lead to different values of the material parameters and, for instance, ambiguity regarding which values are correct and their interpretability. Nonetheless, there is still room for improvement in selecting the energetic component of the constitutive model, which can fix the issue. In the end, all constitutive models capture the diffusion-deformation process in a reasonable way and offer valuable insights into the coupled problem.

\section{Conclusion}
\label{sec:conclusion}

This study presented a detailed classification and analysis of various nonlinear models that depict the diffusion-deformation process in hydrogels caused by non-reactive solvent absorption. We have consolidated these theories into a unified framework, demonstrating that, despite not being evident, all theories follow equivalent thermodynamic arguments. For instance, while having a common set of governing equations, each model showcases differences in the enforcement of incompressibility and formulation of the constitutive model — particularly in terms of the free energy function's components, mainly at the energetic level.
At present, further research appears necessary to conveniently account for the energetic component, ultimately aiming for a cohesive and unified constitutive model.

Different numerical approaches adopted by various researchers to solve these models were analyzed.  Our implemented numerical methods presented reasonable predictions for the diffusion-deformation process. However, our findings suggest that the superiority of one strategy over another remains inconclusive. The selection of an appropriate model should be grounded in a comprehensive understanding of the hydrogel's composition and must be validated experimentally. Notice that some of the theories here introduced have been extended to account for temperature variations, chemical reactions, and damage (refer to, e.g., \cite{Sain2018CuringPolymers,Mao2018DamageGels,Konica2020OxidationGels,Hajikhani2021chemomechanics} for more details). So, they represent a solid foundation for the study of elastomeric materials.

We emphasized on discerning the differences among the leading models in the literature and verifying whether the results presented by different authors are still valid in light of open-source general-purpose software available nowadays, such as FEniCS. 
To this end, an important part of this work is Section \ref{sec:numerics} with the mathematical classification and resulting numerical discretization and algorithms. These allowed us to carefully design mathematical formulations, which were then implemented and used to study the previously mentioned models.

Advances in automated solution techniques for the finite element method (FEM), such as FEniCS, provide the user with a streamlined approach for solving systems of partial differential equations compared to traditional model development. Unlike models implemented on commercial software (e.g., Abaqus), the implementation of our models grants users considerable control over several components.  

The simulation results indicate that there is not a one-size-fits-all model compatible with the parameters across all scenarios. Depending on the specific problem configuration being simulated, adjustments to one or multiple parameters are necessary to guarantee the numerical stability of the solution. Our observations in Section \ref{sec:Sim_results} revealed a strong dependency of the numerical solution's convergence on the diffusion coefficient and bulk modulus value. In fact, very different diffusion coefficients for each example featured in this study were required. Nonetheless, once a suitable value for the material parameters is identified, the numerical solution achieves reliable accuracy and robustness, as evidenced by the convergence analysis. Here, a consistent number of Newton iterations was displayed along the simulation time to reach absolute and relative errors smaller than $10^{-10}$. Moreover, concerning the spatial and temporal discretization, quadratic and linear convergence orders, respectively, for the FEM and Euler schemes for both the monolithic and staggered approaches were observed.

Furthermore, \textit{constitutive model III} stands to be the more fragile in terms of numerical stability.
Because \textit{constitutive model III} directly penalizes the material compressibility, it calls for a numerical solution using a mixed formulation where the stress becomes a primary variable as it is standard in the literature dealing with incompressible materials (see, e.g., \cite{Brink1996MixedIncompressible} or \cite{Pantuso1997StabilityMixedIncompressible}).

Another observation was regarding the application of boundary conditions. Depending on the specific simulation scenario, it might be necessary to adopt a time-ramping strategy to prevent numerical instabilities, which is well known in continuum mechanics and due to the mathematical functional framework in order to have compatible conditions. However, these mathematical assumptions might not be met in many engineering applications. Therefore, one must be careful with any assumption made while numerically solving the problem at hand to avoid unphysical numerical results.

As an overall outcome, in Section \ref{sec:Sim_results}, we found that each model presents diverse deformation states and solute concentrations within the hydrogel, highlighting the complexity of the investigated problem. It is evident that pinpointing a single appropriate model to describe the diffusion-deformation of hydrogels remains a challenge, given the different calibration mechanisms each offers. 
These differences underscore the need for more comprehensive experimental data to reconcile these theoretical distinctions with actual observations. 
With the models' validation, we can transition from merely describing the process to predicting hydrogel behavior, thereby using the model for designing new materials or optimizing the mechanical properties of existing ones.

Some efforts have been made to validate the diffusion-deformation process of hydrogels as predicted by the coupled model. For example, \cite{Chen2020LinearHydrogelsExp} performed a validation of a very similar theory as presented by \cite{Bouklas2012Linear-NL} with a major focus on the linear theory. Several measurements of the strain experienced by a gelatin-glycerol-water hydrogel under free-swelling conditions are reported, and the best-fitting parameters are identified. Neither the time evolution of the diffusion process nor the internal stress were investigated. \cite{Bosnjak2020ExperimentsHydrogels,Alkhoury2022ExperimentsThermoHydrogels} performed an experimental work to validate an extension of the original models presented by \cite{Chester2015Abaqus}. The extended models account for the viscoelastic response of elastomeric gels under isothermal and non-isothermal conditions. Again, the transient behavior of the diffusion process is neglected, and only the stresses are computed for the steady state under external loading conditions. Consequently, no study has been carried out to validate the models as presented in this study.

\backmatter

\bmhead{Supplementary information}
No supplementary information was produced from this research.

\bmhead{Acknowledgments}
MM acknowledges the funding of Regione Lazio (project n. A0375-2020-36756, Progetti di Gruppi di Ricerca 2020, POR FESR LAZIO 2014) and the Italian National Group for Mathematical Physics GNFM-INdAM.

\section*{Declarations}

\textbf{Funding:}
The authors declare that no funds, grants, or other support were received during the preparation of this manuscript.

\noindent\textbf{Conflict of interest/Competing interests:}
On behalf of all authors, the corresponding author states that there is no conflict of interest.

\noindent\textbf{Code availability:}
Codes to reproduce the results presented in this manuscript can be freely accessed at the following link: \url{https://doi.org/10.25835/5v49yfk0}. 








\newpage
\begin{appendices}

\section{Alternative forms: the notion of active chemical potential}\label{sec:alternative-forms}

The total deformation gradient ${\bf F}$ depends on fluid concentration $c_R$ from equation \eqref{eq:F-tot}. To better distnguish between elasti- and fluid-related effects, the stress power can be reformulated as:
\begin{equation}\label{eq:stress-power}
    {\bf P} : \dot{\bf F} = J_f {\bf P}_e : \dot{\bf F}_e + \dfrac{1}{3} \text{tr} ( {\bf M}_e ) \dot{J}_f,
\end{equation}
where ${\bf P}_e$ represents the PK1 stress tensor from the intermediate to the current configuration (also referred to as the elastic PK1) and ${\bf M}_e$ is the Mandel stress tensor in the intermediate configuration\begin{footnote}{In the state-of-the-art, two different stress measures can be found in the present context, namely ${\bf P}_e^R=J_f{\bf P}_e$ and ${\bf M}_e^R=J_f{\bf M}_e$, \citep{Chester2010DiffDeform}. These definitions are analogous but the corresponding stress measures refer to the reference configuration.}\end{footnote}:
\begin{equation} \label{eq:new-stress}
    {\bf P}_e = J_f^{-2/3} {\bf P} = J_e \bm{\sigma} {\bf F}_e^{-T}\quad \text{and} \quad {\bf M}_e = J_e {\bf F}_e^{T} \bm{\sigma} {\bf F}_e^{-T} = {\bf F}_e^{T}{\bf P}_e.
\end{equation}
At this standpoint, a mean normal pressure $\bar{p}$ can be introduced,
\begin{equation} \label{eq:bar-p}
    \bar{p} = - \frac{1}{3} \text{tr} ( {\bf M}_e ) = - \frac{1}{3} J_e \text{tr} ( \bm{\sigma} ) = - \frac{1}{3} {\bf P}_e : {\bf F}_e\, ,
\end{equation}
representing the pull-back of the hydrostatic pressure from the current to the intermediate configuration. Finally, upon enforcing the kinematic constraint defined in equation \eqref{eq:kinematic_constraint}, the stress-power equation \eqref{eq:stress-power} may be written as:
\begin{equation}\label{eq:stress-power-final}
    {\bf P} : \dot{\bf F} = {\bf P}_e : \dot{\bf F}_e - \bar{p} \dot{J}_f = {\bf P}_e : \dot{\bf F}_e - \Omega \bar{p} \dot{c}_R \, .
\end{equation}

Hence, the local form of the second law of thermodynamics, equation \eqref{eq:secondlaw-0}, reads also as:
\begin{equation}\label{eq:secondlaw-0-new}
    {\bf P}_e : \dot{\bf F}_e + (\mu - \Omega \bar{p}) \dot{c}_R - {\bf J}_R \cdot \text{Grad}(\mu) - \dot{\psi}_R \geq 0 \,,
\end{equation}
from which it appears convenient defining the free energy density function as function of the elastic deformation ${\bf F}_e$ and fluid concentration $c_R$, that is 
\begin{equation}\label{eq:Psi-R-e}
    \psi_R = \Psi_R^e({\bf F}_e,c_R)\, .
\end{equation} 
By defining the active chemical potential $\mu_{act}$ as:
\begin{equation} \label{eq:mu-act-def}
    \mu_{act} = \mu - \Omega\bar{p}\, ,
\end{equation}
equation \eqref{eq:secondlaw-0-new} yields:
\begin{equation}\label{eq:secondlaw-final}
    J_f {\bf P}_e : \dot{\bf F}_e + \mu_{act} \dot{c}_R - {\bf J}_R \cdot \text{Grad}(\mu) - \frac{\partial \Psi_R^e}{\partial {\bf F}_e} : \dot{\bf F}_e - \frac{\partial \Psi_R^e}{\partial c_R} : \dot{c}_R \geq 0 \, ,
\end{equation}
from which the following constitutive choices follow for the \textbf{elastic PK1 stress tensor}:
\begin{equation} \label{eq:gen_PK1-el}
    \mathbf{P}_e = J_f^{-1} \dfrac{\partial \Psi_R^e}{\partial \mathbf{F}_e}\, ,
\end{equation}
and \textbf{the active chemical potential}:
\begin{equation}\label{eq:gen_mu-act}
    \mu_{act} = \frac{\partial \Psi_R^e}{\partial c_R}\, .
\end{equation}

It is straightforward to show that constitutive choices in equations \eqref{eq:gen_PK1} and \eqref{eq:gen_mu} are equivalent to the ones in equations \eqref{eq:gen_PK1-el} and \eqref{eq:gen_mu-act} given that:
\begin{equation} \label{eq:rel-Psi}
   \Psi_R^e({\bf F}_e, c_R)=\Psi_R(\overbrace{(1+\Omega c_R)^{1/3}}^{\displaystyle J_f^{1/3}}{\bf F}_e,c_R)\, .
\end{equation}
In fact, by employing the chain rule and considering equation \eqref{eq:rel-Psi}, it results from equation \eqref{eq:gen_PK1-el}:
\begin{equation} \label{eq:P-eq}
  {\bf P}_e = J_f^{-1}\frac{\partial \Psi_R^e}{\partial {\bf F}_e} = J_f^{-1}\left(J_f^{1/3} \frac{\partial \Psi_R}{\partial {\bf F}} \right)=  J_f^{-2/3} \frac{\partial \Psi_R}{\partial {\bf F}} \, ,
\end{equation}
where it follows that ${\bf P} = \partial \Psi_R/\partial {\bf F}$ by definition (cf. equation \eqref{eq:new-stress}), recovering the constitutive relationship for the PK1 stress tensor in equation \eqref{eq:gen_PK1}. 

Moreover, it results from equation\eqref{eq:bar-p}, \eqref{eq:mu-act-def} and \eqref{eq:gen_mu-act} that:
\begin{equation}
  \mu = \frac{\partial \Psi_R^e}{\partial c_R} - \frac{1}{3} \Omega {\bf P}_e : {\bf F}_e \, ,
\end{equation}
where, considering equation \eqref{eq:rel-Psi}, the first term reads also as (accounting for the dependency ${\bf F}={\bf F}(c_R)$ from equation \eqref{eq:F-tot}):
\begin{equation}
  \frac{\partial \Psi_R^e}{\partial c_R} = \frac{\partial \Psi_R}{\partial {\bf F}} : \frac{\partial {\bf F}}{\partial c_R} +  \frac{\partial \Psi_R}{\partial c_R}  \, ,
\end{equation}
while the second term as (with equation \eqref{eq:P-eq}):
\begin{equation}
\frac{1}{3} \Omega {\bf P}_e : {\bf F}_e = \frac{1}{3} \Omega J_f^{-1} \frac{\partial \Psi_R^e}{\partial {\bf F}_e} : {\bf F}_e = \frac{\partial \Psi_R}{\partial {\bf F}} : \left( \frac{1}{3} \Omega J_f^{-2/3} {\bf F}_e \right) = \frac{\partial \Psi_R}{\partial {\bf F}} : \frac{\partial {\bf F}}{\partial c_R}  \, .
\end{equation}
Hence, since:
\begin{equation}
  \mu = \frac{\partial \Psi_R^e}{\partial c_R} - \frac{1}{3} \Omega {\bf P}_e : {\bf F}_e = \left(\frac{\partial \Psi_R}{\partial {\bf F}} : \frac{\partial {\bf F}}{\partial c_R} +  \frac{\partial \Psi_R}{\partial c_R}\right) -\left( \frac{\partial \Psi_R}{\partial {\bf F}} : \frac{\partial {\bf F}}{\partial c_R}\right) = \frac{\partial \Psi_R}{\partial c_R}   \, ,
\end{equation}
the chemical potential computed from equation \eqref{eq:gen_mu-act} is equivalent to the one from equation \eqref{eq:gen_mu}, given that equation \eqref{eq:rel-Psi} holds true.




\end{appendices}


\bibliography{sn-bibliography}

\end{document}